\documentclass[aps,prb,reprint,superscriptaddress]{revtex4-1}
\usepackage[dvipdfmx]{graphicx} 
\usepackage{url}
\usepackage{dcolumn}
\usepackage{bm}
\usepackage{braket}
\usepackage{amsmath}
\usepackage{here}   
\usepackage{amsmath}
\usepackage{amsfonts}   
\usepackage{ulem}
\usepackage{xcolor}
\definecolor{green}{rgb}{0,0.7,0.3}

\begin{document}
\draft
\title{Quantum mechanically driven structural-spin glass \\ 
in two dimensions at finite temperature}
\author{Chisa Hotta$^{1*}$ , Kazumasa Ueda$^{1}$, and Masatoshi Imada$^{2,3}$ }
\address{$^1$ Department of Basic Science, University of Tokyo, Tokyo, 153-8902, Japan}
\email[]{chisa@phys.c.u-tokyo.ac.jp}
\address{$^2$ Toyota Physical and Chemical Research Institute, Yokomichi, Nagakute, Aichi 480-1192, Japan.}
\address{$^3$ Research Institute for Science and Engineering, Waseda University, Okubo, Shinjuku-ku, Tokyo 169-8555, Japan.}
\date{today}
\begin{abstract}
In magnetic materials, spins sometimes freeze into spatially disordered glassy states. 
Glass forming liquids or structural glasses are found very often in three dimensions. 
However, in two dimensions(2D) it is believed that both spin glass and structural glass can never exist at a finite temperature because they are destroyed by thermal fluctuations. 
Using a large-scale quantum Monte Carlo simulation, we discover a quantum-mechanically driven 2D glass phase at finite temperatures. 
Our platform is an Ising spin model with a quantum transverse field on a frustrated triangular lattice. 
How the present glass phase is formed is understood by the following three steps. 
First, by the interplay of geometrical frustration and quantum fluctuation, 
part of the spins spontaneously form an antiferromagnetic honeycomb spin-superstructure. 
Then, small randomness in the bond interaction works as a relevant perturbation to this superstructure and breaks it up into {\it domains}, making it a structural glass. 
The glassiness of the superstructure, in turn, generates an emergent random magnetic field 
acting on the remaining fluctuating spins and freezes them. 
The shape of domains thus formed depends sensitively on the quenching process, 
which is one of the characteristic features of glass, 
originating from a multi-valley free-energy landscape. 
The present system consists only of {\it a single} bistable Ising degree of freedom, 
which naturally does not become a structural glass alone nor a spin glass alone. 
Nevertheless, a glass having both types of nature emerges in the form of coexisting two-component glasses, 
algebraic structural-glass and long-range ordered spin-glass. 
This new concept of glass-forming mechanism opens a way to realize functional glasses even 
in low dimensional systems. 
\end{abstract}
\maketitle
\narrowtext
\section{Introduction} 
\label{sec:intro}
Structural glasses are non-crystalline amorphous materials used in our daily life such 
as window glasses and plastics\cite{berthier11}. 
From an academic point of view, understanding the nature of glass or glass-like behavior 
is a topic that attracts interest in many fields including material science, biochemistry, and information networks. 

A glass phase is conceptually defined as a frozen, thermodynamically stable, 
and disordered configuration of particles or molecules. 
However, there had been a long-standing debate on whether such a phase exists or not, 
since it is practically impossible to exclude a possibility that the glassy state 
finally relaxes, namely remains unfrozen, after an enormous timescale beyond the measurements. 

At the same time, it is believed that the glass phase has a multi-valley free energy landscape with enormous 
numbers of nearly degenerate minima. 
These minima represent the states having completely different configurations of particles in a continuum space, 
and to which minimum the equilibrated state reaches depends sensitively on quenching or cooling processes. 
This non-reproducibility is one of the features of glass. 
It can also be referred to as a non-ergodicity of dynamics. 
Realizing or identifying a glass defined in the above context is not easy; 
for example, if there is a random external potential in the system, 
particles are trapped in {\it a single potential valley} and freeze. 
However, this state cannot be regarded as glass, since there is no multi-valley in free energy, 
and the configuration of particles is uniquely determined by the types of potential. 
Although substantial theoretical progress has been made\cite{berthier11,parisi17} 
with a successful example in a mean-field theory in infinite dimensions\cite{kurchan12}, 
identification of a structural glass phase satisfying the above-mentioned definition in realistic spatial dimensions $d=1$ to $3$, 
especially in $d \leq 2$, 
is still a big challenge. 
\par
On the other hand, 
there exist several studies exploring models on discrete lattices that reproduce essential features of 
structural glasses in a continuum space\cite{biroli01,ciamarra03}. 
In a model of interacting particles that can occupy lattice sites, 
the {\it discrete} translational invariance 
can be lost and the particles freeze by randomly occupying lattice sites. 
This freezing mimics a vitrified charge density wave, 
where the electron density shows an irregular spatial pattern in crystalline solids. 
We classify them all together as lattice-structural glass. 
Although there exist reports on possible experimental realizations of charge glasses 
in 2D organic solids\cite{kagawa13,hashimoto17}, 
the existence of structural glass phase in 2D at nonzero temperatures 
is theoretically unlikely\cite{berthier19}. 
\par
Another important type of glass in solids is a spin glass (SG) defined on periodic lattices\cite{binder-young}. 
Historically, finding the SG in theory 
is as challenging as finding a structural glass transition. 
For the Edwards Anderson (EA) model built as an idealized platform for SG\cite{ea1975}, 
there had been a controversy over fifty years. 
It finally converged to an overall consensus that the SG phase can exist at a nonzero temperature in 3D 
\cite{ogielski85,katzgraber06,bray-moore85,bhatt-young88,bhatt85,kawashima96,palassini99,mari-campbell99,ballesteros00,nakamura10}, 
whereas it is absent in 2D\cite{young83,parisi98,mcmillan83,houdayer01}. 
Notice that if we place an {\it on-site} random magnetic field, 
a frozen thermodynamically stable disordered orientational configuration of spins 
can easily occur, which looks like a SG. 
We, however, exclude such state from a SG\cite{binder-young}, 
since it does not retain the multi-valley free energy structure characteristic of the glass, 
similarly to the aforementioned particle system with random potential. 
To avoid complications, it is natural to confine ourselves to systems that originally preserve a time-reversal symmetry (TRS),  or equivalently, the symmetry about turning over all the spins simultaneously. 
In such a case, the SG phase can be detected by the spontaneously breaking the TRS of spins in the same context 
that the structural glass appears by spontaneously breaking the translational symmetry of locations of particles. 
Notice, however, that the TRS breaking is not necessarily essential for the breaking of ergodicity in SG; 
it is known that the SG happens in the presence of a {\it uniform} external field at least 
in $d\ge 4$\cite{almeida78,larson13,baity-jesi14,holler20,paga21}. 
Experimentally, the SG transition is identified as a cusp in the uniform magnetic susceptibility 
together with the divergent nonlinear susceptibility\cite{suzuki77,gunnarsson91,hasenbusch2008,baity-jesi13} 
in materials such as dilute metallic alloy\cite{cannella72} and Y$_2$Mo$_2$O$_7$\cite{gingras97}. 
\par
\begin{figure}[t]
\includegraphics[width=8.5cm]{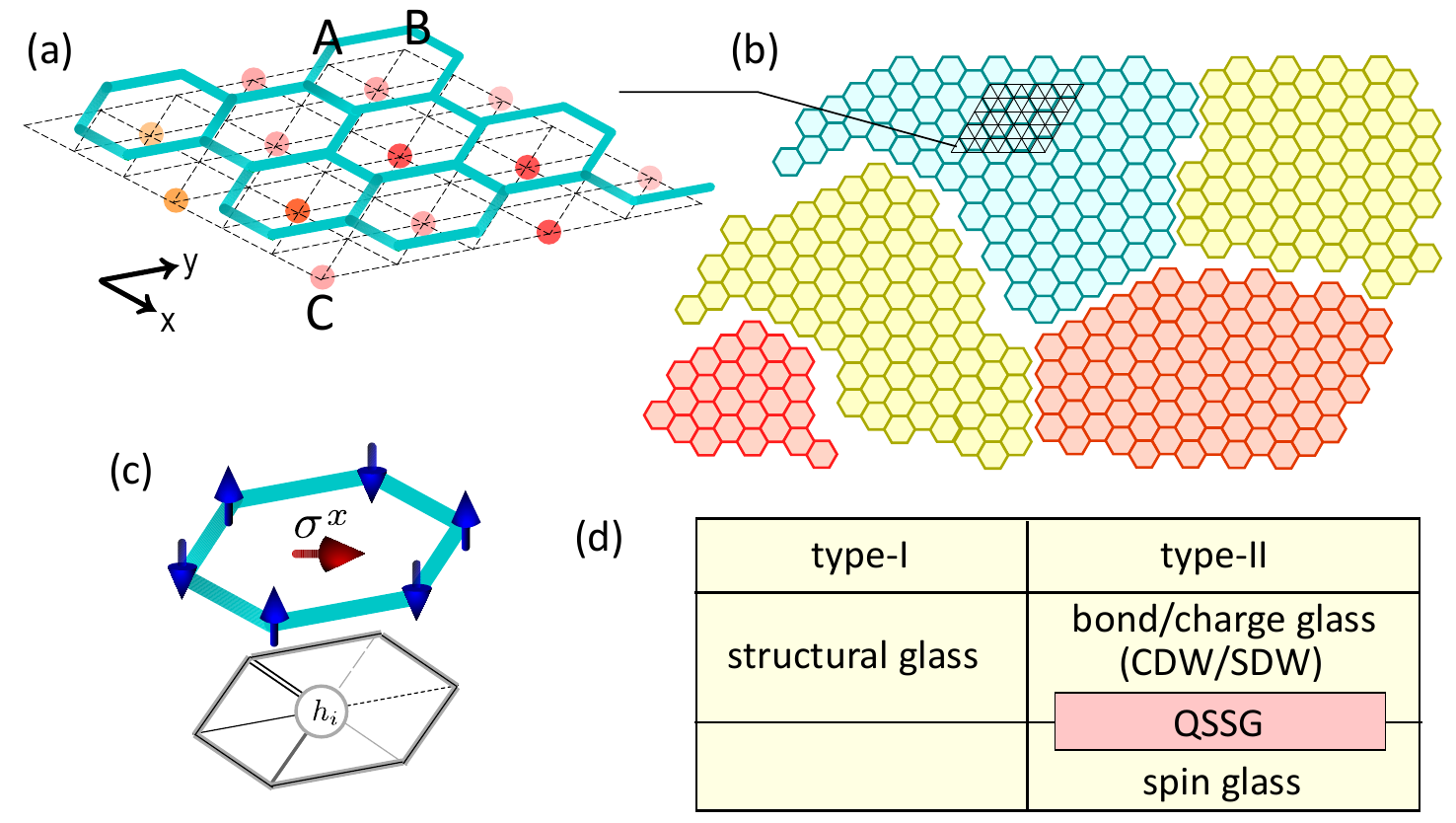}
\caption{(a) Triangular lattice $L\times L$ cluster on which our bond random transverse Ising model is built. 
The blue bonds mark the honeycomb superstructure of the A-B sublattice antiferromagnetic ordering 
in the clock phase, where the C-spins align in the $x$-direction to gain the transverse field.  
(b) Vitrification of the honeycomb structure by breaking up into domains in the QSSG phase. 
Inside each domain, the honeycomb structure is kept. 
Numerous patterns of domains are possible in the equilibrated state for a given set of bond randomness. 
(c) Hexagonal unit of the QSSG phase inside the domain with emergent random field $h_i$ at the center. 
(d) Classification of glasses: type-I in the continuum and type-II on a periodic lattice. 
CDW, SDW, and 3QSSG stand for charge density wave, spin density wave, and quantum structural-spin glass, respectively.
}
\label{f1}
\end{figure}
\par
In this way, both structural and spin glasses 
defined in continuum and on discrete lattices are hardly realized at finite temperature 
in spatial dimensions $d\le 2$. 
In three or higher dimensions, a larger contact area or coordination number of particles or spins 
helps them to act on each other as stable disordered mean fields necessary for the freezing, 
but in $d\le 2$, glasses are easily destroyed by fluctuations. 
Therefore, a realization of the glass in low-spatial dimensions remains a fundamental challenge in physics. 
At the same time, it widens possibilities of the functionality of devices, 
since in lower dimensions, it is simpler to manipulate and synthesize realistic systems. 

In this paper, we show numerical evidence that there can be a quantum-mechanically driven 
structural-spin glass (QSSG) in 2D even at finite temperatures. 
We choose the Ising model with a transverse field on a triangular lattice as a platform for our calculations. 
In this model, we introduce the bond randomness which preserves the TRS. 
Bond randomness is believed not to cause a 2D SG at nonzero temperatures, 
and further, it has no reason to cause a structural glass. 
However, in our quantum phase at low temperature, a behavior characteristic of glass appears. 

Figures~\ref{f1}(a)-\ref{f1}(c) illustrate our central idea consisting of the following three steps: 
(1) When the system does not have any randomness in our model, 
a magnetic super-lattice honeycomb structure called the clock order 
is spontaneously induced by quantum fluctuations (Fig.~\ref{f1}(a)). 
(2) Next, we introduce small bond randomness. 
Although the TRS is preserved, it works as an emergent ``on-bond random field" 
for the super-lattice and converts the clock order to a lattice-structural glass 
by accompanying a domain formation (Fig.~\ref{f1}(b)). 
(3) The lattice-structural disorder in the QSSG phase produces an effective 
``on-site random magnetic field" (Fig.~\ref{f1}(c)) which acts on the spins and generates an SG. 
We show that such {\it emergent} ``random fields" 
essentially differs from {\it an external random field} 
in that, they realize a multi-valley minimum of free energy. 
Although the structural glass and SG can be hardly realized separately in 2D, 
when combined, they stabilize each other and form a QSSG at nonzero temperature. 

We remind here that the transverse-field Ising model is not only a canonical model for 
quantum computer science\cite{annealing}, but also describes a wide class of systems or materials 
represented by coupled particles on a lattice; 
each localized particle can quantum mechanically tunnel in a bistable structure from 
one potential minimum to the other, such as the hydrogen bonding\cite{hydrogen} 
and organic dimer ferroelectrics\cite{ch,naka,peter}. 
The concept of QSSG may thus apply to materials other than magnets. 
\par
We end our discussion by classifying the types of glasses as shown in Fig.\ref{f1}(d). 
The conventional structural glass in continuum space is denoted as type-I. 
As mentioned earlier, the structural glass is also defined on a periodic lattice, 
which we categorize as type-II structural glass. 
To our knowledge, there are few established examples of type-II structural glass. 

The spin/charge density waves are widely observed emergent phases of matter of electrons in solids, 
and when the super-lattice order of these density waves 
are vitrified, they become a type-II structural glass. 
Possible candidates would be the glassy phase of manganese oxides\cite{Tokura} 
and $\theta$-(BEDT-TTF)$_2$RbZn(SCN)$_4$\cite{kagawa13,hashimoto17}. 
So far, the origin of their glassy behavior is not experimentally clarified yet. 
The SG is also a type-II glass. 

In terms of this classification, our finding, the QSSG, is a combination of type-II structural glass and a type-II SG. 
The QSSG differs from previously studied cooperative paramagnets such 
as spin ice\cite{spinice,spinice2}, the classical SG\cite{binder-young}, long-range entangled spin liquids\cite{sl}, 
and the random singlet phase\cite{imada87,vojta10,kawamura14-1,kawamura15, kimuchi18,shao18,sheng19}, 
because the QSSG is a consequence of the synergy of the emergent structural degrees of freedom 
with the spins. 
Most importantly, these two show different types of vitrification, 
the algebraic quasi-long ranged and long-range ordered ones, respectively, 
as we clarify in this paper. 
\par 
The paper is organized as follows: 
we introduce the model in \S.\ref{sec:modelsystem}, 
and in \S.\ref{sec:qssg} the main numerical results are presented, 
starting from the phase diagram and disclosing the SG quantities. 
Those who want to grab the overall results can first view \S.\ref{sec:modelsystem} and 
\S.\ref{sec:qssg} F. 
Then, in \S. \ref{sec:domain}, we show the results indicating the formation of domains, 
followed by the discussion on the physical implications of our findings \S.\ref{sec:discussion}, 
and finally the paper is summarized in \S.\ref{sec:summary}. 
%
\section{Model system} 
\label{sec:modelsystem}
We consider an antiferromagnetic transverse Ising model 
on a triangular lattice with random nearest neighbor exchange interaction, $J_{ij}$($>0$), 
whose Hamiltonian reads
\begin{align}
H=&\sum_{\langle i,j\rangle}J_{ij} \sigma_{i}^z\sigma_{j}^z
+\sum_{i=1}^N \Gamma \sigma_i^x,
\label{ham}
\end{align}
where $\sigma_i^\alpha$ ($\alpha=x,y,z$) is the Pauli operator on the site $i$ in $N$-site system, 
and $\Gamma$ is the transverse field that flips the spins up and down. 
The Ising interaction $J_{ij}$ obeys the bond-independent uniform distribution in the range 
$[J-R/2, J+R/2]$ with an antiferromagnetic mean value $J=1$, which is our energy unit. 
The distribution of randomness in $J_{ij}$ does not have any spatial correlation. 
\par
The two characteristic energy scales of bond interactions are the width of the distribution of randomness 
$R$ and the average of bond interactions, $J=\bar{J_{ij}}$. 
Historically, the SG has been studied extensively in the standard EA models 
(which has the same form of the Hamiltonian (\ref{ham}) at $\Gamma=0$) defined on a bipartite lattice 
where the interactions $J_{ij}$ are quenched and are 
randomly distributed typically following a Gaussian distribution around 
the mean value $J$ with a width $R$. 
The EA model is speculated to have the SG order only when the condition $|J| \lesssim R/2$ is satisfied\cite{Hasenbusch2007}.
Here, it is important to suppress regular magnetic orders such as the antiferromagnetic ones to stabilize the SG, which is the reason why a large $R/|J|$ is required on bipartite lattices. 
\par
We pursuit an alternative route to realize the SG, 
by keeping $|J|\gg R/2$ and by introducing a geometrical frustration to suppress regular magnetic orderings, 
because small randomness may be more easily realized in practical experimental conditions. 
A triangular lattice classical antiferromagnetic Ising model ($R=\Gamma=0$ in Eq.(\ref{ham})) 
behaves paramagnetic down to zero temperature because of a geometrical frustration effect
\cite{wannier}. 
This paramagnet is strongly correlated and always satisfies the condition of having one or two up spins on 
all the triangle elements. 
However, this local condition is not uniquely satisfied, and  
there appears a highly degenerate lowest-energy manifold of states, 
contributing to an order-$N$ residual entropy. 
Among them, some nontrivial types of orders can be selected when a small amount of fluctuations is introduced, 
which is called the ``order-by-disorder" effect \cite{villain80}. 
The transverse field proportional to $\Gamma$ in the Hamiltonian (\ref{ham}) 
works as such quantum mechanical fluctuations. 
\par
Computational details to simulate Eq.(\ref{ham}) are the following:
we perform a continuous imaginary time quantum Monte Carlo (QMC) simulation on 
a $N=L\times L$ cluster (see the broken lines in Fig.\ref{f1}(a)) 
for $L=12, 18, 24, 36, 48, 60$, and for $L=96$ for some selected quantities, 
with periodic boundary conditions in both directions\cite{ct-qmc}. 
A random sampling of $J_{ij}$ is taken typically over 40 samples, 
and for each sample, more than 10 replicas are calculated. 
Thanks to the simplicity of the model, the size scaling analysis 
can be performed up to $N\sim {\cal O}(1000)$, 
which fulfills a criterion required to conclude the existence of SG phase 
as experienced in numbers of classical Monte Carlo studies 
\cite{ogielski85,katzgraber06,bray-moore85,bhatt-young88,bhatt85,kawashima96,mari-campbell99,nakamura10,young83,parisi98,mcmillan83,houdayer01}.
We checked that the system equilibrates to a set of states that exhibits a single peak 
in their energy histgram\cite{mitsumoto21} (see Appendix\ref{app:relaxation}). 
\par
Our model has an important advantage in computational tractability. 
For other quantum models such as frustrated Heisenberg models, 
QMC suffers a serious sign problem, and only a few methods like 
exact diagonalization (ED) or variational wave-function methods such 
as conventional variational Monte Carlo (VMC), 
density matrix renormalization group (DMRG), and tensor networks (TN) apply to the lattice Hamiltonians. 
However, the available sizes are $N \lesssim 30$ in ED. 
In 2D, keeping an aspect ratio of the cluster close to 1 is 
important to pursue a proper size scaling analysis, 
whereas in DMRG often a one-dimensional-like cluster is chosen in favor of an open boundary condition 
with a maximal width still being roughly 14-site. 
In TN and VMC, the wave functions are assumed to be periodic 
to keep the number of parameters amenable to practically available computer power, 
and thus describing a random system remains a challenge. 
Another disadvantage for DMRG and TN is that they tend to choose a minimally entangled quantum states, 
and a quantum entanglement beyond the area law is hardly simulated. 
It is not clear whether a finite-temperature glassy 
quantum state can keep the scale of its entanglement within an area law. 
In contrast, the conventional path integral QMC for Eq.(\ref{ham}) with randomness, which we employ in the present study does not cause 
a sign problem and is essentially exact within the statistical error. 
Its computational cost scales linearly with the size $N$ so that we can afford a large-scale calculation. 
\par
The order parameter of the SG can be defined by a replica overlap, 
\begin{equation}
q_{\alpha\beta}=\frac{1}{N} \sum_i \sigma^z_{i;\alpha} \sigma^z_{i;\beta} \;
= \frac{1}{N} \sum_i q_{i;\alpha\beta},
\label{eq:replica}
\end{equation}
where we add another index to the spins as $\sigma_{i;\alpha}$, 
meaning that it belong to an $\alpha$-th replica, 
and introduce a local replica overlap, $q_{i;\alpha\beta}=\sigma^z_{i;\alpha} \sigma^z_{i;\beta}$. 
By preparing several replicas with the same configuration of randomness and by independently performing simulations, 
we take $\langle \cdots \rangle$ as the ensemble average together with the average over 
all choices of replicas pairs, $\alpha$ and $\beta$. 
One can see that if $\langle q_{\alpha\beta}^2\rangle \ne 0$ there is a freezing of configurations of spins. 
The equal imaginary-time SG susceptibility in the absense of dominant magnetic correlation defined 
without replicas is related to $q_{\alpha\beta}$ as 
\begin{equation}
\chi_{\rm SG}^0=\frac{1}{N} \sum_{i,j=1}^N \overline{ \langle \sigma_i^z\sigma_j^z \rangle ^2 }
= N  \overline{ \langle q_{\alpha\beta}^2\rangle}. 
\label{chisg0}
\end{equation}
Here, the average over samples of different random distributions, $\overline{ \cdots }$, are taken 
after $\langle \cdots \rangle$. 
We often abbreviate $\overline{ \cdots }$ for simplicity in the following. 
Equation (\ref{chisg0}) is applied to the standard SG phase without any competing magnetic orderings. 
To include the case of coexistent SG and magnetic orders, one needs to subtract 
the extra term coming from the finite averaged values $\langle \sigma_{i;\alpha} \rangle \ne 0$ 
or $\langle q_{\alpha\beta} \rangle \ne 0$, 
to properly extract the fluctuation about the ordered components. 
In addition, the competing magnetic order is not necessarily a spatially uniform one. 
Taking them into account, the susceptibility needs to be redefined as 
\begin{eqnarray}
\chi_{\rm SG}(\bm k)&=&\frac{1}{N} \sum_{i,j=1}^N e^{i\bm k (\bm r_i-\bm r_j)} 
\Big( \langle \sigma_i^z\sigma_j^z \rangle ^2 - \langle \sigma_i^z\rangle^2\langle \sigma_j^z \rangle ^2 \Big)
\label{chisgk}
\end{eqnarray}
and its $\bm k=0$ component yields the uniform SG susceptibility, 
\begin{equation}
\chi_{\rm SG}= N \Big( \langle q_{\alpha\beta}^2\rangle 
- \langle q_{\alpha\beta}\rangle^2 \Big) . 
\label{chisg}
\end{equation}
Notice that we may also find 
$\lim_{N\rightarrow\infty} \chi_{\rm SG}/N>0$ for a regular(non-random) magnetic ordering. 
Therefore, we need to carefully exclude this possibility to conclude the presence of SG order. 
We will show later that the regular (not SG) order is absent in our case 
by examining the spin correlation function. 
Details of deriving Eq.(\ref{chisg}) and other SG susceptibilies are discussed in Appendix \ref{app:chisg}. 
\par
For the evaluation of Berezinskii-Kosterlitz-Thouless(BKT) phase and 
a magnetically ordered phase on a triangular lattice called clock phase, we introduce 
a sublattice magnetization, $m_{\rm sub}$, as 
\begin{equation}
m_{\rm sub}  \equiv e^{i\theta}(m_A{\rm e}^{i 4\pi/3} + m_B {\rm e}^{-i 4\pi/3} + m_C )/\sqrt{3}, 
\label{eq:msub}
\end{equation}
where $m_{\rm sub} \ne 0$ indicates that some sort of three-sublattice magnetic structure is present. 
One can distribute the magnetization to the three sublattices in an arbitrary manner. 
Representative ones are $(m_A, m_B, m_C)\propto (1,-1,0)$ and $(1,1,-1)$, which are 
described by $\theta=n+\frac{1}{2}\pi$ and $n \pi$ with integer-$n$. 
These two states are partial order and ferrimagnetic order, respectively, 
and the former is realized in our case\cite{isakov03}. 
The susceptibility of the sublattice magnetization is given as 
\begin{equation}
\chi_{\rm sub}= L^2 \frac{\langle m_{\rm sub}^2\rangle }{k_BT}. 
\label{eq:chisub}
\end{equation}
%
\begin{figure}[tbp]
\includegraphics[width=8cm]{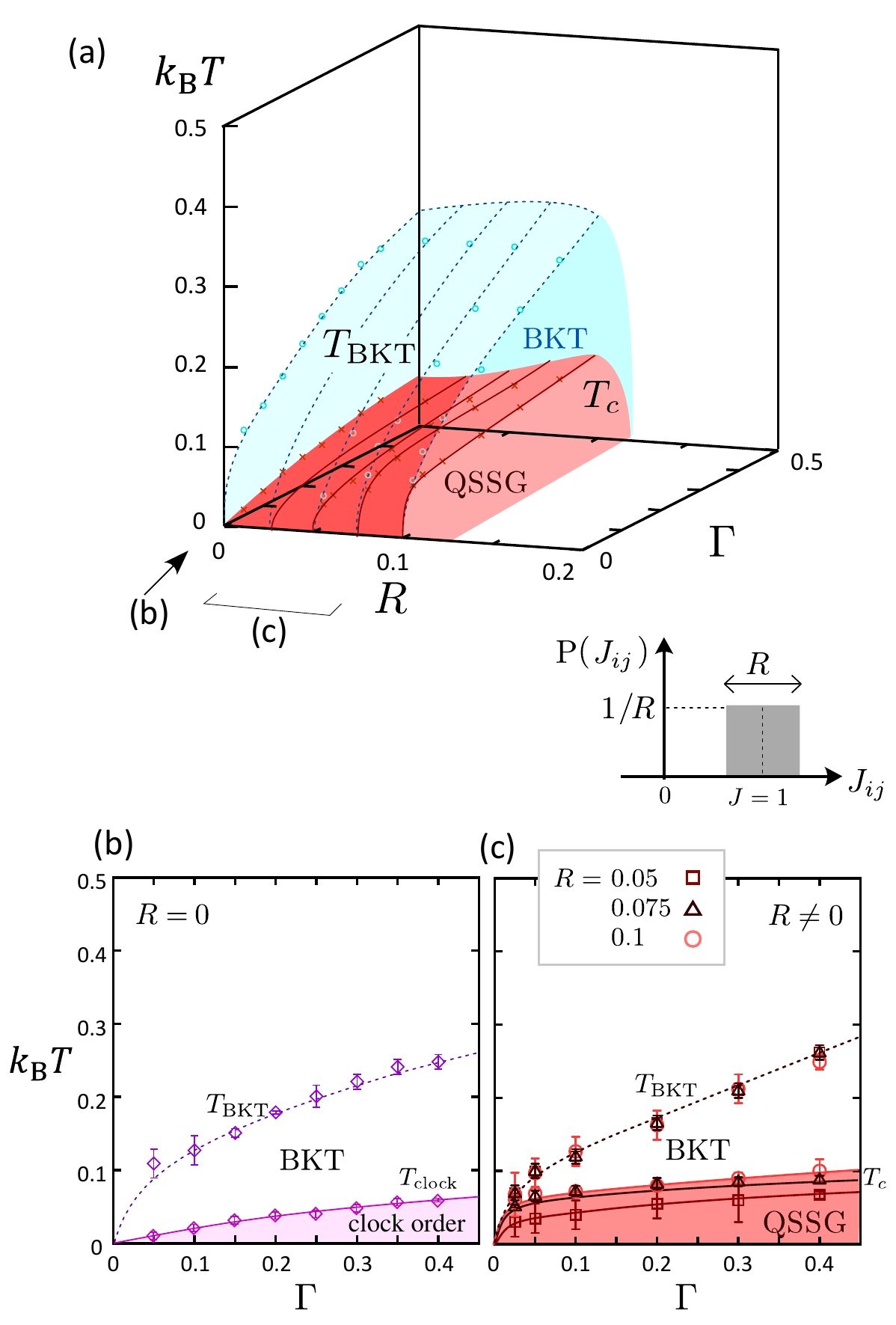}
\caption{
Phase diagram of the bond random transverse Ising model, 
(a) plotted for $\Gamma$, $R$ and $T$, and its crosssections 
(b,c) at $R=0$ and $R>0$. 
Blue and red planes are the BKT transition $T_{\rm BKT}$ and QSSG transition $T_c$, respectively. 
Inset: Rectangular probability distribution of random exchange interaction $J$. 
We determine $T_c$ from the finite size scaling exponent $\eta_{\rm SG}$ of replica peak in Fig.\ref{f5}(d) 
and $T_{\rm BKT}$ from the exponent $\eta$ of $\langle m_{\rm sub}^2\rangle$ in Fig.~\ref{f3}(b). 
}
\label{f2}
\end{figure}
\begin{figure}[tbp]
\includegraphics[width=9cm]{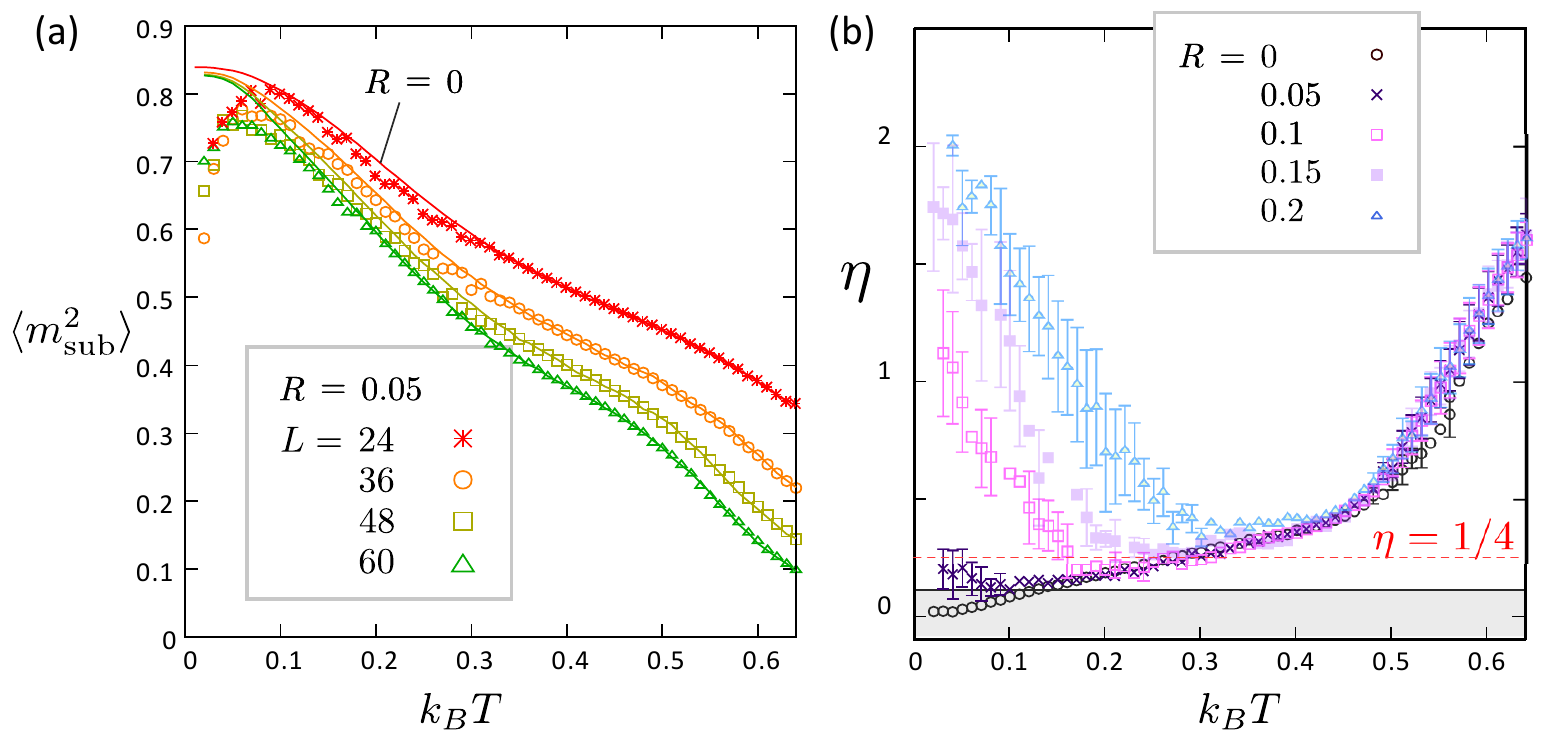}
\caption{(a) Square of sublattice magnetization $\langle m_{\rm sub}^2\rangle$ evaluated for 
$R=0$ (solid lines) and $R=0.05$ (symbols) for $L=24,36,48,60$. 
Finite $\langle m_{\rm sub}^2\rangle$ indicates the divergence of $\chi_{\rm sub}$. 
(b) Exponent $\eta$ of $\langle m_{\rm sub}^2\rangle$ as a function of temperature for 
$R=0, 0.05, 0.1, 0.15, 0.2$. 
The shaded region is $\eta <1/9$ that corresponds to the clock 
long range order, which is not realized for $R>0$. 
}
\label{f3}
\end{figure}
%

\section{Quantum structural-spin glass phase} 
\label{sec:qssg}
In this section, we first present the phase diagram in \S.\ref{sec:qssg}A and 
exclude the possibility of a regular magnetic long range order in the QSSG phase in \S.\ref{sec:qssg}B. 
In \S.\ref{sec:qssg}C-E we show the calculated results to evidence the existence of QSSG in the phase diagram, 
and finally, we explain in \S.\ref{sec:qssg}F how these results consistently verify Fig.~\ref{f1} 
from a unified perspective. 
\subsection{Phase diagram}
Our central result is shown in the phase diagram in Fig.~\ref{f2} obtained from our numerical simulation. 
Here, we find a QSSG phase at low but finite temperatures. 

Before going into the details, let us first consider two limiting cases, $\Gamma=0$ and $R=0$, 
separately to develop physical intuition. 
The starting point is a uniform classical Ising model, $R=\Gamma=0$. 
As mentioned above, the system comprises a huge number of degenerate states and remains paramagnetic down to $T=0$\cite{wannier}. 
\par
At $R \ne 0$, the huge degeneracy is transformed into a set of quasi-degenerate random configurations 
of a classical SG, but the SG is formed {\it only at zero temperature}\cite{imada87,imada87S}; 
it is manifested in a critical divergence of the uniform susceptibility and 
a sublinear criticality of the specific heat toward $T=0$ as demonstrated previously\cite{imada87S}. 
We confirmed the absence of SG order at $T>0$ 
by the classical Monte Carlo calculation in Appendix \ref{app:clchisg}. 
\par 
If we keep $R=0$ and introduce $\Gamma\ne 0$, a frustrated transverse Ising model is realized, 
whose phase diagram is shown in Fig.~\ref{f2}(b). 
The huge classical degeneracy is lifted \cite{villain80} and the three-sublattice clock order 
appears at $T\le T_{\rm clock}$ as is illustrated in Fig.~\ref{f1}(a)~\cite{moessner00}, 
where a superlattice containing three sites in a unit cell labeled by A, B and C emerges. 
On the honeycomb lattice sites (A and B in Fig.~\ref{f1}(a)), spins antiferromagnetically order to gain the energy $J_{ij}$. In other words, the $z$-component of magnetizations at the sites A and B denoted by $m_A$ and $m_B$ take $m_A=-m_B=1$ or $-1$, 
which are doubly degenerate because of the TRS breaking. 
On the other hand, the spins at the center of a hexagon (C site) align in $-x$ direction 
(namely, $m_C=0$ characterized by $\langle \sigma_j^x\rangle \ne 0$) to gain the energy $\Gamma>0$. 
This superstructure breaks the original translational symmetry of the triangular lattice, 
implying that there is a three-fold degeneracy about which of the three sublattices to assign A/B/C-spins. 
Namely, the A-B-C configuration in Fig.~\ref{f1}(a) has the same energy as the configurations obtained by exchanging the locations of B and C spins or A and C spins. 
Then the total degeneracy of the ground state arising from the breakings of the TRS and the translational symmetry is six-fold. 
The long-range order of the clock phase at finite temperature in 2D 
is enabled by this discrete nature of the symmetry breaking. 
The order parameter of the phase is a square of three-sublattice magnetization, $\langle m_{\rm sub}^2\rangle \ne 0$, 
defined in Eq.(\ref{eq:msub}). 

The QSSG is a phase developed from the clock phase by the introduction of {\it small bond randomness}, 
which is the main result of this paper (see \S.\ref{sec:sg} and \S.\ref{sec:replica}). 
We show numerical evidence that the clock order is sensitively destroyed by the randomness and become glass. 

At $R=0$, it is known that the critical BKT phase emerges at $T_{\rm clock}<T < T_{\rm BKT}$~\cite{blankschtein84,moessner00}. 
The number of the degeneracy remaining is six-fold, which is large enough
to stabilize this critical phase originally proposed for the
system with continuous symmetry such as the XY model. 
The BKT phase survives at $R\ne 0$, and plays a role as an underlying backbone structure of QSSG, 
protecting the domain structure of glass from thermal flucuation, 
which we will discuss in detail in \S.\ref{sec:sigx}. 

\subsection{Destruction of clock order by randomness}
\par
Figure~\ref{f3}(a) shows $\langle m_{\rm sub}^2\rangle$ at $R=0$ (solid lines) and $R=0.05$ (symbols) for several $L$. 
The BKT and the clock phases found in Ref.[\onlinecite{isakov03}] are characterized by the 
corresponding three-sublattice susceptibility given in Eq.(\ref{eq:chisub}) and Appendix \ref{app:subsus}, 
Fig.~\ref{fa4}. 
Above and below $T_{\rm BKT}$, $\chi_{\rm sub}$ decays exponentially and algebraically, respectively, 
with system size $L(\rightarrow\infty)$. 
Therefore, by analyzing $\langle m_{\rm sub}^2\rangle \propto L^{-\eta}$ as a function of temperature, 
we obtain a BKT transition temperature, $T_{\rm BKT}$, in Fig.~\ref{f2}(b) 
at which $\eta$ crosses 1/4 (see Fig.~\ref{f3}(b) with $R=0$). 
In further lowering the temperature, $\eta$ becomes smaller than 1/9, which signals the onset 
of the three-sublattice long range order or a clock phase (see Ref.[\onlinecite{isakov03}] for details). 
\par
In introducing $R>0$, $\langle m_{\rm sub}^2\rangle$ conicides overall with the one at $R=0$, 
except that there appears a drop in $\langle m_{\rm sub}^2\rangle$ at $k_BT\lesssim 0.1$. 
This drop indicates the break down of the clock long range order. 
In fact, the value of $\eta$ evaluated for $R>0$ starts to show an upturn at low temperature, 
and does not go below $\eta=1/9$ (see Fig.~\ref{f3}(b)). 
The upturn of $\eta$ corresponding to the drop in $\langle m_{\rm sub}^2\rangle$, 
does not mean that the system goes back to the paramagnetic phase, 
but that the standard scaling of $\eta$ breaks down. 
Notice that even though the clock phase disappears, the BKT transition is still present 
for relatively small randomness $R\lesssim 0.15$. 
\begin{figure*}[tbp]
\begin{center}
\includegraphics[width=17cm]{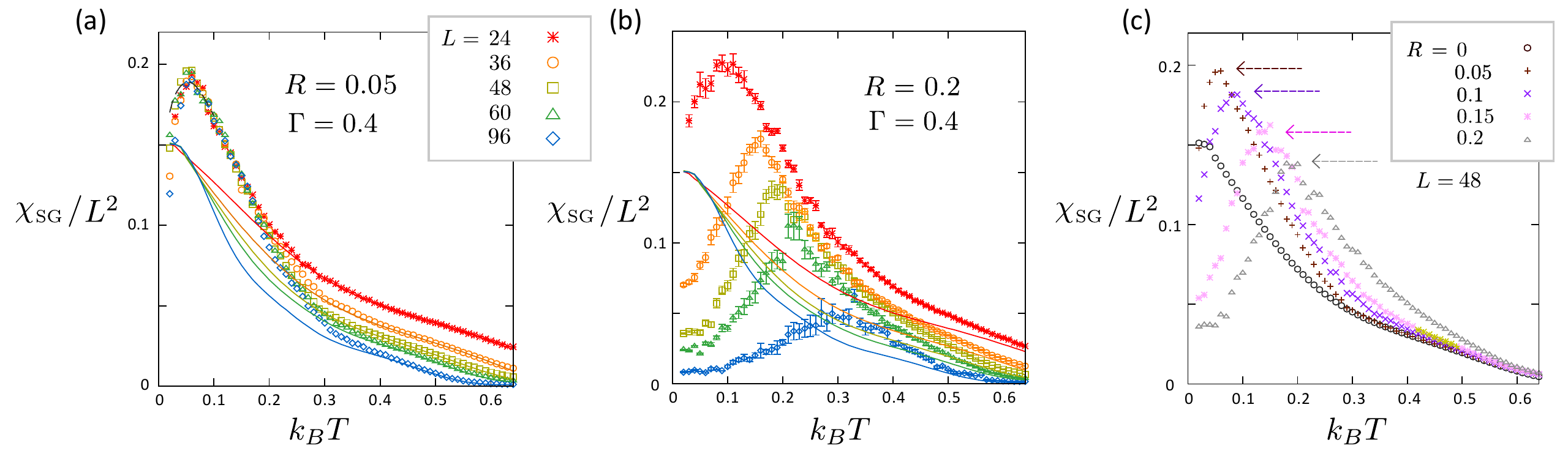}
\end{center}
\caption{
Uniform SG susceptibility density $\chi_{\rm SG}/L^2$ for several system sizes $L$ at $\Gamma=0.4$. 
In general, $\chi_{\rm SG}/L^2 > 0$ for $L\rightarrow \infty$ indicates a finite SG order parameter. 
(a) Comparison between the results at $R=0$ (lower five data points with solid lines) and $R=0.05$ (upper five data) 
for $L=24,36,48,60,96$. 
Broken line at $k_BT<0.1$ is obtained as a $L\rightarrow\infty$ data from the size scaling for $R=0.05$. 
(b) Size dependence at $R=0.2$ (together with $R=0$ data in solid lines). 
(c) Comparison of $R=0, 0.05, 0.1, 0.15, 0.2$ data for $L=48$. 
Arrows indicate the approximate peak positions. 
}
\label{f4}
\end{figure*}
\begin{figure*}[tbp]
\begin{center}
\includegraphics[width=17cm]{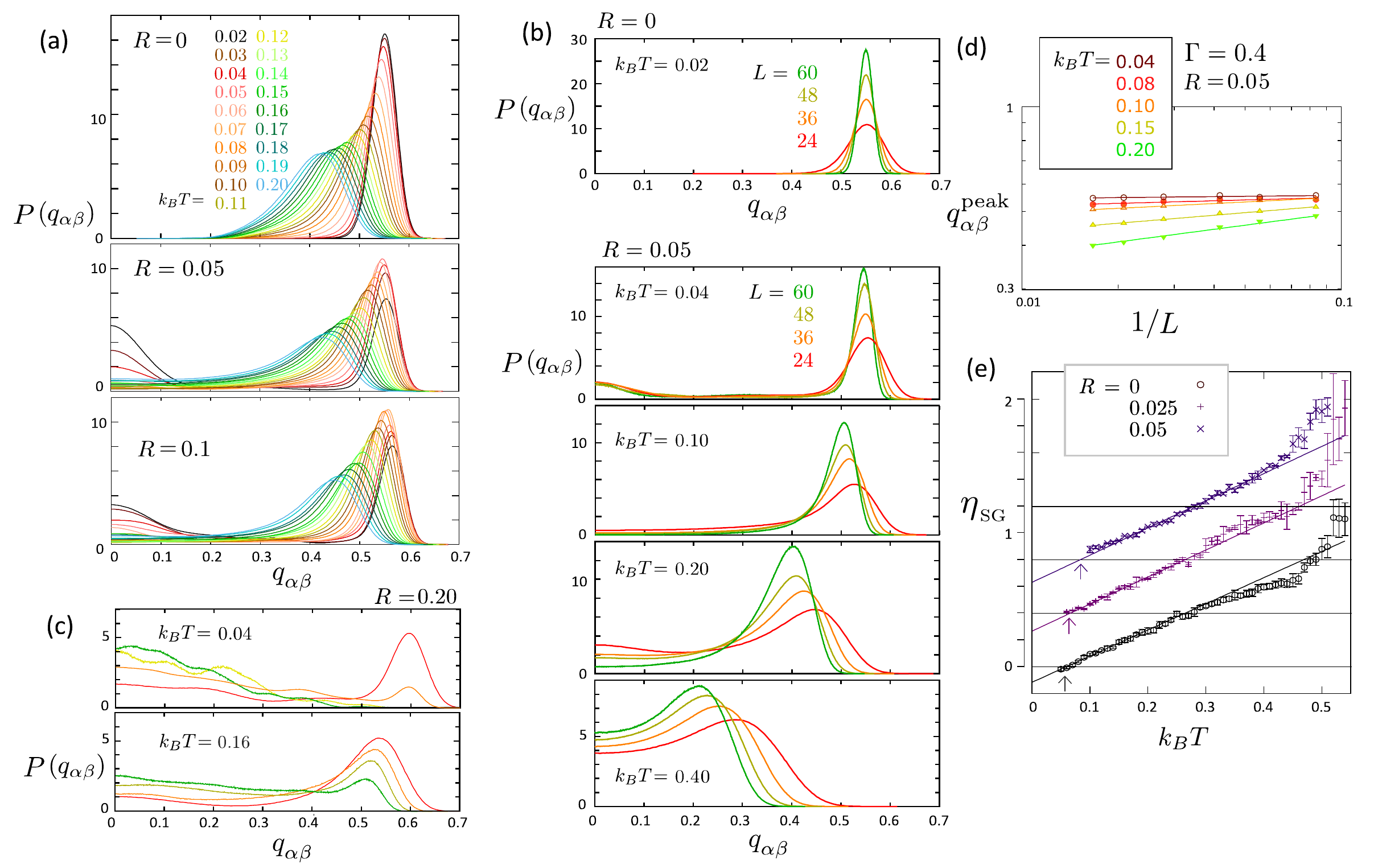}
\end{center}
\caption{(a,b,c) Distribution of the replica overlap $P(q_{\alpha\beta})$. 
In panel (a) the variation of peaks at different temperatures $k_BT=0.02-0.2$ 
is examined for $R=0, 0.05$, and 0.1 inside the QSSG phase at $L=36$. 
The top two panels in (b) shows the size dependences of $P(q_{\alpha\beta})$ for 
$(R,k_BT)=(0,0.02)$ inside the clock ordered phase, 
and $(0.05, 0.04)$ in the QSSG phase. 
In the lower three panels, the ones at $R=0.05$ and at higher temperature region $k_BT=0.1,0.2,0.4$ 
are shown for comparison. 
Panel (c) shows $P(q_{\alpha\beta})$ at $(R,k_BT)=(0.2,0.04)$ and $(0.2,0.16)$ 
both in the paramagnetic phase for comparison. 
The color notation is the same as panel (b). 
(d) $1/L$-dependence of the peak position of $P(q_{\alpha\beta})$ 
at around $q_{\alpha\beta}\simeq 0.55$ at $R=0.05$ 
for different temperatures. Solid lines are the results fitted in powers as, 
$(1/L)^{\eta_{\rm SG}/2}$. 
At $k_BT \ge 0.07$, the peak position approaches zero in the thermodynamic limit, 
while in the QSSG phase ($k_BT=0.04$) 
the peak sustains its nonzero position within the error. 
(e) Power $\eta_{\rm SG}$ evaluated at temperatures above the QSSG phase are 
 shown as a function of $k_BT$ for several $R\ne 0$ with offsets. 
When $\eta_{\rm SG}$ reaches zero, the peak position does not change with $L$, which 
marks the transition temperature $T_c$ to the QSSG phase. 
The linear fit of $\eta_{\rm SG}$ shown in solid lines is used to evaluate $T_c$ indicated by arrows. 
}
\label{f5}
\end{figure*}
\begin{figure}[tbp]
\begin{center}
\includegraphics[width=8cm]{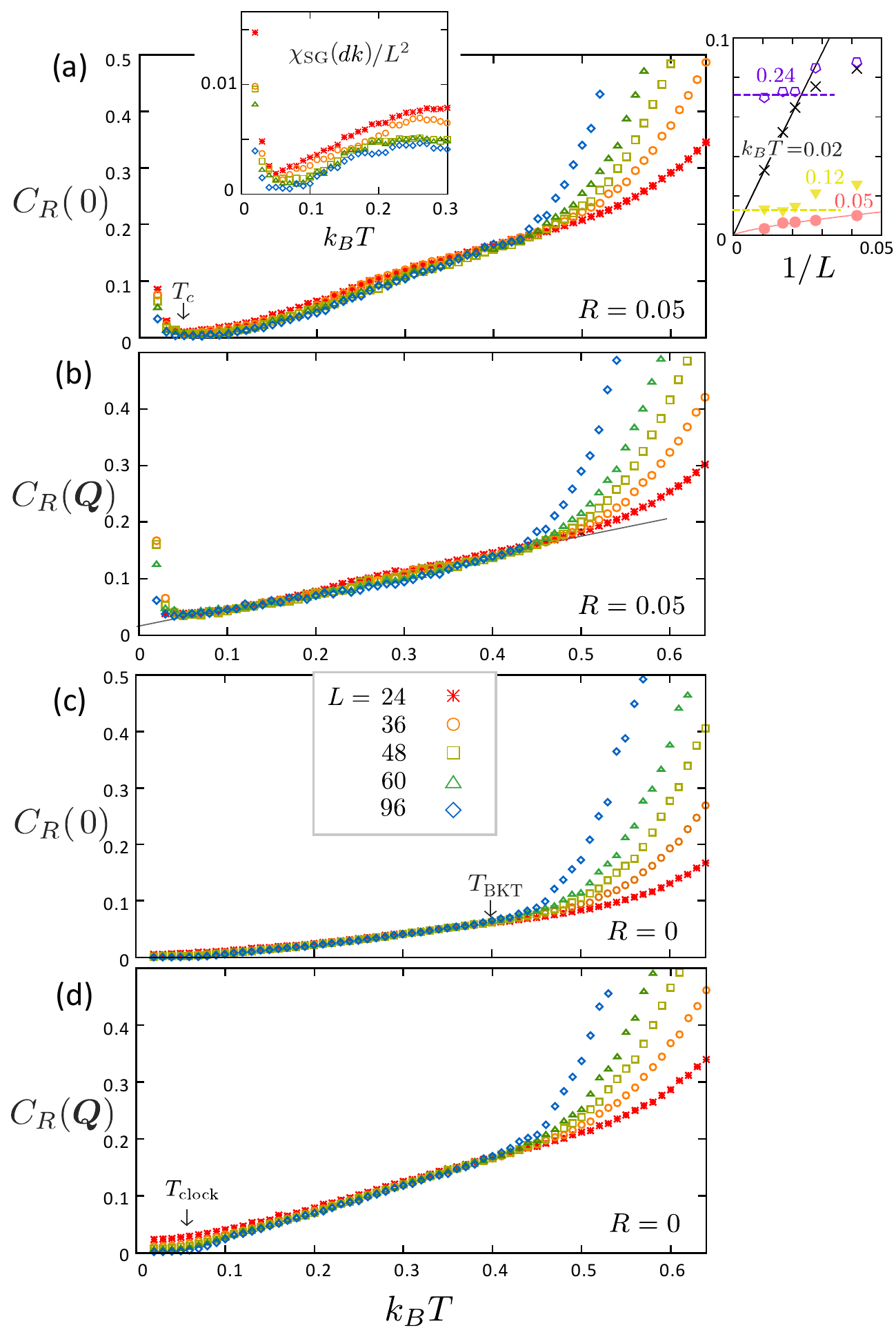}
\end{center}
\caption{
Correlation ratio $C_R(\bm k=0)$ and $C_R(\bm k=\bm Q)$ for (a)(b)$R=0.05$ and (c)(d)$R=0$ 
as functions of temperature for $L=24,36,48,60,96$. 
The ones in panel (a) approaches zero at $T\rightarrow +T_c$. 
Inset(left) of panel (a) shows $\chi_{\rm SG}(dk)$ at $R=0.05$, 
and the right inset shows the size-scaling of $C_R(0)$ for QSSG phase($k_BT=0.02,0.05)$ 
which goes to zero at $L\rightarrow\infty$, 
and BKT phase ($k_BT=0.12,0.24$) which shows almost no $L$-dependence at $L\gtrsim 48$. 
}
\label{fadd1}
\end{figure}
%
\subsection{Quantum spin glass susceptibility}
\label{sec:sg}
The low-temperature phase at $0<R\lesssim 0.15$ no longer has a regular magnetic long-range order. 
We call this phase a QSSG. 
Figures \ref{f4}(a) and \ref{f4}(b) show the uniform SG susceptibility density (see Eq.(\ref{chisg})) 
$\chi_{\rm SG}/L^2$ as a function of temperature for several different system sizes. 
There we directly compare the data of $R=0$ and $R=0.05/0.2$ in the same plot. 
\par
At $R=0$, we find a gradual increase in $\chi_{\rm SG}/L^2$ on decreasing temperature, which 
finally converges to a unique curve for all $L$ 
in the clock phase at $T\le T_{\rm clock}$ (see Fig.~\ref{f2}). 
Our results confirm $\lim_{L\rightarrow\infty} \chi_{\rm SG}/L^2 =0$ for $T >T_{\rm clock}$. 
In the clock phase, there exists a three-sublattice long range order of spins 
shown in Fig.~\ref{f1}(c), where A and B sublattices form an antiferromagnetic long range order, 
which contributes to finite $\overline{\langle \sigma_i\sigma_j\rangle ^2}$ in the thermodynamic limit, 
naturally yielding $\lim_{L\rightarrow\infty} \chi_{\rm SG}/L^2 \ne 0$, 
although it is not a glass phase but a periodically ordered phase. 
\par
The intrinsic contribution of glassiness of spins to $\chi_{\rm SG}/L^2$ 
can thus be examined by the difference between the $R >0$ and $R=0$ data for each $L$. 
Below $k_BT\lesssim 0.3$, there is a significant enhancement in $\chi_{\rm SG}/L^2$ 
attributed to the introduction of $R$. 
Since the difference between $R>0$ and $R=0$ is larger for larger $L$, 
there is an additional frozen component of spins that leads to $\chi_{\rm SG}(R\ne 0)-\chi_{\rm SG}(R=0) \propto L^2$. 
At higher temperatures, all the $R>0$ data smoothly extrapolates to the $R=0$ slope. 
\par
At $R=0.05$, $\chi_{\rm SG}/L^2$ shows similar peak at $k_BT\sim 0.06$ for all $L$ 
(see the broken line showing a $L\rightarrow\infty$ profile). 
We find only a very small size dependence already at $k_BT<0.2$, 
which is because of the nearly scale-invariant property of $q_{\alpha\beta}$ (see Appendix \ref{app:corratio}) 
due to the criticality of the BKT phase we discuss shortly. 
Contrastingly, for $R=0.2$ the peak height decreases, and the peak position shifts to a higher temperature for larger $L$, 
indicating that this peak disappears in the thermodynamic limit. 
\par
We also show in Fig.~\ref{f4}(c) the $R$-dependence of $\chi_{\rm SG}/L^2$ for $L=48$. 
At $R\gtrsim 0.15$ the peak-hight decreases significantly where we no longer expect 
an intrinsic glassy behavior. 
It is consistent with the parameter range where the data in panel (b) starts 
to show a rapid size dependence. 
%
\subsection{Replica overlap}
\label{sec:replica}
For a precise evaluation of a glass phase, 
examining the behavior of a replica overlap given by Eq.(\ref{eq:replica}) is useful. 
This quantity measures how much the two replicas, 
i.e. the systems with the same bond randomness but undergo different MC simulation runs, 
resemble each other in their spin configurations at each snapshot of MCS. 
If the two replicas have similar spin configuration 
$q_{\alpha\beta}$ acquires a nonzero value, while if they are not alike we find $q_{\alpha\beta}=0$. 
Therefore, the distribution function of $q_{\alpha\beta}$ denoted as $P(q_{\alpha\beta})$ 
for $q_{\alpha\beta}=[0,1]$ 
gives an information on what kind of state the system belongs to. 
A single peak of $P(q_{\alpha\beta})$ at $q_{\alpha\beta}=0$ 
indicates that the spin configuration of the replicas do not resemble each other and the state is paramagnetic. 
A single peak of $P(q_{\alpha\beta})$ at $q_{\alpha\beta}\ne 0$ in the absence of disorder 
indicates that there is a regular long range order which has a spatially periodic structure; 
the clock phase at $R=0$ corresponds to this case 
\footnote{Since for $R=0$ we have translational symmetry, we identified A/B/C sublattices for each replica to take the proper overlap between the three sublattices. }
Whereas peak at $q_{\alpha\beta} >0$ together with a finite $P(q_{\alpha\beta}=0)>0$ 
indicate the possibility that the spins are in a glass phase 
having a multi-valley free energy landscape; 
the two replicas can become either similar ($q_{\alpha\beta} \ne 0$) 
or not resemble at all ($q_{\alpha\beta} =0$), 
depending on whether they belong to the same/nearby valley or not. 
The 3D SG of an EA model \cite{kawashima96,marinari98,katzgraber01} shows a continuous profile of 
nonzero $P(q_{\alpha\beta})$ ranging from $q_{\alpha\beta} =0$ toward the peak 
at finite $q_{\alpha\beta}$. 
In the replica symmetry breaking (RSB) theory, 
separate distinct peaks at $q_{\alpha\beta}=0$ and $\ne 0$ is the signature of the 
one-step RSB\cite{castellani2005}. 
\par
Figure \ref{f5}(a) shows how $P(q_{\alpha\beta})$ varies with temperature at $L=36$ and $R=0, 0.05, 0.1$. 
When $R=0$, there is a single peak at $q_{\alpha\beta}^{\rm peak} \ne 0$ 
that indicates the three-sublattice correlation, 
which shifts from $q_{\alpha\beta}^{\rm peak}\sim 0.55$ to smaller values at higher temperature. 
When $R=0.05/0.1$ there appears another peak at $k_BT\lesssim 0.06/0.1$ at around 
$q_{\alpha\beta}\sim 0$. 
\par
To see whether these peaks sustain in the thermodynamic limit, we plot in Fig.~\ref{f5}(b) 
the size dependence for those in the clock phase $(R,k_BT)=(0,0.02)$, 
QSSG phase $(0.05,0.04)$ and critical or paramagnetic phases at $R=0.05$ and $k_BT=0.1,0.2,0.4$. 
In both the clock and QSSG phases, the peak at $q_{\alpha\beta}^{\rm peak}\sim 0.55$ remain robust and 
develops with increasing $L$. 
The peak at $q_{\alpha\beta}\sim 0$ is small but also keeps a finite weight almost independent of $L$. 
Therefore, according to the standard definition, the QSSG phase is a SG phase. 
\par
The two peak structures can generally occur when there is phase separation, 
or a metastable excited state coexisting with the magnetically ordered ground state. 
We checked numerically that such a possibility is excluded; 
in Appendix \ref{app:relaxation}, the energy distribution of the QMC simulation shows a sharp 
single peak that exactly extrapolates to the $\delta$-function at $L\rightarrow \infty$, 
indicating that QSSG is a uniform thermodynamic phase. 
Generally, metastability is observed in the vicinity of the first-order transition 
in glass-forming liquids, while here, the first-order transition is absent. 
\par
Figure~\ref{f5}(c) shows $P(q_{\alpha\beta})$ for a strong randomness, $R=0.2$. 
The peak at $q_{\alpha\beta}^{\rm peak}$ tends to disappear at large $L$ 
for both $k_BT=0.04$ and 0.16 
and we find a gradual development of broad $q_{\alpha\beta}=0$-peak 
signaling the paramagnmetic phase with strong disorder. 
\par
In the paramagnetic phases the peak position $q_{\alpha\beta}^{\rm peak}$ gradually shifts toward zero 
with increasing $L$, which can be analyzed as in Figs.~\ref{f5}(d) and \ref{f5}(e); 
we fit the obtained data according to $q_{\alpha\beta}^{\rm peak}\propto L^{-\eta_{\rm SG}/2}$, 
and plot $\eta_{\rm SG}$ as a function of temperature for different $R$'s.  
They behave linearly with $k_BT$ over a wide temperature range. 
One can extract $T_c$ for each $R$ that gives $\eta_{\rm SG}=0$ (arrows in Fig.~\ref{f5}(e)), 
the temperature at which the peak position does not change with $L$. 
We show the results for the values of $T_c$ in the phase diagrams in Fig.~\ref{f2}. 
This temperature can be recognized as the onset of the QSSG phase, 
which agrees well with the peak temperature of $\chi_{\rm SG}/L^2$. 

\subsection{Correlation ratio}
\label{sec:corratio}
As one of the standard ways to identify the phase transition, 
we introduce a correlation ratio given as 
\begin{equation}
C_R(\bm k)= \chi_{\rm SG}(\bm k+d\bm k)\big/ \chi_{\rm SG}(\bm k), 
\label{eq:corratio}
\end{equation}
where $\bm k+d\bm k$ is the nearest neighbor wave number from $\bm k$ 
of a finite cluster with $dk=2\pi/L$. 
It quantifies how sharp the peak of the structure factor of $q_{\alpha\beta}$ 
at $\bm k$ is, 
and is known as a good measure to pin down the continuous phase transition point\cite{kaul15}. 
If the long range order of $q_{\alpha\beta}$ characterized by the $\delta$-function Bragg peak 
is expected, we find $C_R(0)\rightarrow 0$, 
while it approaches 1 if the long-range order associated with $q_{\alpha\beta}$ is absent. 
It was shown that $C_R$ exhibits crossings among different $L$'s 
at the standard second-order transition, 
and the $L$-dependence of the crossing point is small, 
implying that the transition point can be identified for a calculation with relatively small $L$. 
However, our system does not exhibit a second order transition but a BKT transition. 
In the critical BKT phase, $C_R$ behaves nearly $L$-independent. 
(see Appendix \ref{app:corratio} for the formulation support.)
\par
There are two peaks in $\chi_{\rm SG}(\bm k)$ 
at $\bm k=0$ and $\bm k=\bm Q=(\pm \frac{\pi}{3},\mp\frac{\pi}{3})$; 
the hight of the former is 2-4 times the latter. 
Figures~\ref{fadd1}(a) and \ref{fadd1}(b) show $C_R(0)$ and $C_R(\bm Q)$ when $R=0.05$. 
In the BKT phase at $T_c<T<T_{\rm BKT}$, the $L$-dependence is almost lost for $L\gtrsim 48$
(see the right-inset of Fig.~\ref{fadd1}(a)). 
For the temperature dependence, we see a nearly linear line approaching 
$C_R(0)\rightarrow 0$ at $T\rightarrow +T_c$, 
which signals the phase transition point from the BKT to the QSSG phase. 
Meanwhile, $C_R(\bm Q)$ remains finite down to zero temperature. 
The opposite behavior is found for the clock phase at $R=0$ shown in Figs.~\ref{fadd1}(c) and \ref{fadd1}(d). 
$C_R(\bm Q)\rightarrow 0$ at $T\rightarrow +T_{\rm clock}$ whereas $C_R(0)$ remain finite down to $T=0$. 
\par
At $T<T_c$ we find a slight upturn of $C_R(0)$ which shows a clear $L$-dependence. 
This is another clear sign of QSSG transition. 
Still, the QSSG phase has a SG order because $\lim_{L\rightarrow\infty} C_R\rightarrow 0$ 
(see the right inset of Fig.~\ref{fadd1}(a)). 
When the system enters a different phase from the BKT phase, the scale-free character is lost, 
which is observed both at $T<T_c$ and $T>T_{\rm BKT}$. 
The similar behavior is observed 
in other models exhibiting multi-BKT transitions\cite{matsubara14}. 
\subsection{Unveiling unified understanding from the SG quantities}
\label{sec:summary_qssg}
The essential feature of QSSG is that it is built from two types of degrees of freedom. 
The two degrees of freedom emerge from the single spin degrees of freedom 
and distribute uniformly in space (see Fig.~\ref{f1}), 
and with the aid of quenched randomness, 
coorperatively generate two distinct glasses which we discuss in the following. 
\par
We first summarize the list of our conclusions on the QSSG: 
it is [I] the long-range SG order at finite temperature in 2D, and 
[II] the spatially uniform coexistence of the two-component glasses: 
one component is a rigidly long-range ordered SG 
and the other is the algebraic structural-glass 
which is characterized by an anomalous power-law decay of SG correlation on top of the former long-range order. 
Such coexistence of long-range order and the algebraic correlation in a spatially 
uniform phase has never been observed in nature. 
\par
Now, we relate [I] and [II] to the supporting numerical results. 
We perform the standard analysis on three quantities. 
The first quantity is the distribution function $P(q_{\alpha\beta})$ of the replica overlap, Eq.(\ref{eq:replica}). 
If the nonzero probability of the overlap exists, it is indicative of freezing of spins 
and further if the nonzero probability occurs at both zero and nonzero $q_{\alpha\beta}$, it evidences a SG order. 
The second quantity is the wavenumber-dependent connected 
SG susceptibility $\chi_{\rm SG}(\bm k)$, Eq.(\ref{chisgk}). 
If $\chi_{\rm SG}(\bm k=0)/N$ remains nonzero in the thermodynamic limit, 
it is another evidence of the SG order. These two quantities must be consistent with each other. 
The third evidence of the SG order can be obtained from the correlation ratio $C_R$, Eq.(\ref{eq:corratio}). 
If the long-range order of the associated susceptibility exists, it converges to zero in the thermodynamic limit. 
At the continuous transition point, $C_R$ crosses at a finite value for all the asymptotically large system sizes: 
in the ordered phase $C_R$ decreases with increasing $L$ while it increases in the non-ordered phase. 
The crossing point does not move already from very small system sizes. 
Note that $C_R$ is then a scale-invariant quantity. 
\par
By using these three quantities, we have shown the following numerical results. 
\begin{itemize}
\vspace{-2mm}
\item[(1)] There exists nonzero critical temperature $T_c$, 
below which two peaks appear in $P(q_{\alpha\beta})$ at $q_{\alpha\beta}\sim 0.55$ and 
$q_{\alpha\beta}=0$. 
Based on the following (4)-(6) we conclude that they are the spatially uniform coexistence of 
the SG-ordered component and the algebraically-decaying component, respectively. 
\vspace{-2mm}
\item[(2)] At $T\lesssim T_c$, we find $\lim_{N\rightarrow\infty} \chi_{\rm SG}(\bm k=0)/N>0$, 
signaling the SG long-range order. It slightly drops with further decreasing temperature from $T_c$ but remains nonzero. 
\vspace{-2mm}
\item[(3)] At $T<T_c$, $C_R$ decreases with increasing $L$ as in the usual long-range ordered phase, 
which indicates the existence of the spatially uniform SG order, 
while $C_R$ increases for $T> T_{\rm BKT}$. 
In the intermediate region, $T_c<T< T_{\rm BKT}$, the system size dependence vanishes, 
which is the character of the critical phase. 
This means that the conventional critical point of the continuous transition 
is extended to the critical nonzero temperature window. 
\vspace{-2mm}
\end{itemize}
The results displayed in Appendix \ref{app:corratio} additionally show the following (4)-(6).
\begin{itemize}
\vspace{-2mm}
\item[(4)] $\chi_{\rm SG}(\bm k)$ at nonzero $k$ and its Fourier transform 
(i.e., $q_{i;\alpha\beta}q_{j;\alpha\beta}$ in real space) 
consistently exhibit the power-law correlation of SG, 
indicating the algebraic component of the glass. 
Note that power-law decay in real space generates a power-law decay in the momentum space. 
\vspace{-2mm}
\item[(5)] 
On top of (4), at $T<T_c$, $\chi_{\rm SG}(\bm k)$ deviates from the power-law fitting function 
at small nonzero $0<k/\pi<0.1$, because the additional weight in $\chi_{\rm SG}(\bm k)$ appears in this range, 
simultaneously with a slight reduction of $\chi_{\rm SG}(\bm k=0)$.
\vspace{-2mm}
\item[(6)] 
In $P(q_{\alpha\beta})$, the $q_{\alpha\beta}\sim 0.55$ peak component transfers to the $q_{\alpha\beta}=0$ component 
with decreasing temperature at $T< T_c$, but the weights of these two components both remain nonzero at any $T< T_c$. 
\end{itemize}
\vspace{-2mm}
(1)-(3) consistently support the existence of the SG order; 
since (1) and (2) are related by $\chi_{\rm SG}(\bm k=0)=2\big(\int_0^1 q^2 P(q)dq -(\int_0^1 qP(q)dq )^2\big)$, 
the drop of $\chi_{\rm SG}(\bm k=0)$  at $T<T_c$ is attributed to the emergent $P(q_{\alpha\beta}=0)$ peak. 
(3) guarantees (2) in that the $\chi_{\rm SG}(\bm k=0)$-peak converges to the $\delta$-function. 
\par
At the same time, (4)-(6) indicate the coexistence of two different characters of glass, 
the ordered SG and the algebraic glass. 
The former contributes to the nonzero $\chi_{\rm SG}(\bm k=0)/N$ and $P(q_{\alpha\beta}\sim 0.55)$ weight. 
The latter contributes to $\chi_{\rm SG}(\bm k\ne 0)/N$ and $P(q_{\alpha\beta}=0)$ weight. 
This classification is deduced from the observation that 
the former features continue from the higher temperature BKT phase, 
whereas the latter emerge only at $T<T_c$. 
\par
As mentioned earlier, spin degrees of freedom breaks up into two; the staggered spins forming a honeycomb lattice, 
and the transverse spins at the center of the hexagon. 
The honeycomb-lattice-vitrification (emergent domains) is the origin of the algebraic component of glass, 
and the randomness of staggered and transverse spins is the origin of the standard uniform SG order. 
This scenario is supported by Appendix \ref{app:corratio} 
and by another series of results presented in the next section. 
\par
To be short, 
because a substantial weight is transfered from $\chi_{\rm SG} (k=0)$ to $\chi_{\rm SG}(k>0)$ in (5) at $T<T_c$, 
its Fourier transform, i.e., the real-space $q_{i;\alpha\beta}q_{j;\alpha\beta}$ correlation, 
shows a significant drop at long distances (see Fig.~\ref{faadd2}(c)), 
while it keeps an algebraic BKT-like decay at short distances. 
This agrees with the scenario of emergent domains at $T<T_c$, 
since the domain state preserves the short range correlation 
but the long distance correlation is lost due to the ensemble average of domains. 
At the same time, the two replicas with different domain locations no longer resemble on the whole 
and yields a $P(q_{\alpha\beta}=0)$ peak, in consistency with (1) and (6). 
\par
One might suspect the possible phase separation in real space is related to the coexistence of two glasses. 
However, we can fully exclude it from the additional data showing a single peak in the energy distribution 
of quantum Monte Carlo simulation (see Appendix \ref{app:relaxation} and Fig.~\ref{faadd1}). 
From a more general point of view, metastability is a character of first order transition. 
The continuous nature of our SG transition is supported by 
the scale-free quantity $C_R$ and the scaling quantity $\eta$. 
%
\begin{figure*}[tbp]
\includegraphics[width=16cm]{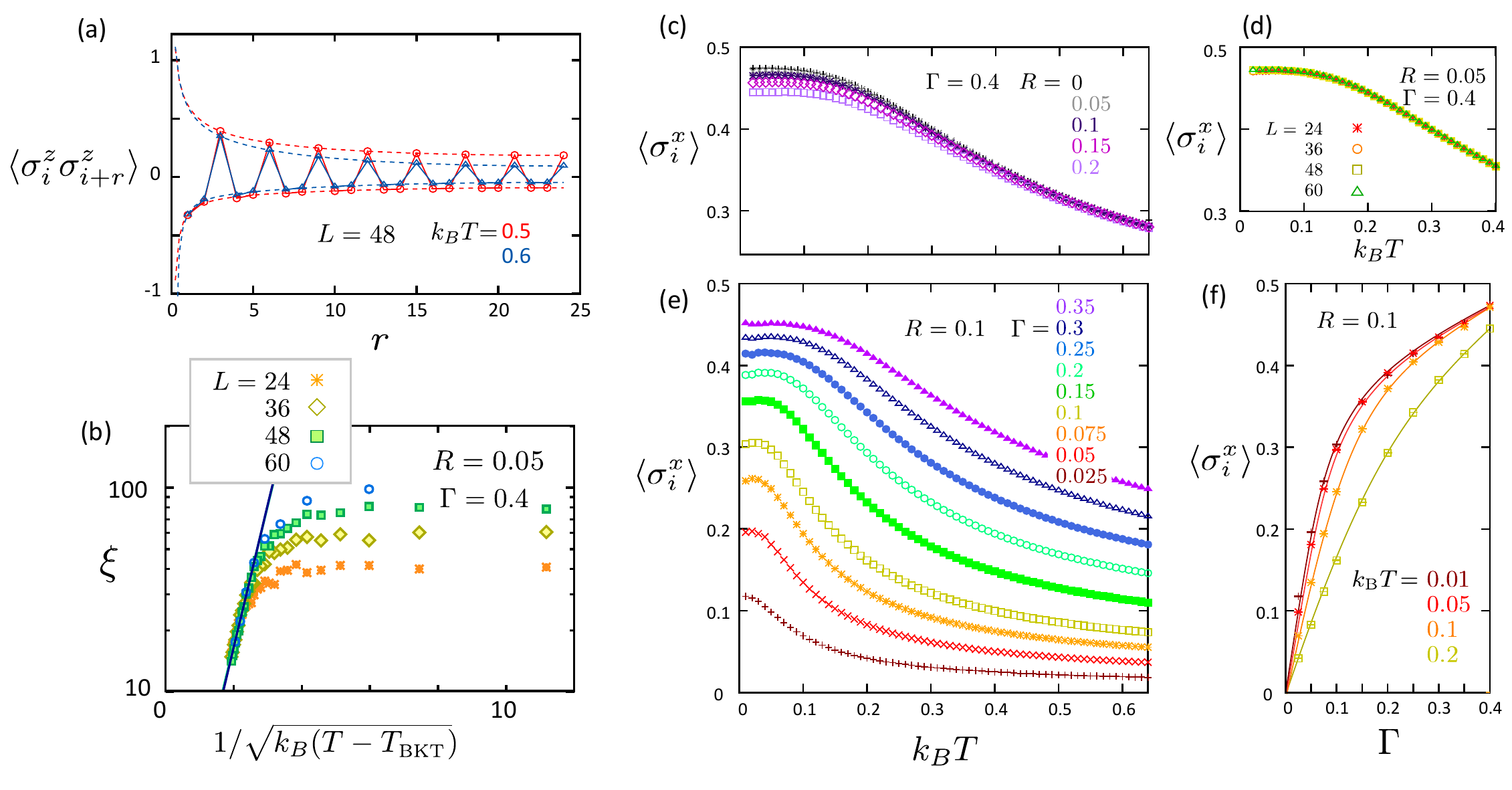}
\caption{
(a) Spin-spin correlation function $\langle \sigma_i^z\sigma_{i+r}^z \rangle$ 
as function of $r$ for $R=0.05$ and $k_BT=$0.5 and 0.6 above the BKT transition. 
Broken lines are Eq.(\ref{corr}). 
(b) $\xi$ plotted as a function of $1/\sqrt{k_B(T-T_{\rm BKT})}$ 
at $R=0.05$ and $\Gamma=0.4$. 
Solid line is the fitted results following Eq.(\protect\ref{eq:xi}) with $k_{\rm B}T_{\rm BKT}=0.371$. 
(c-f) Transverse magnetization $\langle \sigma_i^x \rangle$ 
evaluated for (c) $\Gamma=0.4$ with different $R$, 
(d) $\Gamma=0.4$, $R=0.05$ with different $L$, 
(e) $R=0.1$ for several $\Gamma$, 
(f) $\Gamma$-dependence with $R=0.1$ at low temperatures. 
Since the size dependence is negligible, we take $L=36$.
}
\label{f6}
\end{figure*}
\section{Formation of Domains} 
\label{sec:domain}
The QSSG phase is not simply a random freezing of spins. 
Through a power-law spin-spin correlation in the BKT phase at higher temperatures, 
the algebraic replica-overlap correlation of the QSSG is protected. 
We clarify the details of real-space magnetic properties in both phases in the present section. 
\subsection{ BKT transition }
The QSSG no longer exists when the BKT phase disappears, 
meaning that the BKT transition plays an essential role in stabilizing QSSG. 
The finite-size scaling performed in Appendix \ref{app:bkt} guarantees that the BKT phase sustains at $0<R\lesssim 0.15$. 
Near the BKT transition temperature, the correlation length follows 
\begin{equation}
\xi \propto {\rm exp}(c (T-T_{\rm BKT})^{-1/2}),\;\; c:{\rm constant}, 
\label{eq:xi}
\end{equation}
which diverges with $T\rightarrow T_{\rm BKT}$. 
This is indeed confirmed by directly evaluating the spin-spin correlation function, 
which is expected to obey the following form for size $L$ along the $x$ or 
$y$-directions around the periodic phase boundary, 
\begin{equation}
\langle \sigma_i^z\sigma_{i+r}^z \rangle 
\propto r^{-c_1} {\rm e}^{-r/\xi} + (L-r)^{-c_1} {\rm e}^{-(L-r)/\xi}. 
\label{corr}
\end{equation}
Figure \ref{f6}(a) shows the representative behavior of $\langle \sigma_i^z\sigma_{i+r}^z\rangle$ 
obtained at the temperature above the BKT transition. 
The broken lines obtained by the fitting with the formula of Eq.(\ref{corr}) 
show a good agreement with the data. 
The values of $\xi$ evaluated from this fitting are shown in Fig.~\ref{f6}(b). 
With increasing system size, $\xi$ asymptotically approaches Eq.(\ref{eq:xi}), 
and the transition temperature evaluated, $T_{\rm BKT}=0.371$, 
is in a good agreement with the one independently evaluated from the Binder ratio (Appendix \ref{app:bkt}). 

\subsection{ Quantum transverse magnetization}
\label{sec:sigx}
So far we focused on how the longitudinal $z$-component of spin behaves at $R>0$. 
Figures \ref{f6}(c)-(f) show the transverse magnetization $\langle \sigma^x_i\rangle$ for different parameters. 
Although we plot the $L$-dependence only for $R=0.05$ and $\Gamma=0.4$ in Fig.~\ref{f6}(d), 
$\langle \sigma^x_i\rangle$'s for all different $L$ have the same value within ${\cal O}(10^{-4})$. 
By comparing the data for different $\Gamma$ and $R$ in Figs.~\ref{f6}(c) and \ref{f6}(e), 
we see that $\langle \sigma^x_i\rangle$ is determined by $\Gamma$ and $k_BT$. 
In Fig.~\ref{f6}(f), there is a sharp increase of $\langle \sigma^x_i\rangle$ from zero at $\Gamma=0$ 
to $\langle \sigma^x_i\rangle\sim 0.1$ already at $\Gamma\sim 0.02$, indicating that 
the QSSG phase is supported by strong quantum fluctuations represented by 
$\langle \sigma^x_i\rangle$. 
This fact is contrary to the previous consensus that the quantum fluctuations will destroy SG\cite{guo94}.
Contrastingly to $\sigma^z_i$-related properties, 
$\langle \sigma^x_i\rangle$ is insensitive to the bond randomness $R$. 
\\
\begin{figure*}[tbp]
\includegraphics[width=18cm]{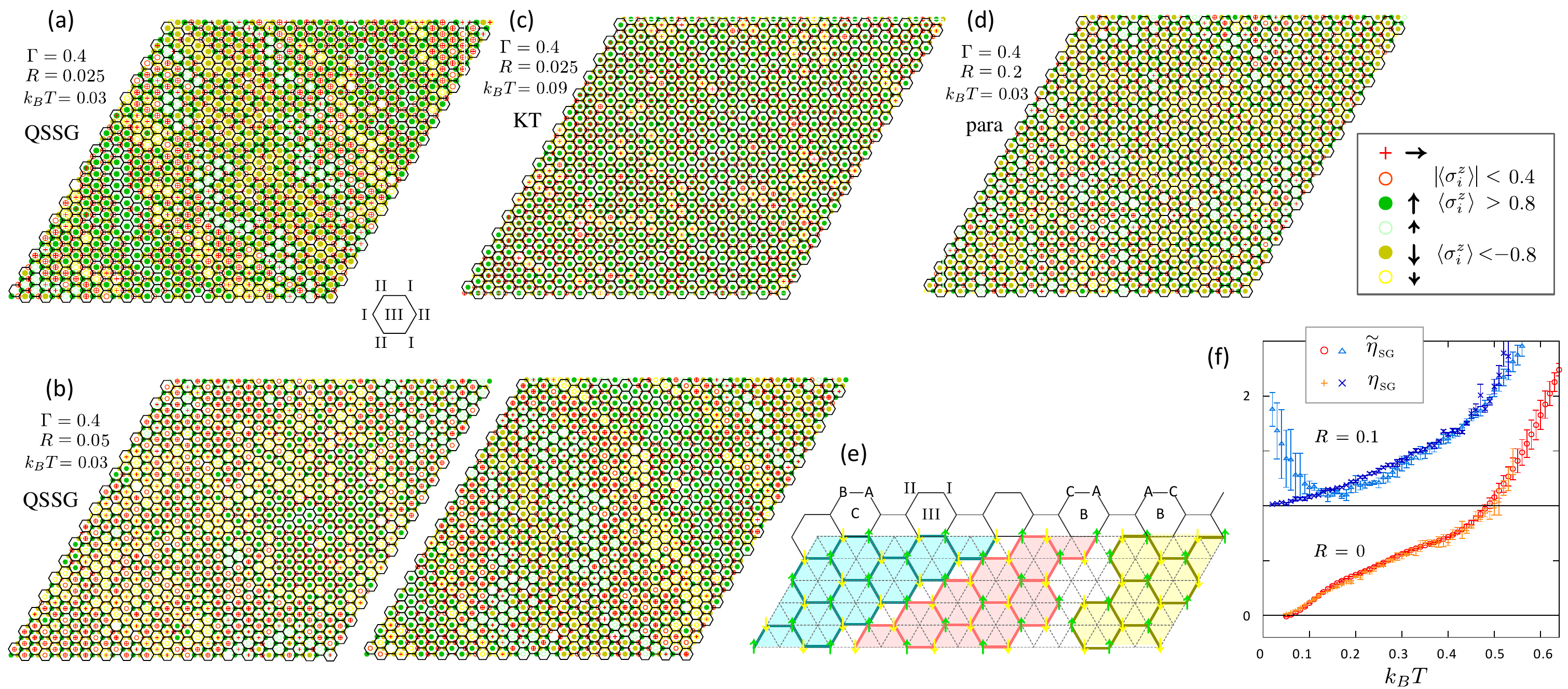}
\caption{
(a-d) Typical snapshots of the spin configuration for QSSG (a,b), KT (c), and paramagnetic (d) phases 
for $L=48$. 
Green/yellow/red symbols are up/down/transversal spins, and 
filled and open symbols indicate large ($>0.4$) and small ($<0.4$) amplitudes of $\langle \sigma_i^z\rangle$. 
I/II/III is the sublattice indices fixed throughout the system and the honeycomb frame is written along I-II sublattices for a guide to the eye. 
When one focuses on sublattice-III at the center of the hexagons, 
red, yellow, and green colored hexagon belongs to the clock order domain with I-II sublattices formed by 
AB, BC, and CA spins, which we call C, A, and B domains, respectively, according to panel (e). 
The two snapshots in panel (b) are obtained for the same temperature $T=0.03$ inside the QSSG phase and the same bond-randomness. 
(e) Schematic structure of domains and their domain boundaries. 
The upper part solid lines are the regular I-II sublattice written in the same manner as (a)-(d), 
and the bold honeycomb lines on the main part are the local bond orders of A(up)-B(down) spins. 
The domains align in a particular order in that they favor 
the exchange of assignment of spins on the I/II/III sublattices 
in either C$\leftrightarrow$A or C$\leftrightarrow$B, when they go to the nearby domain. 
(f) Comparison of $\tilde\eta_{\rm SG}$ obrained from $\chi_{\rm SG} \propto L^{2-\tilde\eta_{\rm SG}}$
and $\eta_{\rm SG}$ obtained from the peak position of the replica overlap 
(see Fig.\ref{f5}(d)) at $R=0$ and $R=0.1$ with offsets. 
The two exponents coincide at $R=0$ but deviate at $R=0.1$ at low temperatures. 
}
\label{f7}
\end{figure*}
\subsection{ Domains }
\label{sec:domain}
Although the long range clock order is easily destroyed at $R\ne 0$, 
an underlying short-ranged but well-developed three sublattice 
structure $(m_A, m_B, m_C)\propto (1,-1,0)$ survives in the QSSG phase, 
and $\langle \sigma_i^x\rangle$ takes a substantially large value, 
which originates from the $m_C\sim 0$ site. 
Therefore, the glassy behavior detected in the emergent peak of $\chi_{SG}$ at $R >0$ and 
the multiple peak structures in the replica overlap $P(\langle q_{\alpha\eta}\rangle )$ 
are not ascribed to a simple SG. 

We visualize the spin configurations by taking several snapshots in the Monte Carlo runs 
after the system reaches an equilibrium in Figs.~\ref{f7}(a)-\ref{f7}(d). 
Here, local spin states are marked according to six characteristic configurations; 
we assign ``+" to the spins pointing in the $x$-direction, 
i.e. they flip more than ten times along the imaginary $\tau$-axis. 
We also classify spins pointing mainly in the $z$-direction according to their 
$\langle \sigma_i^z\rangle$ values as explained in the legend. 
We assign lattice indices I,II, and III for a regular three sublattice sites and draw honeycomb lines along I and II-sublattices as guides for the eyes, where III is at the center of the hexagon. 
It helps us to see which of the A/B/C spins occupy I, II, and III sublattices. 
\par
In the QSSG phase in Figs.~\ref{f7}(a) and \ref{f7}(b), the spins form loose domain structures which 
are discriminated by the color code of spins occupying sublattice III. 
Figure ~\ref{f7}(e) shows the schematic structure of domains. 
We particularly find that the large up (green bullet) and down (yellow bullet) spins frequently form neighbors, 
and the ``+" spins are surrounded by them; 
they form $(m_A, m_B, m_C)\propto(1,-1,0)$ and contribute to $\chi_{\rm sub}$. 
However, there is often a misfit in the position of these A/B/C spins among I, II, and III sublattice sites, 
and accordingly, several domains are formed. 
Besides the stripe-like domains, the island-type domain is also found. 
Most importantly, {\it for the same bond-disorder configuration} thoroughly different domain structures appear 
one by one on different replicas, e.g. the structures of two panels in Fig.~\ref{f7}(b). 
The domains are {\it not pinned to particular locations}, which is a feature 
similar to the type-I structural glass realized in a uniform continuum. 
This point crucially differs from a spin-freezing caused by the random field as 
we discuss later in Sec.~\ref{sec:discussion}. 
\par
At higher temperatures, $T_c<T<T_{\rm KT}$ as in Fig.~\ref{f7}(c), 
this kind of domains become unstable even at small $R$. 
We also show in Fig.~\ref{f7}(d) an example for large randomness $R=0.2$ and at low temperature $k_BT=0.03$; 
the state is outside the QSSG phase and we find that 
although the system should suffer a large amount of bond disorder, 
the domains no longer exist and at the same time, 
the transverse ``$+$" configuration of spins is suppressed. 
These results indicate that the formation of loose domains consisting of three-sublattice 
local structure is intrinsic to the formation of the QSSG phase. 

We finally see how the emergence of domains manifests in the physical quantities. 
Figure~\ref{f7}(f) compares the exponent $\tilde\eta_{\rm SG}$ and $\eta_{\rm SG}$ evaluated by 
$\chi_{\rm SG}/L \propto (1/L)^{\tilde\eta_{\rm SG}}$ and $q_{\alpha\beta}^{\rm peak} \propto (1/L)^{\eta_{\rm SG}/2}$, 
where the latter is presented already in Fig.~\ref{f5}(d). 
When $R=0$, the two agrees well throughout the whole temperature range. 
However for $R=0.1$, $\tilde\eta_{\rm SG}$ starts to become smaller than $\eta_{\rm SG}$ below $T_{\rm BKT}$, 
and shows an upturn below $T_c\sim 0.1$. 
This can be understood as follows; since $\chi_{\rm SG} \propto \langle q_{\alpha\beta}^2\rangle$ 
(see Eq.(\ref{chisg})), 
the two exponents should be the same as far as the distribution $P(\langle q_{\alpha\beta}\rangle)$ has a single peak structure, 
where $\langle q_{\alpha\beta}^2\rangle$ is represented by $(q_{\alpha\beta}^{\rm peak})^2$. 
When $R>0$, $\chi_{\rm SG}$ in Fig.~\ref{f4}(a) becomes less $L$-dependent below $T_{\rm BKT}$ 
compared to those of $R=0$, which suppresses $\tilde\eta_{\rm SG}$ from $\eta_{\rm SG}$. 
At $T\sim T_c$, the extra peak at $q_{\alpha\beta} \sim 0$ appears because of an emergent domain structure. 
Since $\chi_{\rm SG}$ has contributions from both peaks, 
their size scaling breaks down which results in an upturn of $\tilde\eta_{\rm SG}$ 
and a drop of $\chi_{\rm SG}$. 
For this reason, the QSSG phase can be determined more accurately from $q_{\alpha\beta}^{\rm peak}\ne 0$ then from $\chi_{\rm SG}$. 
All the numerical results consistently support the existence of QSSG. 

\section{Discussion} 
\label{sec:discussion}
Our numerical results show several characteristic features of QSSG, 
which are summarized into the following points, {\it (1)} and {\it (2)}.  
Let us consider a QSSG state realized 
for a given set of randomly distributed $J_{ij}$, 
namely for a certain type of quenched randomness. 
First, focusing on a single replica, we find that 
{\it (1-1) a domain structure of a three-sublattice superlattice is clearly visible in a snapshot of MCS 
if $R$ is small} (namely, $R\ll J, \Gamma$), 
and at the same time by examining the details of the spin configurations e.g. the one in 
Fig.~\ref{f7}(a), we see that 
{\it (1-2) there seems to be an underlying rule of how to choose the shape of domain boundaries 
and the relative orientations of spins belonging to neighboring domains, 
since the domains are apparently correlated with each other and 
spin correlations are frozen over long distances. } 

Second, for the same type of quenched randomness, 
one can prepare a set of replicas that are thermalized by different QMC processes from different initial states; 
their equilibrium states may have different spatial spin configurations 
which equivalently contribute to thermodynamics. 
By comparing snapshots of different replicas, e.g. those in Fig.~\ref{f7}(b), 
{\it (2) a variety of thoroughly different domain structures are observed. } 
This indicates that the system has a multi-valley free-energy landscape, 
and is consistent with a possible one-step RSB observed in Fig.~\ref{f5}(b), 
which is a feature common to structural glass \cite{castellani2005} 
or to spin glass with $p(\ge 3)$-body interaction in infinite dimensions\cite{cugliandolo1997}. 

Our Hamiltonian Eq.(\ref{ham}) is a typical spin Hamiltonian with bond randomness, 
and for this reason, we so far focused on the nature of randomness of spins 
by analyzing the spin-glass susceptibility and replica overlaps, 
which are the standard methods used to establish the existence of SG. 
Whereas in the following, we disclose the underlying physics suggested from the observations {\it (1)} and {\it (2)}, 
which are {\it not} straightforwardly understood from these conventional methods. 
Before that, we first add some comments to {\it (1)}. 

Item {\it (1-1)} implies that our QSSG has a qualitative difference from an ordinary SG. 
To be more precise, the spin configurations in QSSG are not fully random within 
a short lengthscale but keeps an overall three-sublattice superstructure, 
which manifests as a visible domain structure over that length scale. 
Such domains apparently differ from ``domains" or droplets in SG; 
the term ``domain" had been often used to describe 
a fictitiously configured spatial region that contains finite numbers of spins, 
to examine the instability of SG\cite{fisher88}. 
However, in reality, a spatially frozen spin configuration of SG is 
not trivially visible\cite{hartmann2002} and 
is magnetically structureless, namely do not have any distinct peak structure in their Fourier component. 
For our structured domains to appear,
the mechanism of a free energy gain is more easily and
intuitively understood, since the intra-domain super-
structure and at the same time domain boundaries can
be clearly identified for small $R$. 
The former comes from the overall energy gain common to the one in the nonrandom case that drives a spontaneous symmetry breaking into the clock phase. 
The latter should be the gain to optimize a randomized part of the Ising bond energy. 
The vitrification in QSSG is a configurational disordering or 
a translational symmetry breaking of a superlattice, 
which maintains short-ranged but well developed clock order correlation if $R$ is small. 
In an ordinary structural glass, the average coordination number is found to be 
approximately the same as that of the crystal of the corresponding spatial dimensions, 
which implies an atomic-scale-short-range order. 
Compared to that case, the structural-order correlation, namely a clock correlation in QSSG 
is tunable by controlling $R$. 

Coming back to the spin degrees of freedom, 
item {\it (1-2)} shows that domains are not independent with each other but 
the spins belonging to different domains are strongly correlated and are frozen, 
which is the source of the divergence of $\chi_{\rm SG}$ and the RSB 
at finite temperature, which may resemble the feature expected for an ordinary SG. 

Now, to clarify the origin of the features {\it (1)} and {\it (2)}, 
it is useful to start from the nonrandom $R=0$ state, 
and to consider that the long-range ordered clock phase is converted to QSSG at $R\ne 0$, 
since the regular superstructure of the clock phase 
naturally continues to the local superstructure inside the domains of QSSG. 

For the understanding of how such domains can be formed, 
we propose that Imry and Ma's argument\cite{imry-ma} 
about the formation of ferromagnetic domains by a random field 
in the simplest classical ferromagnet explains the essential part of the mechanism. 
They showed phenomenologically that {\it a small but nonzero static random field }
destroys a classical ferromagnetic long-range order and split it into 
well identified large domains at dimensions $d\le 2$, 
which was proved rigorously later on\cite{aizenman89}. 
Of course, Imry-Ma's case requires a random field that destroys a TRS and differs from our case with a TRS Hamiltonian. 
Nevertheless, there is the following reason why the Imry-Ma mechanism essentially applies. 

The honeycomb superstructure at $R=0$ has a three-fold degeneracy, 
since the three ways of assigning (A,B,C), (B,C,A), and (C,A,B) sites to (I, II, III) sublattices are 
energetically equivalent because of the translational symmetry. 
At $R\ne 0$, this degeneracy becomes incomplete, and at each local part of the system 
one of these three with the lowest energy density is selected. 
This {\it local energetics} has two aspects; 
(i) the honeycomb superstructure recognizes the bond randomness as a random ``field'', 
and wants to select a unique AB sublattice from the three patterns having a largest 
$\sum_{\langle i,j\rangle} J_{ij}$ over some lengthscale. 
At the same time, 
(ii) the exchange interactions acting on a C site from the spins on the neighboring A and B sites 
do not perfectly cancel anymore at $R\ne 0$. They work as a random longitudinal mean field on site-$i$ (C sites), 
$h_i= \sum_{j\in {\rm A,B}} J_{ij} \langle \sigma_j^z \rangle$ 
(see Fig.\ref{f1}(c)), and spins slightly cant off $x$-direction, while still keeping a 
considerably large $x$-component to maximize the energy gain including those 
from the transverse field. 
Therefore, the coupling of local three-fold state to the random bond interactions is, 
apart from the number of degrees of freedom being three instead of two, 
equivalent to the Imry-Ma's coupling of the order parameter (spin) with the random field, 
and split the superstructure into domains at $d=2$. 
Following Imry-Ma, according to the central limit theorem for fluctuations, 
the energy gain of choosing a particular type of domain of size $\ell$ 
from among the three has the energy gain of the order $O(\sqrt{S})$ with $S=\ell^d$, 
which can easily overwhelm the energy loss caused at the domain boundary scaled by $\propto \ell^{d-1}$. 

Despite the equivalency in the driving force to form domains, 
our domains have an important feature that qualitatively differs from Imry-Ma's domain. 
Again we focus on the TRS our Hamiltonian has. 
In Imry-Ma's case, 
the orientation of spins in each domain 
with the lowest energy is uniquely determined by the loss of TRS due to random fields. 
Hence, it does not cause a multi-valley free energy landscape characteristic of glass\cite{binder-young} 
although it yields $\langle q_{\alpha\beta}\rangle \ne 0$\cite{bray-moore,nattermann}. 
Whereas, the present case has an underlying complexity that allows for a multi-valley structure. 
Suppose that {\it at least one} of the domains are energetically isolated from the other part, 
namely if all the ABC-spins within a single domain or all the spins outside that domain 
might be turned over altogether without the energy loss, 
the degeneracy in the ground state would remain because of the TRS. 
A more important difference lies at the domain boundary; 
Imry-Ma's boundary for weak random fields is a ``hard domain boundary'' 
whose energy cost comes from the mismatch of interactions 
only 
along a thin domain boundary, 
and is determined solely by a domain-lengthscale $\ell$ as $\ell^{d-1}$. 
By contrast, the energy cost of our boundary 
depends on the choice of six different patterns of 
three-sublattice structures on both sides; 
how the path of boundaries is chosen are tightly correlated with the choice of these patterns. 
Resultantly, there can be various choices of shapes 
of domains and patterns of domains that may nontrivially 
yield a highly degenerate free energy. 

As shown in Fig.~\ref{f7}(e), 
the (I,II,III) sublattices filled with (A,B,C) spins on the left-hand side domain becomes (A,C,B) on the central domain and (C,A,B) on the right-hand side domain, 
exchanging the spins in a manner of C$\leftrightarrow$B and C$\leftrightarrow$A, respectively. 
The reason why the system chooses such configuration is as follows; 
Suppose that we create a domain wall (hard domain / one lattice spacing) consisting {\it only} of A and B spins. 
One can consider numerous low-energy patterns that the Ising energy density, 
namely, energy divided by the number of bonds after summing up the A-A, B-B, and A-B bond energies along the boundary, 
is nearly the same as the Ising energy density of the uniform clock phase without the boundary. 
However, this energy counting is not taking account of the energy gain by $\Gamma$ relevant to C-spins. 
Since to form such (A,B) domains, the density of C-spins around the domain wall is reduced, 
and the energy gain by $\Gamma$ is lost, 
they are energetically unfavorable compared to the case of a uniform clock state.  
In this way, to minimize the energy loss in creating a boundary, one needs to organize the
spins to prevent the decrease of the number of C-spins. This forces the two adjacent
domains to share either of A or B spin on one of the I/II/III sublattices as much as possible, 
as illustrated in Fig.~\ref{f7}(e) between domains colored with blue and pink.

At the same time, to gain the energy from randomness, ``soft domain walls" are formed as
we see in the snapshot in Fig.~\ref{f7}(b), with substantial proximity of the domain-wall region to
deep inside of the domains. 
This involved and soft structure seems to be 
the origin of the multi-valley landscape. In total, the patterns in the domain boundary separating the
neighboring domains supports the rigid correlations between the neighboring domains.

Let us interpret the above considerations on QSSG in 
the context of the conventional theories for SG. 
There had been a long-standing debate on which of the pictures, 
replica symmetry breaking(RSB) and a phenomenological droplet theory would properly capture the essence of SG. 
According to the droplet theory\cite{fisher88}, 
the energy cost of turning over all the spins belonging to a size-$\ell$ droplet scales as $\ell^\theta$, 
which is called domain-wall energy. 
The energy cost inside the droplet-domain away from the boundary is zero because of the TRS. 
There is an inequality $\theta \leq (d-1)/2$ and the upper bound corresponds to the aforementioned ``hard domain'' limit. 
However, this ``hard domain" is not necessarily realistic. 
The growth of spin-spin correlation over distances may soften the boundary and makes $\theta$ smaller. 
In 3D classical SG, a positive $\theta \sim 0.2$\cite{bray-moore84,huse85} was proposed to give a nonzero $T_c$, 
whereas in 2D, the reported $\theta <0$\cite{mcmillan84} implies that the domain-wall energy scales to zero in the thermodynamic limit, and we no longer have an SG transition at nonzero temperatures within this scheme. 
The droplet theory predicts the absence of RSB, and resultantly 
$P(q_{\alpha\beta})$ is expeced to have peaks at some $\pm q_{\alpha\beta}\ne 0$ and $P(q_{\alpha\beta}=0)=0$. 
\par
The counterpart replica theory~\cite{parisi79,parisi83,parisi80,parisi80-2,parisi80-3} 
for SG shows a RSB at $d=\infty$~\cite{SK1978,SK1975},  
which gives peaks in $P(q_{\alpha\beta})$ at both $q_{\alpha\beta}=0$ and $q_{\alpha\beta}\ne 0$. 
The theory asserts that, when the replica symmetry is broken, the energy cost to turn over a finite fraction of the spins stays 
a finite constant value at large system sizes implying $\theta\sim 0$. 
The 3D SG at nonzero temperature is still controversial; 
the most recent numerical studies find RSB behavior \cite{alvarez_ba_os2010} 
while the droplet behavior could only be recovered after a crossover for a very large system size 
often unreachable, which is difficult to exclude. 
\par
The droplet and replica theories have similarities each with different aspects of QSSG. 
First, one can apply the idea of a droplet to our clear and visible domain structure. 
A stable correlation of spins belonging to different domains is bridged by the optimal choice of spin patterns at the boundaries, 
which will yield a finite energy cost of turning over all the spins inside our domain. 
We thus find small but nonzero $\theta>0$ even though we are in 2D, which guarantees 
existence of domains at finite temperature. 
At the same time, totally different domain structures 
suggest the existence of numerous quasi-degenerate states, 
which contribute to the $\langle q_{\alpha\beta}\rangle=0$ replica peak. 
This point is different from the droplet theory, 
but shows essential consistency with the 1S-RSB in the replica theory. 
Although these two theories give some clues to understanding the vitrification mechanism, 
QSSG is not the subclass of conventional SG, since both theories contradict in other detailed aspects. 
\par
Finally, we point out that 
a 2D Ising model with a transverse field at zero temperature can be mapped to 
a 3D classical Ising model in the thermodynamic limit as the path integral formalism tells 
by introducing the imaginary time direction as the third dimension~\cite{dutta}.
This supports the existence of an SG phase at $T=0$ in 2D if the SG phase exists in a classical 3D EA model, 
although our model is not exactly equivalent to the 3D classical EA model in that the types of random bonds are different, 
and that our model gives uniform interaction along the imaginary time direction. 
The present results propose a way to stabilize the SG even at nonzero temperatures making use of this third dimension. 

\section{Summary and Outlook}
\label{sec:summary}
We have examined the Ising model in a transverse field on a triangular lattice with small randomness. 
Although the spin-glass phase in two dimensions has been argued as unstable at nonzero temperatures in most of the literature, 
we found the first example of such a glass phase. 
The synergy effect to have a structural glass and a spin glass works as an efficient glass former. 

We first consider a nontrivial honeycomb superlattice structure called ``clock order" 
that emerges due to {\it the order-by-disorder effect} in the absence of randomness. 
Because the triangular lattice possesses a strong geometrical frustration, the classical Ising spins do not order at all in a presence of substantial antiferromagnetic interactions. 
The quantum fluctuations induced by a transverse field release this frustration and the spins partially order antiferromagnetically and form a honeycomb superstructure. 
The rest of the spins located at the center of the honeycomb hexagon continue fluctuating 
(forming an off-diagonal long-range order). 
This process gives rise to two qualitatively different emergent degrees of freedom, 
a superstructure, and a quantum spin, out of a single Ising degree of freedom. 
\par
The bond randomness work as an emergent random field to this clock order, 
and vitrify it to structural glass. 
The structural glass, in turn, stabilizes the spin glass and the spin glass reinforces the structural glass. 
Such a synergy effect can overwhelm the effect of large fluctuations caused by the low dimensionality, 
and stabilize a glass. 
\par 
The mechanism of the glass stabilization relies on the fact that any long-range order can be destroyed by {\it a nonzero random ``field" conjugate to the order parameter} in 2D. 
In our case, bond randomness is a ``field" that breaks the structural bond order of spins. 
In general, since random fields uniquely determine the freezing pattern, 
it does not straightforwardly yield a glass. 
However, on top of that, we have a glassy correlation between the structural bond order and transverse spins. 
They work together and lead to a replica symmetry breaking not ever found 
in an SG for finite dimensions. 
\par
The SG transition is identified as the divergence of uniform SG susceptibility, 
namely a finite $\chi_{\rm SG}(k=0)/N >0$ in the thermodynamic limit. 
The correlation ratio $C_R$ which is the scale-invariant property that measures 
the peak height of $\chi_{\rm SG}(k=0)$ against the peak width, 
captures alternatively the existence of SG order as $C_R\rightarrow 0$. 
The QSSG satisfies both conditions for SG long range order. 
\par
At the same time, an unconventional feature arises as 
the emergent weight of $\chi_{\rm SG}(k>0)$ 
transferred from the $\chi_{\rm SG}(k=0)$  peak at $T<T_c$. 
This weight comes from the quasi-long-range ordered elements of glass, 
which is related to the domains growing on top of 
the underlying power-law correlation characteristic of the BKT phase. 
This contribution superimposes the power-law decaying correlation to the replica overlap parameter. 
Indeed, the replica overlap distribution $P(q_{\alpha\beta})$ starts to show two peaks at 
$q_{\alpha\beta}=0$ and $q_{\alpha\beta}\ne 0$. 
The former peak found at $T<T_c$ indicates that the two replicas 
no longer resemble because of the domain structures giving diffusive $\chi_{\rm SG}(k>0)$ elements, 
while the $q_{\alpha\beta}\ne 0$-peak continuing from the higher temperature 
contribute to the uniform SG of $\chi_{\rm SG}(k=0)$. 
\par
Our data fully supports the picture that the QSSG phase is a synergy of these two types of the glasses, 
namely, the uniform long-range SG order 
and the glass dominated by the structural domains on top of 
the power-law spin-spin correlation of the BKT phase. 
This duality has a tight connection with the coexisting structural and spin glasses. 
\par
Recently in quantum spin models,  
the existence of quantum spin liquids is established~\cite{kitaev2006,liao2017,yan2011,depenbrock2012,nishimoto2013,kaneko2014,yasir2016,shijie2019,becca2013,nomura2020}. 
In those cases, the spins are fluctuating and have a long-range quantum entanglement, but are not frozen. 
However, the spin correlation shows a power-law decay in the case of gapless algebraic spin liquids, 
which is apparently the same as the present case. 
Therefore, despite a crucial difference between the present quantum glass and the quantum spin liquids, 
the two may have similar types of quantum entanglement, which is to be clarified in future studies. 
\par 
Although the physics presented here may first seem rather specific to this quantum spin model, 
it can be shared with a far wider class of systems. 
One possible example is a family of order-disorder type dielectric materials 
in which the atoms or molecules form a bistable positional pseudo-spin degree of freedom, 
such as hydrogen-bonded materials\cite{hydrogen}, 
quantum paraelectric materials similar to SrTiO$_3$ or BaTi$_{1-x}$Zr$_x$O$_3$, 
and polarizable molecular 2D solids such as BEDT-TTF compounds\cite{ch,naka,peter}. 
When these atoms or molecules are coupled in a frustrated manner on periodic lattices, 
and if randomness is introduced, they may become a platform of our novel glass. 
On the application side, the 2D glass designed on surfaces or interfaces is potentially important 
since the multi-valley energy-landscape structure of a glass phase can be utilized 
for a future memory device. 
It had been believed that surfaces or interfaces are not favorable for this purpose since they are 2D systems. 
The present glass-forming mechanism may solve this practical issue. 
Finally, the transverse Ising model is at the core of the quantum annealing algorithm 
for quantum computing~\cite{annealing}, 
where our glass-forming mechanism may also be utilized to control the nature of phase transitions 
used for annealing\cite{martin-mayor15,katzgraber14}.

\appendix
\section{Relaxation process of QMC}
\label{app:relaxation}
The typical set of QMC calculations is given after the relaxation of 40,000 MCS and 
took averages over 1,000,000 MCS with a parallel run of 10-20 replicas (which give 45-190 replica overlaps) per each random sample. 
To check the validity of the results, we first calculated the above set of data 
using the random initial configuration of spins for each run. 
Then, we gave another set of the run, starting from the final state of the previous run (equilibrium state reached 
after 1,000,000 MCS). Both sets gave quantitatively good agreement. 
\par
In random/glassy systems in the vicinity of magnetic long range ordering, 
one needs to exclude the possibility that the system is trapped in the metastable state while there is a 
true ordered state as a true thermal equilibrium. Such a situation often happens for systems that undergo 
a first-order transition, and the metastable state which avoids crystallization to the ordered state by 
making use of its high configurational entropy is called supercooled liquid in structural glass. 
\par
\begin{figure}[tbp]
\includegraphics[width=8.5cm]{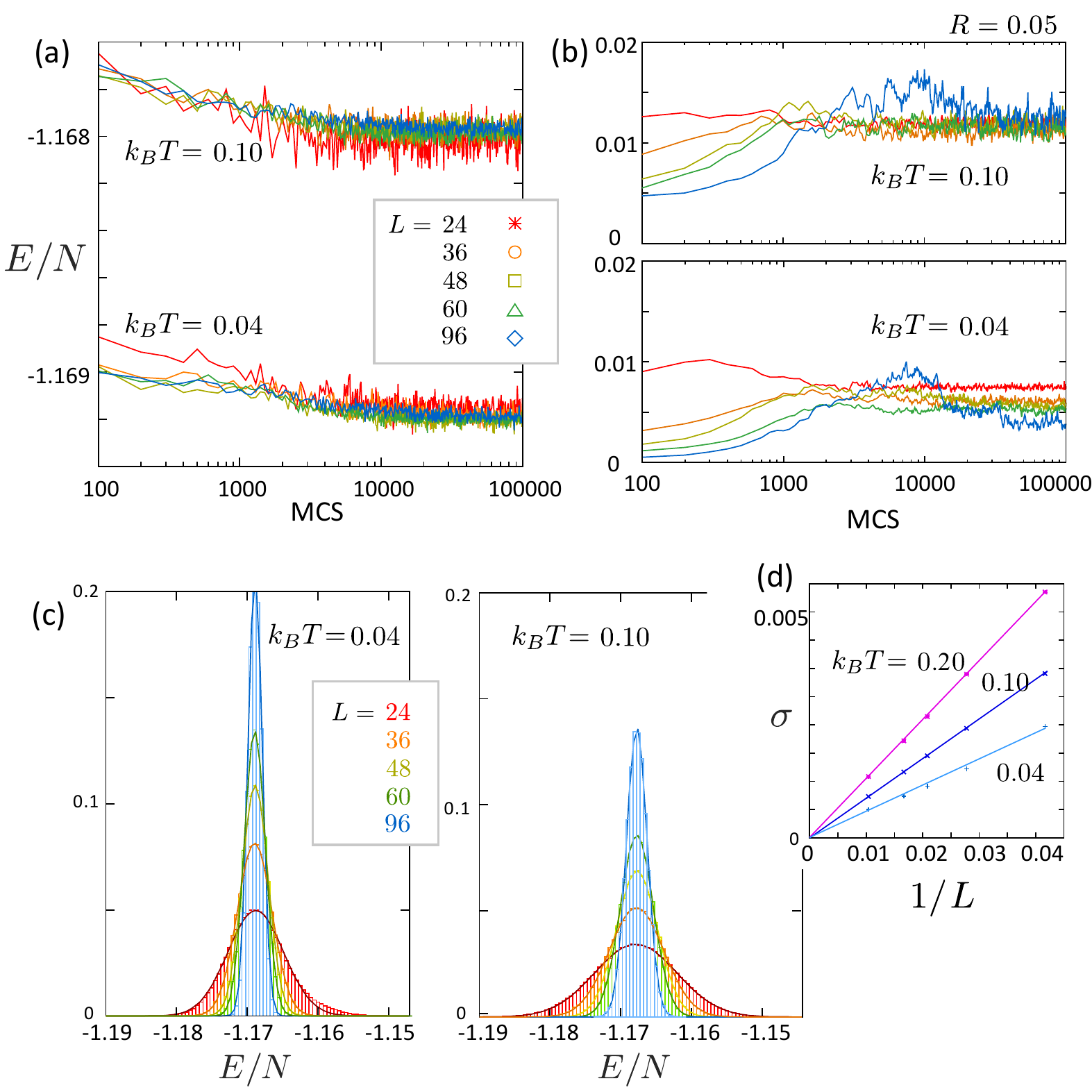}
\caption{(a,b) Relaxation process of energy density $E/N$ and $C_R(\bm k=0)$ as a function of MCS. 
The initial states are prepared as the equilibrium states of $R=0$ of the same temperatures, 
$k_BT=0.04$(clock phase) and $0.1$(BKT phase), and we quenched the system by $R=0.05$, 
with the random average taken for 5-10 samples. 
The data is presented for every 100 MCS (200-500MCS at larger MCS). 
(c) Energy histram of $L=24,36,48,60,96$ 
taken over more than 60,000 MCS after the relaxation process in (a) for two temperatures and are normalized. 
The Gaussian fits are shown in solid lines, and (d) their variance $\sigma$ is plotted 
as a function of $1/L$. 
}
\label{faadd1}
\end{figure}

To confirm that QSSG state is different from the uniform clock phase or other uniform phases at $R=0$, 
we first prepared several equilibrium state of the clock state ($k_BT=0.04$) and the uniform BKT state ($k_BT=0.1$) 
both at $R=0$ by performing a standard QMC calculation. 
Then, we quenched the system by introducing $R=0.05$ and performed a QMC calculation to more than 100,000 MCS. 
The relaxation process is recorded by taking the average of the physical quantities per every 100 MCS. 
In Fig.~\ref{faadd1}(a) the evolution of energy density $E/N$ averaged over about 5-10 random samples are shown for several different 
system sizes. We find that for all system sizes, the energy relaxes to similar values which 
are the same as the one we obtained previously. 
The correlation ratio $C_R(\bm k=0)$ develops systematically to the values shown in Fig.~\ref{fadd1}(a). 
To exclude the possibility of phase separation or coexistent two phases, 
we take an energy histogram after relaxation in Fig.~\ref{faadd1}(c), where we find a single peak structure 
fitted well by the Gaussian. The width of Gaussian plotted against $1/L$ in Fig.~\ref{faadd1}(d) shows that 
these peaks approach the delta function in the thermodynamic limit. 
These results indicate that the QSSG phase we found is in the thermal equilibrium state. 
\section{Spin glass susceptibilities}
\label{app:chisg}
As mentioned in the main text, there are several definitions of spin glass susceptibilities 
depending on the model and the nature of the target phase. 
Here, we overview the derivation of Eq.(\ref{chisg}) that serves as a susceptibility of 
replica overlap $q_{\alpha\beta}$ that signals the ergodicity breaking, 
following Ref.[\onlinecite{parisi89}]. 
Let us introduce a snall positive interaction $\lambda>0$ that couples the spins on two replicas $\alpha$ and $\beta$ as
\begin{eqnarray}
{\mathcal H}_{\alpha+\beta}= {\mathcal H}_\alpha + {\mathcal H}_\beta - \lambda \sum_{i=1}^N \sigma_{i;\alpha}^z\sigma_{i;\beta}^z
\end{eqnarray}
with ${\mathcal H}_\alpha$ being the Hamiltonian of replica-$\alpha$. 
We can compute replica overlap parameter conjugate to $\lambda$ as 
\begin{eqnarray}
&& \langle q_{\alpha\beta} \rangle= \lim_{\lambda\rightarrow +0} \lim_{N\rightarrow \infty} Q_{\alpha\beta}(N,\lambda) 
\nonumber \\
&& Q_{\alpha\beta}(N,\lambda) = \frac{1}{N}\frac{\partial}{\partial\lambda} 
(-k_BT)\ln \big[ {\rm Tr} e^{-\beta {\mathcal H}_{\alpha +\beta} }\big]  
\nonumber \\
&& \rule{15mm}{0mm}=  \frac{1}{N}\sum_{i=1}^N \langle \sigma_{i;\alpha}^z\sigma_{i;\beta}^z \rangle_\lambda,  
\end{eqnarray}
where $\langle \cdots\rangle_\lambda$ is a thermal ensemble average at $\lambda>0$. 
Here, if we take $\lambda\rightarrow +0$ prior to $N\rightarrow\infty$, 
the state can keep the ergodicity at finite size and we find 
$Q_{\alpha\beta}(N<\infty,\lambda\rightarrow 0)=0$. 
Taking $N\rightarrow\infty$ first we can safely detect the breaking of ergodicity by using $q_{\alpha\beta}$. 
The susceptibility about $\lambda$ is naturally derived as
\begin{eqnarray}
\chi_{SG}&=&\frac{\partial Q_{\alpha\beta}(N,\lambda)}{\partial \lambda}\bigg|_{\lambda=0}
\nonumber\\ 
&=& \frac{1}{N} \Big( \sum_{i,j}\langle \sigma_{i;\alpha}^z\sigma_{i;\beta}^z \sigma_{j;\alpha}^z\sigma_{j;\beta}^z \rangle_{\lambda=0}
  - \langle \sigma_{i;\alpha}^z\sigma_{i;\beta}^z \rangle^2_{\lambda=0} \Big)
\nonumber \\
&=&  \frac{1}{N} \sum_{i,j} \Big( \langle \sigma_{i}^z\sigma_{j}^z \rangle^2
     -\langle \sigma_{i}^z\rangle^2\langle \sigma_{j}^z \rangle^2 \Big)
\nonumber \\
&=&  \frac{1}{N}  \sum_{i,j} \big( \langle q_{i;\alpha\beta} q_{j:\alpha\beta} \rangle
     -\langle q_{i;\alpha\beta} \rangle \langle q_{j:\alpha\beta} \rangle \Big), 
\label{eq:appsg}
\end{eqnarray}
where $q_{i;\alpha\beta}=\sigma_{i;\alpha}^z\sigma_{i;\beta}^z$ and 
we dropped the subscript $\alpha,\beta$ when it is replaced by the single-replica average. 
The last third line is obtained by factorizing 
$\langle N^{-1}\sum_{i=1}^N \sigma_{i;\alpha}^z\sigma_{i;\beta}^z \rangle
=N^{-1}\sum_{i=1}^N  \langle \sigma_{i;\alpha}^z \rangle \langle \sigma_{i;\beta}^z \rangle$, 
since the replica's are independent at $\lambda=0$. 
However, since  Eq.(\ref{eq:appsg}) is obtained by taking $\lambda\rightarrow +0$, 
we implicitly assume $\langle q_{i:\alpha\beta} \rangle \ge 0$. 
Indeed, in the numerical simulation for $\lambda=0$ in a finite size system, 
the distribution function $P(q_{\alpha\beta})$ distributes over $\pm q_{\alpha\beta}$ 
and we find $\langle q_{i:\alpha\beta} \rangle=0$ in practice. 
The physically meaningful evaluation of Eq.(\ref{eq:appsg}) is done by 
symmetrizing $P(q_{\alpha\beta})$ and confining ourselves to 
$q_{\alpha\beta}>0$ by a proper normalization. 
In this way, we find Eq.(\ref{eq:appsg}) as an equivalent form of Eq.(\ref{chisg}) we adopt 
in the main calculations in Fig.~\ref{f2}. 
\par
Suppose we are dealing with a glassy phase having a multi-valley landscape in the free energy. 
If the inter-valley potential wall between valleys develops by $\Theta(N)$, 
the state breaks the ergodicity and we find $\chi_{SG}/N >0$. 
In a coexistent phase of ferromagnetic ordering and SG, the definition 
$\chi_{\rm SG}^m=\frac{1}{N} \sum_{i,j=1}^N
\overline{( \langle \sigma_i^z\sigma_j^z \rangle -\langle \sigma_i^z\rangle\langle \sigma_j^z \rangle)^2}$,
is often adopted, which is qualitatively equivalent to Eq.(\ref{chisg}) and give the consistent value with $\chi_{\rm SG}$. 
\par
\begin{figure}[tbp]
\includegraphics[width=8cm]{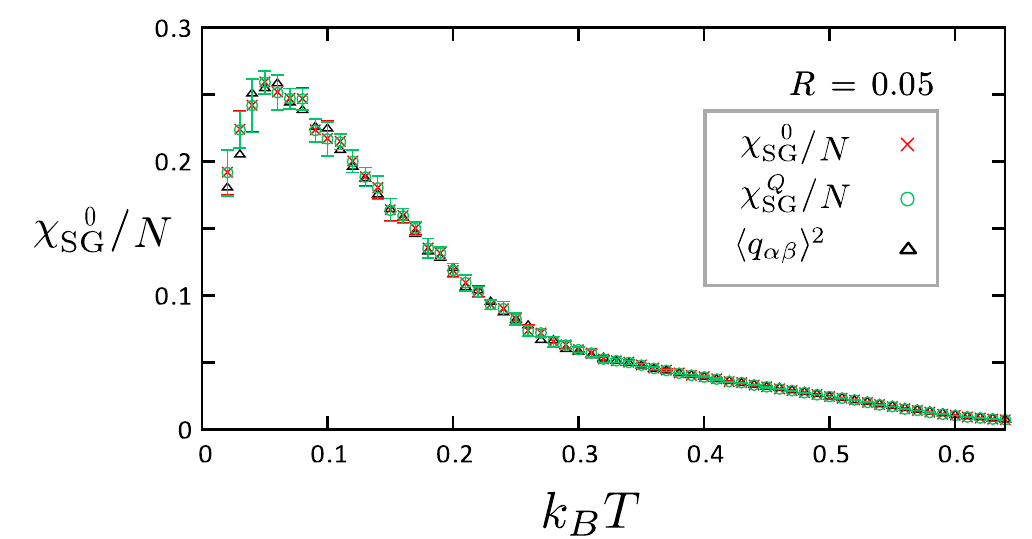}
\caption{Comparison of $\chi_{\rm SG}^0$ following the first definition of Eq.(\ref{chisg0}), 
$\chi_{\rm SG}^{\it Q}$ and $\langle q_{\alpha\beta}^2\rangle$ 
at $R = 0.05$, $\Gamma= 0.4$ and $L=32$. 
}
\label{fa1}
\end{figure}

Since we are dealing with quantum model, 
where the spin confiturations acquire an imaginary time degrees of freedom $\tau=1-(k_BT)^{-1}$, 
we need to confirm whether a {\it quantum} spin glass susceptibility, 
\begin{equation}
\chi_{\rm SG}^{\it Q}=\frac{(k_BT)^2}{N} \sum_{i,j=1}^N 
\overline{\Big\langle \int_0^{(k_BT)^{-1}} d\tau \sigma_i^z(0)\sigma_j^z(\tau) \Big\rangle^2}, 
\label{chisgq}
\end{equation}
behave consistent with $\chi_{\rm SG}^0$. 
As we show in Fig.~\ref{fa1}, $\chi_{\rm SG}^0$ and $\chi_{\rm SG}^{\it Q}$ 
obtained by the spin-spin correlation are almost identical, 
and so as $\langle q_{\alpha\beta}^2\rangle$ evaluated from the replica overlaps. 
The consistency is valid regardless of $L$ and model parameters. 
One can see by comparing Fig.~\ref{fa1} with Fig.~\ref{f2}(a), that 
the profile of $\chi_{\rm SG}$ is perfectly reproduced by reducing 
$\chi_{\rm SG}^0$ by about 20$\%$ over the whole temperature range. 
%
\begin{figure}[tbp]
\includegraphics[width=7.5cm]{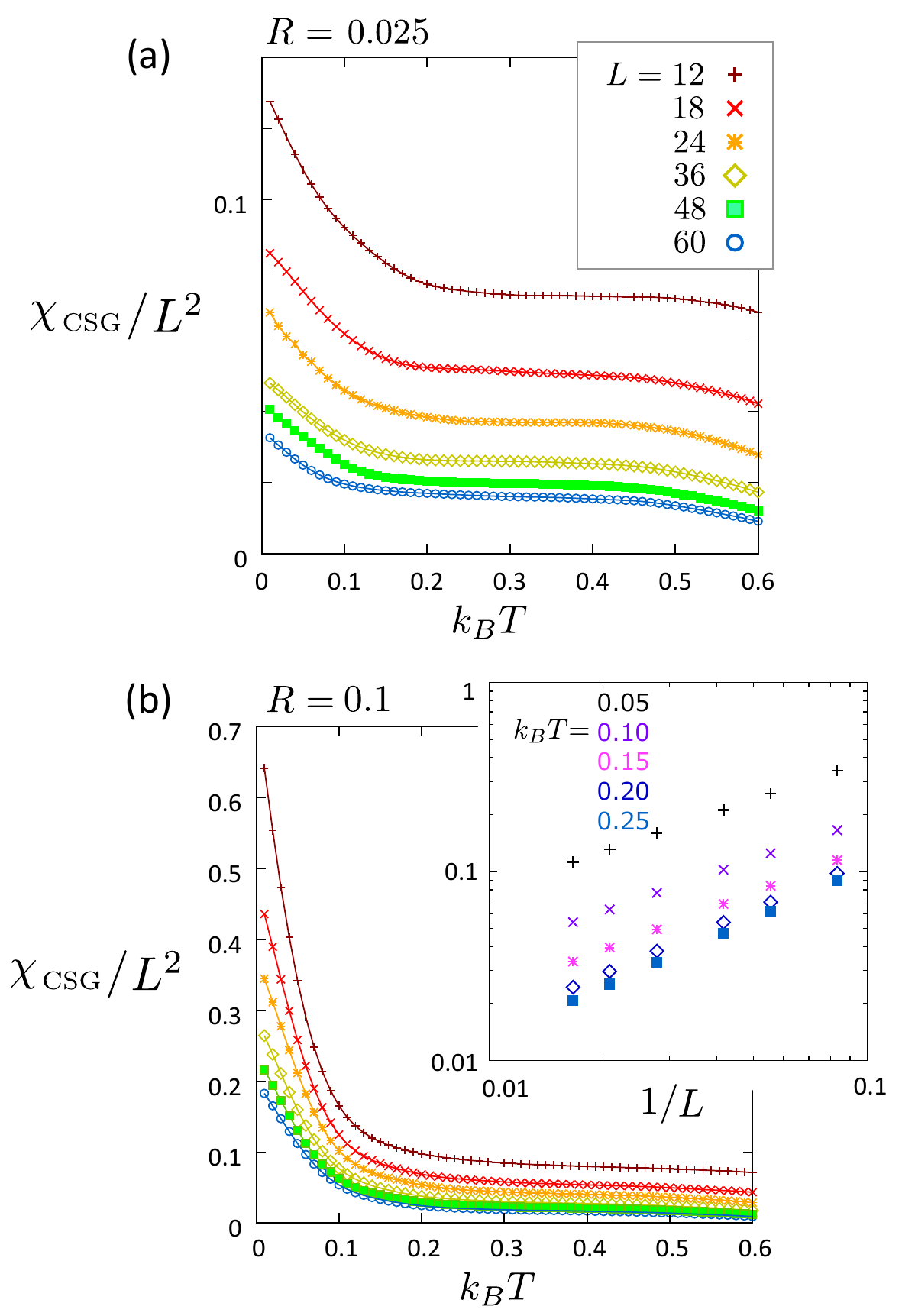}
\caption{(a) Classical spin glass susceptibility $\chi_{\rm CSG}/L^2$ at $\Gamma=0$ 
for (a) $R = 0.025$ and (b) 0.1. The $1/L$ dependence for several 
$k_BT$’s are shown in the inset.
}
\label{fa2}
\end{figure}
\begin{figure}[t]
\includegraphics[width=6.5cm]{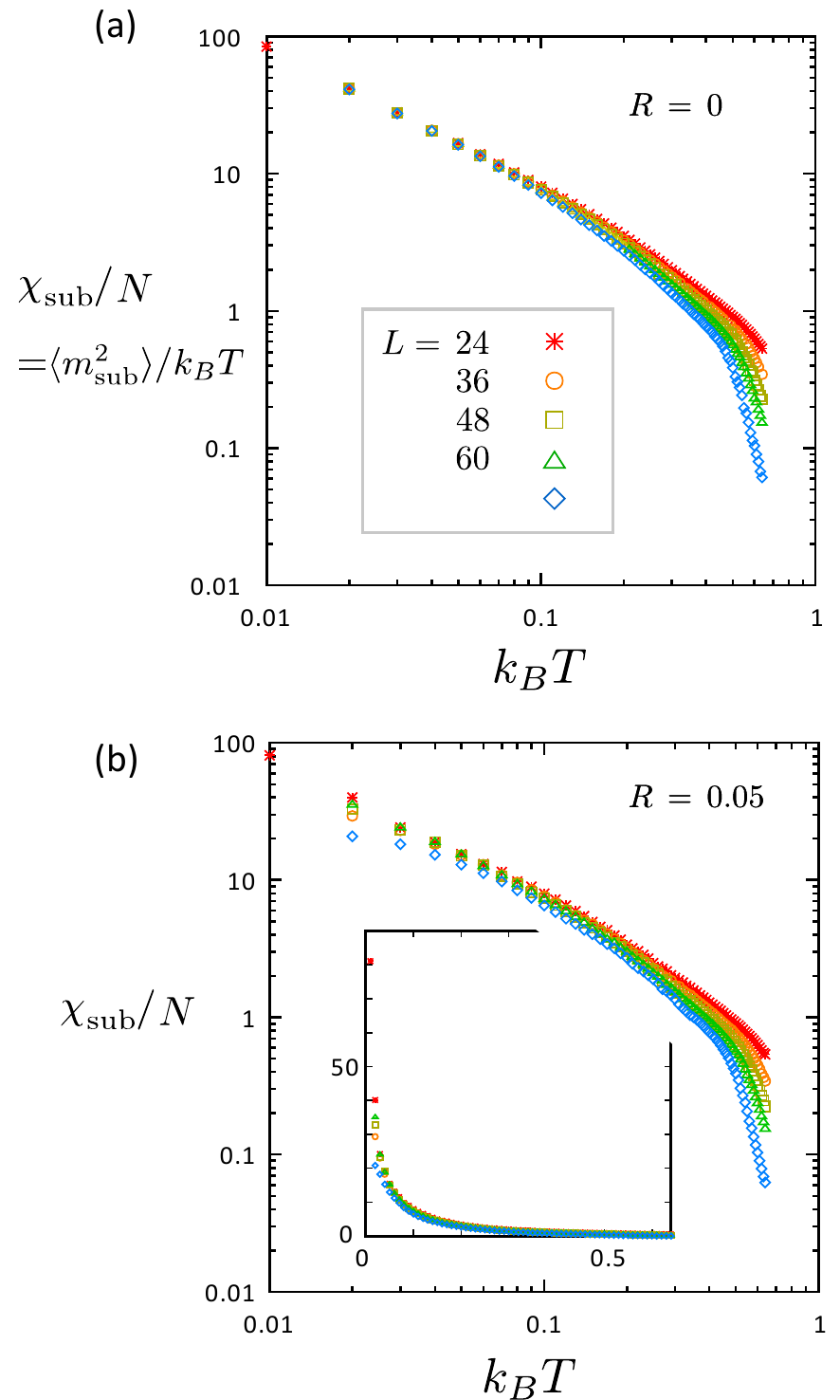}
\caption{Three-sublattice susceptibility in Eq.(\ref{eq:chisub}) for (a) $R=0$ and (b) $R=0.05$ 
with $L=24,36,48,60,96$. 
}
\label{fa4}
\end{figure}
\begin{figure*}[t]
\includegraphics[width=16cm]{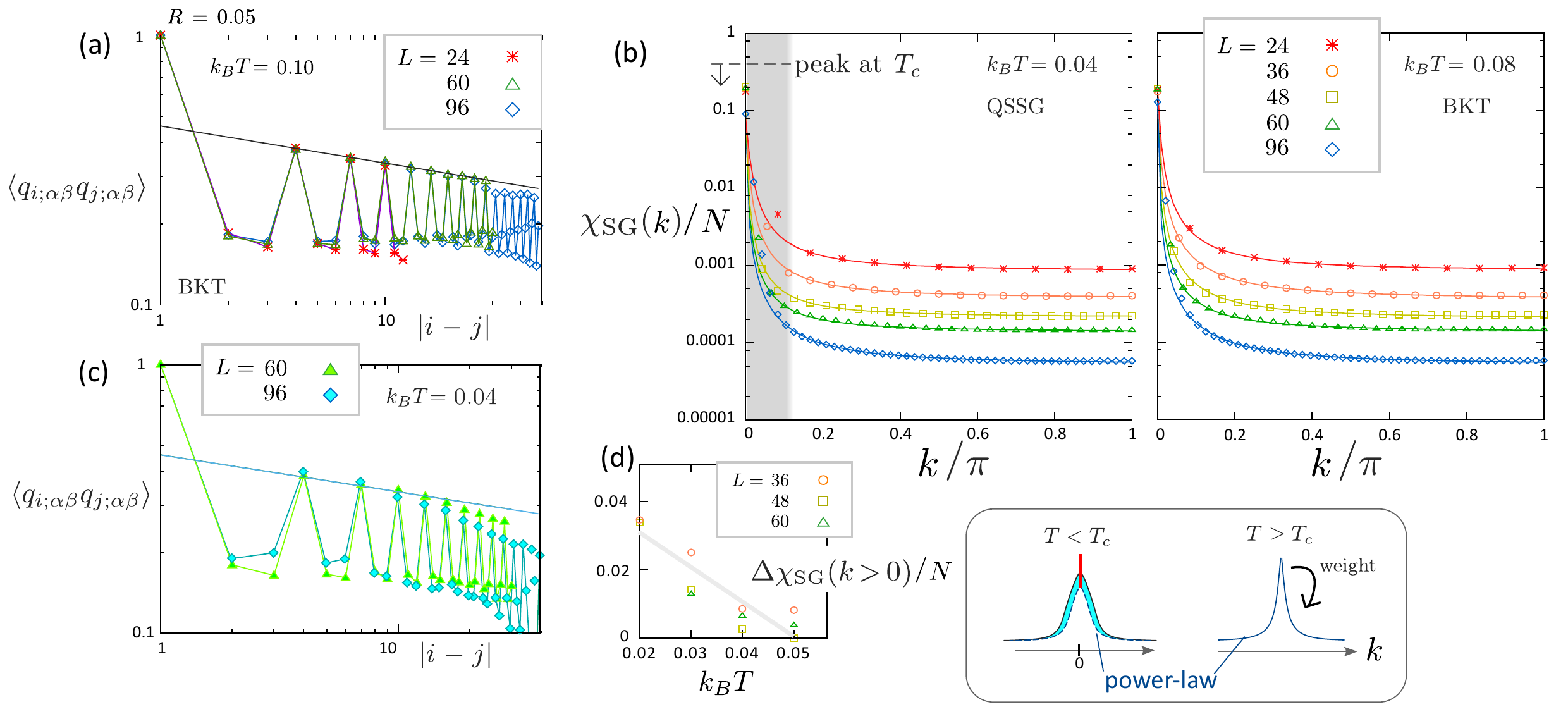}
\caption{(a) Spatial correlation of replica overlap parameter 
$\langle q_{i;\alpha\beta}q_{j;\alpha\beta}\rangle$ 
as a function of $|i-j|/L$ for $L=24,60,96$ at $R=0.05$ and $k_BT=0.1$ in the BKT phase. 
The solid line is the power function $\propto r^{-p}$ with $p=0.137$. 
(b) $\chi_{\rm SG}(k)$ in Eq.(\ref{chisgk}) at $k_BT=0.04$(QSSG phase) and $0.08$(BKT phase) at $R=0.05$ and $\Gamma=0.4$, 
for two directions $\bm k=(k,0)$ and $(0,k)$ averaged. 
At $T<T_c$ peak-height decreases from $T=T_c$ and its weight shifts to the small nonzero-$k$ values. 
The data points at $k\gtrsim 0.1\pi$ are found to be identical between different temperatures below and above $T_c$. 
Solid lines are the power functions  $f(k)=a/(1+(k/\gamma)^b)+{\rm const}$ 
fitted for data off the shaded region; 
in all data for two temperatures, we we are able to adopt common power, $b=1.45$. 
In the BKT phase, a single curve fits all the data well, but for QSSG, the $k\lesssim 0.1\pi$ points 
fall off from the curve (even if we change $b$ or the choice of power functions). 
For $k_BT=0.04$, $k=0$ weight at higher temperature ($T=T_c$) 
is shifted to the small $k\ne 0$ weight (see the bottom panel). 
(c) Correlation $\langle q_{i;\alpha\beta}q_{j;\alpha\beta}\rangle$ at $T<T_c$ and $R=0.05$ 
showing the drop from the power-law decay $\propto r^{-p}$ (the same solid line as panel (a)) 
at long distances, which is the indication of finite $\Delta\chi_{\rm SG}(k>0)/N$. 
(d) Rough estimate of the shifted weight, $\Delta\chi_{\rm SG}(k>0)/N$, 
obtained by subtracting $f(k)$ from $\chi_{\rm SG}(k>0)/N$ at $T<T_c$. 
}
\label{faadd2}
\end{figure*}
\section{Classical spin glass at $\Gamma=0$}
\label{app:clchisg}
To evaluate the properties of the $\Gamma= 0$ limit, namely the classical Ising model on the triangular lattice, 
we separately performed the calculation using the classical exchange Monte Carlo Method. 
Representative results of spin glass susceptibility are shown in Fig.~\ref{fa2}, which we denote $\chi_{\rm CSG}$ 
(the same definition as Eq.(\ref{chisg}) but with classical variables $\sigma_i=\pm 1$ ) 
to clarify that they are obtained in classical calculation. 
For larger $R$, the value of $\chi_{\rm CSG}$ at low temperature is enhanced,
while it always takes the smaller value for larger $L$. 
The power law $1/L$-dependence is shown in the inset, namely
$\chi_{\rm CSG}/L^2 \rightarrow 0$ at $L\rightarrow\infty$ for all values of $R$ down to lowest temperature. 
These results indicate that the spin glass is present only in the ground state. 
It is consistent with the overall consensus on the two-dimensional Ising
model with quenched randomness that the finite temperature spin glass phase cannot exist\cite{young83,parisi98,mcmillan83,houdayer01}. 
\section{Sublattice magnetic susceptibility}
\label{app:subsus}
We show in Fig. \ref{fa4} the three-sublattice susceptibility as a function of temperature for different $L$. 
The data of $R = 0$ and $R = 0.05$ do not differ much except for a 
slight size-dependence at low temperature found in $R =0.05$ which is consistent with Fig.~\ref{f3}(a).  
\section{Details of correlation ratio and $\chi_{\rm SG}(\bm k)$}
\label{app:corratio}
\subsection{Scale-free behavior in the BKT phase}
Suppose that the real space correlation of the replica overlap parameter on site-$i$ and $j$ 
shows a power-law decay with distance as 
$\langle q_{i;\alpha\beta}q_{j;\alpha\beta}\rangle \sim |i-j|^{-p}$ at long distances, 
which is the natural assumption when the system is in the BKT phase. 
Then, one can roughly evaluate the SG susceptibility in 
finite systems of length $L$ as 
\begin{equation}
\chi_{\rm SG}^0(\bm k) \sim \int d\bm r e^{i \bm k\bm r } r^{-p} 
=\int_0^{2\pi}d\theta \int_\epsilon^L rdr r^{-p}e^{ikr\cos\theta}.
\rule{5mm}{0mm}
\end{equation}
In calculating the correlation ratio we choose the shortest wave number $dk=2\pi/L$, 
and for the two specific choices $k=0,\; dk$, we are able to perform the above integral as 
\begin{eqnarray}
&&\chi_{\rm SG}^0(0) \sim \int_\epsilon^L r^{1-p}dr  \sim 
\left\{\begin{array}{ll} L^{2-p} & (p<2) \\
\ln L & (p=2)\\
{\rm const} \rule{5mm}{0mm}& (p>2) 
\end{array}
\right.
\\
&& \chi_{\rm SG}^0(dk) \sim 
\int d\theta \!\int^{dkL\cos\theta} \hspace{-3mm}\!e^{i y} y^{1-p} dy (dk)^{p-2}
\nonumber \\
&&  \rule{12mm}{0mm} \sim 
\left\{\begin{array}{ll} (dk)^{p-2} & (p<2) \\
{\rm const}\times \ln L & (p=2)\\
\chi_{\rm SG}^0(0) \rule{5mm}{0mm} & (p>2) 
\end{array}
\right. 
\end{eqnarray}
These results will roughly give us an estimation about the correlation ratio; 
\begin{eqnarray}
C_R(\bm k=0)\sim 
\left\{\begin{array}{ll} (dk\times L)^{p-2}={\rm const} & (p<2) \\
{\rm const} & (p=2)\\
1 \rule{5mm}{0mm} & (p>2) 
\end{array}
\right. 
\end{eqnarray}
Therefore, $C_R(\bm k=0)$ does not depend on system size $L$. 
Figure~\ref{faadd2}(a) shows the spatial dependence of 
$\langle q_{i;\alpha\beta}q_{j;\alpha\beta}\rangle$ which clearly shows a power-law decay 
at long distances, whose power is given as $p\sim 0.14$. 
This is the reason for the nearly $L$-free behavior of $C_R(\bm k)$ shown in Fig.~\ref{fadd1} 
(the same discussion applies for $\bm k=0$ and $\bm Q$). 

\subsection{$k$-dependence of the correlation ratio}
To clarify the origin of the low-temperature behavior of the correlation ratio, 
we plot $k$-dependence of $\chi_{\rm SG}(k)$ averaged for $\bm k=(0,k)$ and $(k,0)$ 
in Fig.~\ref{faadd2}(b). 
The data is fitted by the power function $f(k)=a/(1+(k/\gamma)^b)+{\rm const}$ shown in solid lines, 
where we chose the optimal value of power $b=1.45$ for all data sets below and above $T_c$. 
In the BKT phase, all data points are fitted by $f(k)$ with the same power. 
However, in the QSSG phase at $k_BT=0.04$, several data points close to $k\sim 0$ but $k\ne 0$ 
(shaded region) 
show increase off $f(k)$ (or equivalently from the ones at $k_BT=0.08$), 
while in most of the region away from these points, i.e. $k \gtrsim 0.1\pi$, 
the temperature dependence is almost negligible. 
In fact, $f(k)$ or other choices of power function with a single peak do not fit the data at $T<T_c$. 
\par
The power function $f(k)$ indicates a robust background BKT-algebraic correlation that sustain at $T<T_c$. 
Then, the natural interpretation of this result is that at $T<T_c$ 
there appears an extra $\chi_{\rm SG}(k>0)$ (shaded region) component on top of the power function. 
At the same time, $\chi_{\rm SG}(k=0)/N$ drops at $T<T_c$ (see Fig.~\ref{f4}); 
in the left panel of Fig.~\ref{faadd2}(b) it is observed as the decrease of the $k=0$ peak 
from peak-value at $T_c$ (dotted line). 
Figure~\ref{faadd2}(c) shows the replica overlap correlation function 
at $k_BT=0.04$ to be compared with panel (a), where we draw the same power-law-fitted solid line. 
The short range correlation is the same from panel (a), 
while there is a decrease from the solid line at long distances, 
and this decrease explains the increase of the $\chi_{\rm SG}(k>0)$-weight. 
\par
In Fig.~\ref{faadd2}(d) we plot a rough estimate of the $k\ne 0$ component off the power function, 
$\Delta\chi_{\rm SG}(k>0)/N$. Its amplitude increases in lowering the temperature, 
and is consistent with the magnitude of the drop of $\chi_{\rm SG}(k=0)/N$. 
\par
The bottom inset of Fig.~\ref{faadd2} shows schematically a change in the peak profile 
below and above $T_c$. 
As we mentioned in the main text, the emergent peak of 
$P(q_{\alpha\beta}=0)>0$ at $T<T_c$ indicates 
that the finite fraction of replica overlaps do not resemble, 
which was ascribed to the emergent domain structures. 
In this Appendix, we additionally showed the relevance of this $q_{\alpha\beta}=0$-peak with the 
drop of spatial correlation only at long distances, 
Namely, $q_{i;\alpha\beta}$ and $q_{j;\alpha\beta}$ are 
algebraically correlated at short distances,
but becomes uncorrelated at long distances. 
It fits with the domain scenario,
since numerous different configurations of domains joining a thermal ensemble average 
rumple $\langle q_{i;\alpha\beta}q_{j;\alpha\beta}\rangle$ at long distances, 
where we naturally expect an algebraic glass behavior. 
\begin{figure}[t]
\includegraphics[width=8.5cm]{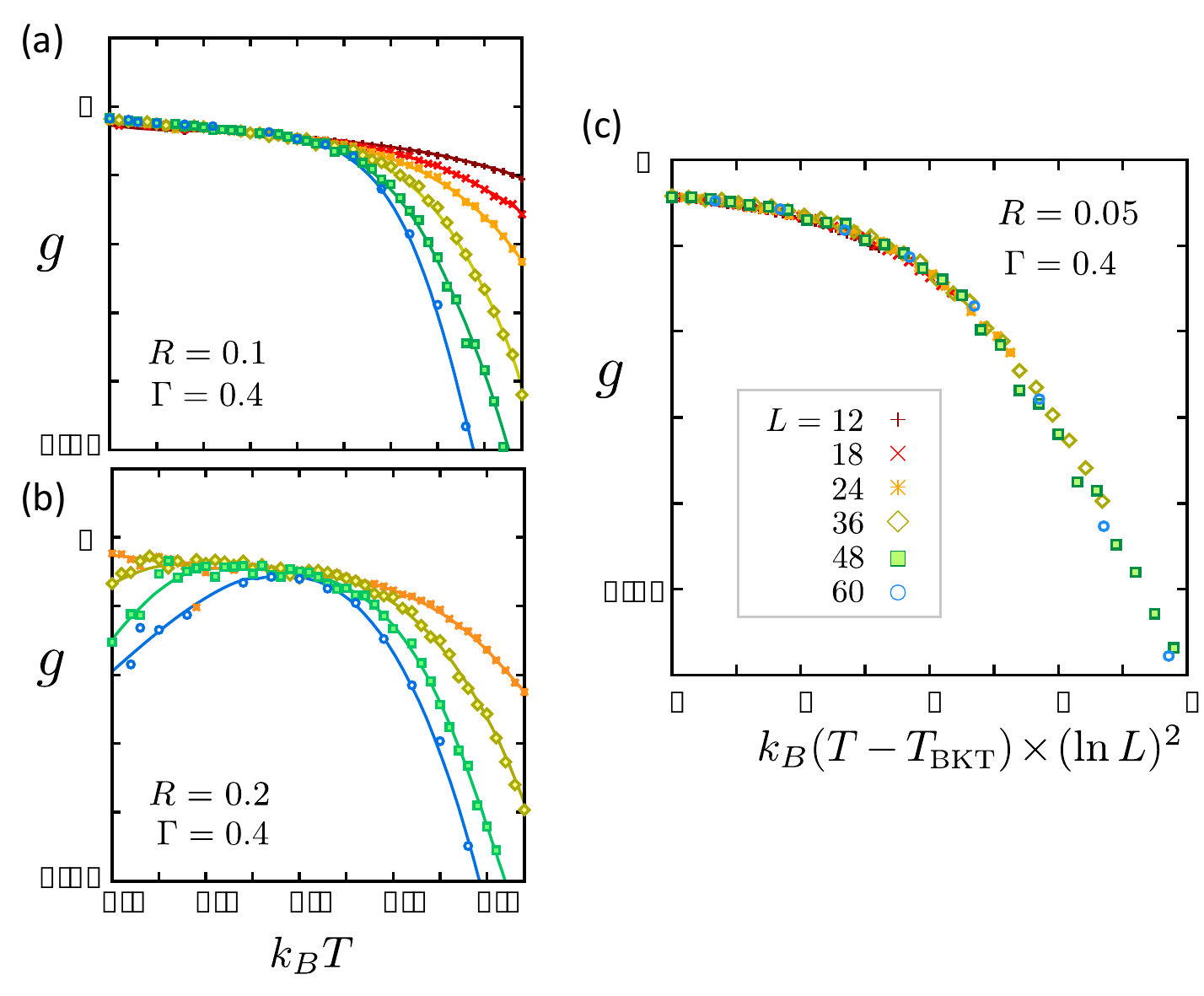}
\caption{
Binder ratio $g$ of the sublattice magnetization $m_{\rm sub}$, 
indicating (a) the presence ($R=0.1, 0.05)$  and 
(b)the absence ($R=0.2$) of the BKT transition at $\Gamma=0.4$. 
(c) Finite size scaling analysis is performed which gives $k_{\rm B}T_{\rm BKT}=0.372$. 
}
\label{fa3}
\end{figure}
\section{BKT transition}
\label{app:bkt}
It is known from Ref.[\onlinecite{isakov03}] that the finite temperature BKT transition takes place 
in the transverse Ising model ($R=0$) when $0<\Gamma/J \lesssim 1.7$. 
Since the BKT transition has a topological nature, it should be insensitive to the small perturbation, 
and thus we expect the BKT phase to extend toward $R \ne 0$. 
In the same way, the clock phase at a lower temperature is protected by the BKT phase just above, 
the QSSG phase that smoothly extends from the clock phase in the same temperature region at $R>0$ should require a BKT phase. 
\par
The BKT transition point can be located in several ways. 
The standard measure is the dimensionless Binder ratio, 
\begin{equation}
g=\frac{1}{2}\Big(3-\frac{\langle m^4_{\rm sub}\rangle}{\langle m_{\rm sub}^2\rangle}\Big), 
\label{binder}
\end{equation}
which has a zero scaling dimension and collapses to a single curve for all $L$ 
in the critical phase at $T\le T_{\rm BKT}$. 
Figures~\ref{fa3}(a) and \ref{fa3}(b) show $g$ as a function of temperature for $R=0.1$ and $0.2$. 
The curves at $R=0.1$ collapse for all different $L$'s below $T \lesssim 0.4$ while the ones at $R=0.2$ do not, 
indicating that the BKT phase disappears somewhere between the two parameter values. 
The transition point is evaluated more precisely by the finite size scaling analysis. 
Since the correlation length follows Eq.(\ref{eq:xi}), 
the Binder ratio should scale as $g= \tilde g \:[ (\ln L)^2 (T-T_{\rm BKT})]$. 
In Fig.~\ref{fa3}(c) we show the collapse of $\tilde g$ which gives $T_{\rm BKT}=0.37(2)$ 
for $R=0.05$ and $\Gamma=0.4$. 
As discussed in Ref.[\onlinecite{isakov03}], at finite $L$ there is a finite correction 
to the scaling and the obtained $T_{\rm BKT}$ is rather an overestimate. 
To locate $T_{\rm BKT}$ in the phase diagram, we instead use the size dependence of the three sublattice magnetic 
susceptibility in Eq.(\ref{eq:chisub}) which follows $\chi_{\rm sub} \propto L^{2-\eta}$, 
and evaluate the transition point at which the critical exponent takes $\eta=1/4$ (see Fig.\ref{f4}(d)). 
The actual behavior of $\chi_{\rm sub}$ is given in Appendix \ref{app:subsus}. 
The obtained $T_{\rm BKT}$ is approximately three quarters the values obtained from $g$ 
and is consistent with the previous studies for $R=0$. 
\begin{acknowledgments}
We thank Hajime Yoshino and Atsushi Ikeda for discussions, 
and Natalia Drichko, Peter Armitage, and Hikaru Kawamura for critical reading of the manuscript. 
This work is supported by 
JSPS KAKENHI Grants  No. JP17K05533, No. JP18H01173, No.JP21K03440, 
and No. 16H06345, 
by the MEXT HPCI Strategic Programs, and by 
the Creation of New Functional Devices and High-Performance Materials 
to Support Next Generation Industries (CDMSI). 
The calculation was done using the facilities of the Supercomputer
Center, the Institute for Solid State Physics, the University of Tokyo, and 
by the supercomputer at Yukawa Institute for Theoretical Physics in Kyoto University. 
M.I. was supported in part by the projects conducted under MEXT Japan named as
``Program for Promoting Research on the Supercomputer
Fugaku" in the subproject, ``Basic Science for Emergence and Functionality
in Quantum Matter: Innovative Strongly-Correlated Electron Science by Integration
of Fugaku and Frontier Experiments". We also thank the support by the RIKEN Advanced
Institute for Computational Science through the HPCI System Research project
(hp190145, and hp200132, hp210163). 
\end{acknowledgments}
\bibliography{qglass_rev}

\begin{thebibliography}{99}%
\makeatletter
\providecommand \@ifxundefined [1]{%
 \@ifx{#1\undefined}
}%
\providecommand \@ifnum [1]{%
 \ifnum #1\expandafter \@firstoftwo
 \else \expandafter \@secondoftwo
 \fi
}%
\providecommand \@ifx [1]{%
 \ifx #1\expandafter \@firstoftwo
 \else \expandafter \@secondoftwo
 \fi
}%
\providecommand \natexlab [1]{#1}%
\providecommand \enquote  [1]{``#1''}%
\providecommand \bibnamefont  [1]{#1}%
\providecommand \bibfnamefont [1]{#1}%
\providecommand \citenamefont [1]{#1}%
\providecommand \href@noop [0]{\@secondoftwo}%
\providecommand \href [0]{\begingroup \@sanitize@url \@href}%
\providecommand \@href[1]{\@@startlink{#1}\@@href}%
\providecommand \@@href[1]{\endgroup#1\@@endlink}%
\providecommand \@sanitize@url [0]{\catcode `\\12\catcode `\$12\catcode
  `\&12\catcode `\#12\catcode `\^12\catcode `\_12\catcode `\%12\relax}%
\providecommand \@@startlink[1]{}%
\providecommand \@@endlink[0]{}%
\providecommand \url  [0]{\begingroup\@sanitize@url \@url }%
\providecommand \@url [1]{\endgroup\@href {#1}{\urlprefix }}%
\providecommand \urlprefix  [0]{URL }%
\providecommand \Eprint [0]{\href }%
\providecommand \doibase [0]{http://dx.doi.org/}%
\providecommand \selectlanguage [0]{\@gobble}%
\providecommand \bibinfo  [0]{\@secondoftwo}%
\providecommand \bibfield  [0]{\@secondoftwo}%
\providecommand \translation [1]{[#1]}%
\providecommand \BibitemOpen [0]{}%
\providecommand \bibitemStop [0]{}%
\providecommand \bibitemNoStop [0]{.\EOS\space}%
\providecommand \EOS [0]{\spacefactor3000\relax}%
\providecommand \BibitemShut  [1]{\csname bibitem#1\endcsname}%
\let\auto@bib@innerbib\@empty
\bibitem [{\citenamefont {Berthier}\ and\ \citenamefont
  {Biroli}(2011)}]{berthier11}%
  \BibitemOpen
  \bibfield  {author} {\bibinfo {author} {\bibfnamefont {L.}~\bibnamefont
  {Berthier}}\ and\ \bibinfo {author} {\bibfnamefont {G.}~\bibnamefont
  {Biroli}},\ }\href {\doibase 10.1103/RevModPhys.83.587} {\bibfield  {journal}
  {\bibinfo  {journal} {Rev. Mod. Phys.}\ }\textbf {\bibinfo {volume} {83}},\
  \bibinfo {pages} {587} (\bibinfo {year} {2011})}\BibitemShut {NoStop}%
\bibitem [{\citenamefont {Charbonneau}\ \emph {et~al.}(2017)\citenamefont
  {Charbonneau}, \citenamefont {Kurchan}, \citenamefont {Parisi}, \citenamefont
  {P.},\ and\ \citenamefont {Zamponi}}]{parisi17}%
  \BibitemOpen
  \bibfield  {author} {\bibinfo {author} {\bibfnamefont {P.}~\bibnamefont
  {Charbonneau}}, \bibinfo {author} {\bibfnamefont {J.}~\bibnamefont
  {Kurchan}}, \bibinfo {author} {\bibfnamefont {U.}~\bibnamefont {Parisi},
  \bibfnamefont {G.}}, \bibinfo {author} {\bibnamefont {P.}},\ and\ \bibinfo
  {author} {\bibfnamefont {F.}~\bibnamefont {Zamponi}},\ }\href@noop {}
  {\bibfield  {journal} {\bibinfo  {journal} {Annu. Rev. Condens. Matter
  Phys.}\ }\textbf {\bibinfo {volume} {8}},\ \bibinfo {pages} {265} (\bibinfo
  {year} {2017})}\BibitemShut {NoStop}%
\bibitem [{\citenamefont {Kurchan}\ \emph {et~al.}(2012)\citenamefont
  {Kurchan}, \citenamefont {Parisi},\ and\ \citenamefont
  {Zamponi}}]{kurchan12}%
  \BibitemOpen
  \bibfield  {author} {\bibinfo {author} {\bibfnamefont {J.}~\bibnamefont
  {Kurchan}}, \bibinfo {author} {\bibfnamefont {G.}~\bibnamefont {Parisi}},\
  and\ \bibinfo {author} {\bibfnamefont {F.}~\bibnamefont {Zamponi}},\ }\href
  {\doibase 10.1088/1742-5468/2012/10/p10012} {\bibfield  {journal} {\bibinfo
  {journal} {Journal of Statistical Mechanics: Theory and Experiment}\ }\textbf
  {\bibinfo {volume} {2012}},\ \bibinfo {pages} {P10012} (\bibinfo {year}
  {2012})}\BibitemShut {NoStop}%
\bibitem [{\citenamefont {Biroli}\ and\ \citenamefont
  {M\'ezard}(2001)}]{biroli01}%
  \BibitemOpen
  \bibfield  {author} {\bibinfo {author} {\bibfnamefont {G.}~\bibnamefont
  {Biroli}}\ and\ \bibinfo {author} {\bibfnamefont {M.}~\bibnamefont
  {M\'ezard}},\ }\href {\doibase 10.1103/PhysRevLett.88.025501} {\bibfield
  {journal} {\bibinfo  {journal} {Phys. Rev. Lett.}\ }\textbf {\bibinfo
  {volume} {88}},\ \bibinfo {pages} {025501} (\bibinfo {year}
  {2001})}\BibitemShut {NoStop}%
\bibitem [{\citenamefont {Ciamarra}\ \emph {et~al.}(2003)\citenamefont
  {Ciamarra}, \citenamefont {Tarzia}, \citenamefont {de~Candia},\ and\
  \citenamefont {Coniglio}}]{ciamarra03}%
  \BibitemOpen
  \bibfield  {author} {\bibinfo {author} {\bibfnamefont {M.~P.}\ \bibnamefont
  {Ciamarra}}, \bibinfo {author} {\bibfnamefont {M.}~\bibnamefont {Tarzia}},
  \bibinfo {author} {\bibfnamefont {A.}~\bibnamefont {de~Candia}},\ and\
  \bibinfo {author} {\bibfnamefont {A.}~\bibnamefont {Coniglio}},\ }\href
  {\doibase 10.1103/PhysRevE.67.057105} {\bibfield  {journal} {\bibinfo
  {journal} {Phys. Rev. E}\ }\textbf {\bibinfo {volume} {67}},\ \bibinfo
  {pages} {057105} (\bibinfo {year} {2003})}\BibitemShut {NoStop}%
\bibitem [{\citenamefont {Kagawa}\ \emph {et~al.}(2013)\citenamefont {Kagawa},
  \citenamefont {Sato}, \citenamefont {Miyagawa}, \citenamefont {Kanoda},
  \citenamefont {Tokura}, \citenamefont {Kobayashi}, \citenamefont {Kumai},\
  and\ \citenamefont {Murakami}}]{kagawa13}%
  \BibitemOpen
  \bibfield  {author} {\bibinfo {author} {\bibfnamefont {F.}~\bibnamefont
  {Kagawa}}, \bibinfo {author} {\bibfnamefont {T.}~\bibnamefont {Sato}},
  \bibinfo {author} {\bibfnamefont {K.}~\bibnamefont {Miyagawa}}, \bibinfo
  {author} {\bibfnamefont {K.}~\bibnamefont {Kanoda}}, \bibinfo {author}
  {\bibfnamefont {Y.}~\bibnamefont {Tokura}}, \bibinfo {author} {\bibfnamefont
  {K.}~\bibnamefont {Kobayashi}}, \bibinfo {author} {\bibfnamefont
  {R.}~\bibnamefont {Kumai}},\ and\ \bibinfo {author} {\bibfnamefont
  {Y.}~\bibnamefont {Murakami}},\ }\href@noop {} {\bibfield  {journal}
  {\bibinfo  {journal} {Nature Phys.}\ }\textbf {\bibinfo {volume} {9}},\
  \bibinfo {pages} {419} (\bibinfo {year} {2013})}\BibitemShut {NoStop}%
\bibitem [{\citenamefont {Sasaki}\ \emph {et~al.}(2017)\citenamefont {Sasaki},
  \citenamefont {Hashimoto}, \citenamefont {Kobayashi}, \citenamefont {Itoh},
  \citenamefont {Iguchi}, \citenamefont {Nishio}, \citenamefont {Ikemoto},
  \citenamefont {Moriwaki}, \citenamefont {Yoneyama}, \citenamefont {Watanabe},
  \citenamefont {Ueda}, \citenamefont {Mori}, \citenamefont {Kobayashi},
  \citenamefont {Kumai}, \citenamefont {Murakami}, \citenamefont {M\"uller},\
  and\ \citenamefont {T.}}]{hashimoto17}%
  \BibitemOpen
  \bibfield  {author} {\bibinfo {author} {\bibfnamefont {S.}~\bibnamefont
  {Sasaki}}, \bibinfo {author} {\bibfnamefont {K.}~\bibnamefont {Hashimoto}},
  \bibinfo {author} {\bibfnamefont {R.}~\bibnamefont {Kobayashi}}, \bibinfo
  {author} {\bibfnamefont {K.}~\bibnamefont {Itoh}}, \bibinfo {author}
  {\bibfnamefont {S.}~\bibnamefont {Iguchi}}, \bibinfo {author} {\bibfnamefont
  {Y.}~\bibnamefont {Nishio}}, \bibinfo {author} {\bibfnamefont
  {Y.}~\bibnamefont {Ikemoto}}, \bibinfo {author} {\bibfnamefont
  {T.}~\bibnamefont {Moriwaki}}, \bibinfo {author} {\bibfnamefont
  {N.}~\bibnamefont {Yoneyama}}, \bibinfo {author} {\bibfnamefont
  {M.}~\bibnamefont {Watanabe}}, \bibinfo {author} {\bibfnamefont
  {A.}~\bibnamefont {Ueda}}, \bibinfo {author} {\bibfnamefont {H.}~\bibnamefont
  {Mori}}, \bibinfo {author} {\bibfnamefont {K.}~\bibnamefont {Kobayashi}},
  \bibinfo {author} {\bibfnamefont {R.}~\bibnamefont {Kumai}}, \bibinfo
  {author} {\bibfnamefont {Y.}~\bibnamefont {Murakami}}, \bibinfo {author}
  {\bibfnamefont {J.}~\bibnamefont {M\"uller}},\ and\ \bibinfo {author}
  {\bibfnamefont {S.}~\bibnamefont {T.}},\ }\href@noop {} {\bibfield  {journal}
  {\bibinfo  {journal} {Science}\ }\textbf {\bibinfo {volume} {357}},\ \bibinfo
  {pages} {1381} (\bibinfo {year} {2017})}\BibitemShut {NoStop}%
\bibitem [{\citenamefont {Berthier}\ \emph {et~al.}(2019)\citenamefont
  {Berthier}, \citenamefont {Charbonneau}, \citenamefont {Ninarello},
  \citenamefont {Ozawa},\ and\ \citenamefont {Yaida}}]{berthier19}%
  \BibitemOpen
  \bibfield  {author} {\bibinfo {author} {\bibfnamefont {L.}~\bibnamefont
  {Berthier}}, \bibinfo {author} {\bibfnamefont {P.}~\bibnamefont
  {Charbonneau}}, \bibinfo {author} {\bibfnamefont {A.}~\bibnamefont
  {Ninarello}}, \bibinfo {author} {\bibfnamefont {M.}~\bibnamefont {Ozawa}},\
  and\ \bibinfo {author} {\bibfnamefont {S.}~\bibnamefont {Yaida}},\ }\href
  {\doibase 10.1038/s41467-019-09512-3} {\bibfield  {journal} {\bibinfo
  {journal} {Nature Communications}\ }\textbf {\bibinfo {volume} {10}},\
  \bibinfo {pages} {1508} (\bibinfo {year} {2019})}\BibitemShut {NoStop}%
\bibitem [{\citenamefont {Binder}\ and\ \citenamefont
  {Young}(1986)}]{binder-young}%
  \BibitemOpen
  \bibfield  {author} {\bibinfo {author} {\bibfnamefont {K.}~\bibnamefont
  {Binder}}\ and\ \bibinfo {author} {\bibfnamefont {A.~P.}\ \bibnamefont
  {Young}},\ }\href {\doibase 10.1103/RevModPhys.58.801} {\bibfield  {journal}
  {\bibinfo  {journal} {Rev. Mod. Phys.}\ }\textbf {\bibinfo {volume} {58}},\
  \bibinfo {pages} {801} (\bibinfo {year} {1986})}\BibitemShut {NoStop}%
\bibitem [{\citenamefont {Edwards}\ and\ \citenamefont
  {Anderson}(1975)}]{ea1975}%
  \BibitemOpen
  \bibfield  {author} {\bibinfo {author} {\bibfnamefont {S.~F.}\ \bibnamefont
  {Edwards}}\ and\ \bibinfo {author} {\bibfnamefont {P.}~\bibnamefont
  {Anderson}},\ }\href@noop {} {\bibfield  {journal} {\bibinfo  {journal}
  {Journal of Physics F: Metal Physics}\ }\textbf {\bibinfo {volume} {5}},\
  \bibinfo {pages} {965} (\bibinfo {year} {1975})}\BibitemShut {NoStop}%
\bibitem [{\citenamefont {Ogielski}\ and\ \citenamefont
  {Morgenstern}(1985)}]{ogielski85}%
  \BibitemOpen
  \bibfield  {author} {\bibinfo {author} {\bibfnamefont {A.~T.}\ \bibnamefont
  {Ogielski}}\ and\ \bibinfo {author} {\bibfnamefont {I.}~\bibnamefont
  {Morgenstern}},\ }\href {\doibase 10.1103/PhysRevLett.54.928} {\bibfield
  {journal} {\bibinfo  {journal} {Phys. Rev. Lett.}\ }\textbf {\bibinfo
  {volume} {54}},\ \bibinfo {pages} {928} (\bibinfo {year} {1985})}\BibitemShut
  {NoStop}%
\bibitem [{\citenamefont {Katzgraber}\ \emph {et~al.}(2006)\citenamefont
  {Katzgraber}, \citenamefont {K\"orner},\ and\ \citenamefont
  {Young}}]{katzgraber06}%
  \BibitemOpen
  \bibfield  {author} {\bibinfo {author} {\bibfnamefont {H.~G.}\ \bibnamefont
  {Katzgraber}}, \bibinfo {author} {\bibfnamefont {M.}~\bibnamefont
  {K\"orner}},\ and\ \bibinfo {author} {\bibfnamefont {A.~P.}\ \bibnamefont
  {Young}},\ }\href {\doibase 10.1103/PhysRevB.73.224432} {\bibfield  {journal}
  {\bibinfo  {journal} {Phys. Rev. B}\ }\textbf {\bibinfo {volume} {73}},\
  \bibinfo {pages} {224432} (\bibinfo {year} {2006})}\BibitemShut {NoStop}%
\bibitem [{\citenamefont {Bray}\ and\ \citenamefont
  {Moore}(1985{\natexlab{a}})}]{bray-moore85}%
  \BibitemOpen
  \bibfield  {author} {\bibinfo {author} {\bibfnamefont {A.~J.}\ \bibnamefont
  {Bray}}\ and\ \bibinfo {author} {\bibfnamefont {M.~A.}\ \bibnamefont
  {Moore}},\ }\href {\doibase 10.1103/PhysRevB.31.631} {\bibfield  {journal}
  {\bibinfo  {journal} {Phys. Rev. B}\ }\textbf {\bibinfo {volume} {31}},\
  \bibinfo {pages} {631} (\bibinfo {year} {1985}{\natexlab{a}})}\BibitemShut
  {NoStop}%
\bibitem [{\citenamefont {Bhatt}\ and\ \citenamefont
  {Young}(1988)}]{bhatt-young88}%
  \BibitemOpen
  \bibfield  {author} {\bibinfo {author} {\bibfnamefont {R.~N.}\ \bibnamefont
  {Bhatt}}\ and\ \bibinfo {author} {\bibfnamefont {A.~P.}\ \bibnamefont
  {Young}},\ }\href {\doibase 10.1103/PhysRevB.37.5606} {\bibfield  {journal}
  {\bibinfo  {journal} {Phys. Rev. B}\ }\textbf {\bibinfo {volume} {37}},\
  \bibinfo {pages} {5606} (\bibinfo {year} {1988})}\BibitemShut {NoStop}%
\bibitem [{\citenamefont {Bhatt}\ and\ \citenamefont {Young}(1985)}]{bhatt85}%
  \BibitemOpen
  \bibfield  {author} {\bibinfo {author} {\bibfnamefont {R.~N.}\ \bibnamefont
  {Bhatt}}\ and\ \bibinfo {author} {\bibfnamefont {A.~P.}\ \bibnamefont
  {Young}},\ }\href {\doibase 10.1103/PhysRevLett.54.924} {\bibfield  {journal}
  {\bibinfo  {journal} {Phys. Rev. Lett.}\ }\textbf {\bibinfo {volume} {54}},\
  \bibinfo {pages} {924} (\bibinfo {year} {1985})}\BibitemShut {NoStop}%
\bibitem [{\citenamefont {Kawashima}\ and\ \citenamefont
  {Young}(1996)}]{kawashima96}%
  \BibitemOpen
  \bibfield  {author} {\bibinfo {author} {\bibfnamefont {N.}~\bibnamefont
  {Kawashima}}\ and\ \bibinfo {author} {\bibfnamefont {A.~P.}\ \bibnamefont
  {Young}},\ }\href {\doibase 10.1103/PhysRevB.53.R484} {\bibfield  {journal}
  {\bibinfo  {journal} {Phys. Rev. B}\ }\textbf {\bibinfo {volume} {53}},\
  \bibinfo {pages} {R484} (\bibinfo {year} {1996})}\BibitemShut {NoStop}%
\bibitem [{\citenamefont {Palassini}\ and\ \citenamefont
  {Caracciolo}(1999)}]{palassini99}%
  \BibitemOpen
  \bibfield  {author} {\bibinfo {author} {\bibfnamefont {M.}~\bibnamefont
  {Palassini}}\ and\ \bibinfo {author} {\bibfnamefont {S.}~\bibnamefont
  {Caracciolo}},\ }\href {\doibase 10.1103/PhysRevLett.82.5128} {\bibfield
  {journal} {\bibinfo  {journal} {Phys. Rev. Lett.}\ }\textbf {\bibinfo
  {volume} {82}},\ \bibinfo {pages} {5128} (\bibinfo {year}
  {1999})}\BibitemShut {NoStop}%
\bibitem [{\citenamefont {Mari}\ and\ \citenamefont
  {Campbell}(1999)}]{mari-campbell99}%
  \BibitemOpen
  \bibfield  {author} {\bibinfo {author} {\bibfnamefont {P.~O.}\ \bibnamefont
  {Mari}}\ and\ \bibinfo {author} {\bibfnamefont {I.~A.}\ \bibnamefont
  {Campbell}},\ }\href {\doibase 10.1103/PhysRevE.59.2653} {\bibfield
  {journal} {\bibinfo  {journal} {Phys. Rev. E}\ }\textbf {\bibinfo {volume}
  {59}},\ \bibinfo {pages} {2653} (\bibinfo {year} {1999})}\BibitemShut
  {NoStop}%
\bibitem [{\citenamefont {Ballesteros}\ \emph {et~al.}(2000)\citenamefont
  {Ballesteros}, \citenamefont {Cruz}, \citenamefont {Fern\'andez},
  \citenamefont {Mart\'{\i}n-Mayor}, \citenamefont {Pech}, \citenamefont
  {Ruiz-Lorenzo}, \citenamefont {Taranc\'on}, \citenamefont {T\'ellez},
  \citenamefont {Ullod},\ and\ \citenamefont {Ungil}}]{ballesteros00}%
  \BibitemOpen
  \bibfield  {author} {\bibinfo {author} {\bibfnamefont {H.~G.}\ \bibnamefont
  {Ballesteros}}, \bibinfo {author} {\bibfnamefont {A.}~\bibnamefont {Cruz}},
  \bibinfo {author} {\bibfnamefont {L.~A.}\ \bibnamefont {Fern\'andez}},
  \bibinfo {author} {\bibfnamefont {V.}~\bibnamefont {Mart\'{\i}n-Mayor}},
  \bibinfo {author} {\bibfnamefont {J.}~\bibnamefont {Pech}}, \bibinfo {author}
  {\bibfnamefont {J.~J.}\ \bibnamefont {Ruiz-Lorenzo}}, \bibinfo {author}
  {\bibfnamefont {A.}~\bibnamefont {Taranc\'on}}, \bibinfo {author}
  {\bibfnamefont {P.}~\bibnamefont {T\'ellez}}, \bibinfo {author}
  {\bibfnamefont {C.~L.}\ \bibnamefont {Ullod}},\ and\ \bibinfo {author}
  {\bibfnamefont {C.}~\bibnamefont {Ungil}},\ }\href {\doibase
  10.1103/PhysRevB.62.14237} {\bibfield  {journal} {\bibinfo  {journal} {Phys.
  Rev. B}\ }\textbf {\bibinfo {volume} {62}},\ \bibinfo {pages} {14237}
  (\bibinfo {year} {2000})}\BibitemShut {NoStop}%
\bibitem [{\citenamefont {Nakamura}(2010)}]{nakamura10}%
  \BibitemOpen
  \bibfield  {author} {\bibinfo {author} {\bibfnamefont {T.}~\bibnamefont
  {Nakamura}},\ }\href {\doibase 10.1103/PhysRevB.82.014427} {\bibfield
  {journal} {\bibinfo  {journal} {Phys. Rev. B}\ }\textbf {\bibinfo {volume}
  {82}},\ \bibinfo {pages} {014427} (\bibinfo {year} {2010})}\BibitemShut
  {NoStop}%
\bibitem [{\citenamefont {Young}(1983)}]{young83}%
  \BibitemOpen
  \bibfield  {author} {\bibinfo {author} {\bibfnamefont {A.~P.}\ \bibnamefont
  {Young}},\ }\href {\doibase 10.1103/PhysRevLett.50.917} {\bibfield  {journal}
  {\bibinfo  {journal} {Phys. Rev. Lett.}\ }\textbf {\bibinfo {volume} {50}},\
  \bibinfo {pages} {917} (\bibinfo {year} {1983})}\BibitemShut {NoStop}%
\bibitem [{\citenamefont {Parisi}\ \emph {et~al.}(1998)\citenamefont {Parisi},
  \citenamefont {Ruiz-Lorenzo},\ and\ \citenamefont {Stariolo}}]{parisi98}%
  \BibitemOpen
  \bibfield  {author} {\bibinfo {author} {\bibfnamefont {G.}~\bibnamefont
  {Parisi}}, \bibinfo {author} {\bibfnamefont {J.~J.}\ \bibnamefont
  {Ruiz-Lorenzo}},\ and\ \bibinfo {author} {\bibfnamefont {D.~A.}\ \bibnamefont
  {Stariolo}},\ }\href@noop {} {\bibfield  {journal} {\bibinfo  {journal} {J.
  Phys. A}\ }\textbf {\bibinfo {volume} {31}},\ \bibinfo {pages} {4657}
  (\bibinfo {year} {1998})}\BibitemShut {NoStop}%
\bibitem [{\citenamefont {McMillan}(1983)}]{mcmillan83}%
  \BibitemOpen
  \bibfield  {author} {\bibinfo {author} {\bibfnamefont {W.~L.}\ \bibnamefont
  {McMillan}},\ }\href {\doibase 10.1103/PhysRevB.28.5216} {\bibfield
  {journal} {\bibinfo  {journal} {Phys. Rev. B}\ }\textbf {\bibinfo {volume}
  {28}},\ \bibinfo {pages} {5216} (\bibinfo {year} {1983})}\BibitemShut
  {NoStop}%
\bibitem [{\citenamefont {Houdayer}(2001)}]{houdayer01}%
  \BibitemOpen
  \bibfield  {author} {\bibinfo {author} {\bibfnamefont {J.}~\bibnamefont
  {Houdayer}},\ }\href@noop {} {\bibfield  {journal} {\bibinfo  {journal} {Eur.
  Phys. J. B}\ }\textbf {\bibinfo {volume} {22}},\ \bibinfo {pages} {479}
  (\bibinfo {year} {2001})}\BibitemShut {NoStop}%
\bibitem [{\citenamefont {Almeida}\ and\ \citenamefont
  {Thouless}(1978)}]{almeida78}%
  \BibitemOpen
  \bibfield  {author} {\bibinfo {author} {\bibfnamefont {J.~R. L.~d.}\
  \bibnamefont {Almeida}}\ and\ \bibinfo {author} {\bibfnamefont {D.~J.}\
  \bibnamefont {Thouless}},\ }\href {\doibase 10.1088/0305-4470/11/5/028}
  {\bibfield  {journal} {\bibinfo  {journal} {Journal of Physics A: Math.
  Gen.}\ }\textbf {\bibinfo {volume} {11}} (\bibinfo {year} {1978}),\
  10.1088/0305-4470/11/5/028}\BibitemShut {NoStop}%
\bibitem [{\citenamefont {Larson}\ \emph {et~al.}(2013)\citenamefont {Larson},
  \citenamefont {Katzgraber}, \citenamefont {Moore},\ and\ \citenamefont
  {Young}}]{larson13}%
  \BibitemOpen
  \bibfield  {author} {\bibinfo {author} {\bibfnamefont {D.}~\bibnamefont
  {Larson}}, \bibinfo {author} {\bibfnamefont {H.~G.}\ \bibnamefont
  {Katzgraber}}, \bibinfo {author} {\bibfnamefont {M.~A.}\ \bibnamefont
  {Moore}},\ and\ \bibinfo {author} {\bibfnamefont {A.~P.}\ \bibnamefont
  {Young}},\ }\href {\doibase 10.1103/PhysRevB.87.024414} {\bibfield  {journal}
  {\bibinfo  {journal} {Phys. Rev. B}\ }\textbf {\bibinfo {volume} {87}},\
  \bibinfo {pages} {024414} (\bibinfo {year} {2013})}\BibitemShut {NoStop}%
\bibitem [{\citenamefont {Baity-Jesi}\ \emph {et~al.}(2014)\citenamefont
  {Baity-Jesi}, \citenamefont {Ba{\~{n}}os}, \citenamefont {Cruz},
  \citenamefont {Fernandez}, \citenamefont {Gil-Narvion}, \citenamefont
  {Gordillo-Guerrero}, \citenamefont {I{\~{n}}iguez}, \citenamefont {Maiorano},
  \citenamefont {Mantovani}, \citenamefont {Marinari}, \citenamefont
  {Martin-Mayor}, \citenamefont {Monforte-Garcia}, \citenamefont
  {Mu{\~{n}}oz~Sudup}, \citenamefont {Navarro}, \citenamefont {Parisi},
  \citenamefont {Perez-Gaviro}, \citenamefont {Pivanti}, \citenamefont
  {Ricci-Tersenghi}, \citenamefont {Ruiz-Lorenzo}, \citenamefont {Schifano},
  \citenamefont {Seoane}, \citenamefont {Tarancon}, \citenamefont
  {Tripiccione},\ and\ \citenamefont {Yllanes}}]{baity-jesi14}%
  \BibitemOpen
  \bibfield  {author} {\bibinfo {author} {\bibfnamefont {M.}~\bibnamefont
  {Baity-Jesi}}, \bibinfo {author} {\bibfnamefont {R.~A.}\ \bibnamefont
  {Ba{\~{n}}os}}, \bibinfo {author} {\bibfnamefont {A.}~\bibnamefont {Cruz}},
  \bibinfo {author} {\bibfnamefont {L.~A.}\ \bibnamefont {Fernandez}}, \bibinfo
  {author} {\bibfnamefont {J.~M.}\ \bibnamefont {Gil-Narvion}}, \bibinfo
  {author} {\bibfnamefont {A.}~\bibnamefont {Gordillo-Guerrero}}, \bibinfo
  {author} {\bibfnamefont {D.}~\bibnamefont {I{\~{n}}iguez}}, \bibinfo {author}
  {\bibfnamefont {A.}~\bibnamefont {Maiorano}}, \bibinfo {author}
  {\bibfnamefont {F.}~\bibnamefont {Mantovani}}, \bibinfo {author}
  {\bibfnamefont {E.}~\bibnamefont {Marinari}}, \bibinfo {author}
  {\bibfnamefont {V.}~\bibnamefont {Martin-Mayor}}, \bibinfo {author}
  {\bibfnamefont {J.}~\bibnamefont {Monforte-Garcia}}, \bibinfo {author}
  {\bibfnamefont {A.}~\bibnamefont {Mu{\~{n}}oz~Sudup}}, \bibinfo {author}
  {\bibfnamefont {D.}~\bibnamefont {Navarro}}, \bibinfo {author} {\bibfnamefont
  {G.}~\bibnamefont {Parisi}}, \bibinfo {author} {\bibfnamefont
  {S.}~\bibnamefont {Perez-Gaviro}}, \bibinfo {author} {\bibfnamefont
  {M.}~\bibnamefont {Pivanti}}, \bibinfo {author} {\bibfnamefont
  {F.}~\bibnamefont {Ricci-Tersenghi}}, \bibinfo {author} {\bibfnamefont
  {J.}~\bibnamefont {Ruiz-Lorenzo}}, \bibinfo {author} {\bibfnamefont
  {S.}~\bibnamefont {Schifano}}, \bibinfo {author} {\bibfnamefont
  {B.}~\bibnamefont {Seoane}}, \bibinfo {author} {\bibfnamefont
  {A.}~\bibnamefont {Tarancon}}, \bibinfo {author} {\bibfnamefont
  {R.}~\bibnamefont {Tripiccione}},\ and\ \bibinfo {author} {\bibfnamefont
  {D.}~\bibnamefont {Yllanes}},\ }\href {\doibase
  10.1088/1742-5468/2014/05/p05014} {\bibfield  {journal} {\bibinfo  {journal}
  {Journal of Statistical Mechanics: Theory and Experiment}\ }\textbf {\bibinfo
  {volume} {2014}},\ \bibinfo {pages} {P05014} (\bibinfo {year}
  {2014})}\BibitemShut {NoStop}%
\bibitem [{\citenamefont {H\"oller}\ and\ \citenamefont
  {Read}(2020)}]{holler20}%
  \BibitemOpen
  \bibfield  {author} {\bibinfo {author} {\bibfnamefont {J.}~\bibnamefont
  {H\"oller}}\ and\ \bibinfo {author} {\bibfnamefont {N.}~\bibnamefont
  {Read}},\ }\href {\doibase 10.1103/PhysRevE.101.042114} {\bibfield  {journal}
  {\bibinfo  {journal} {Phys. Rev. E}\ }\textbf {\bibinfo {volume} {101}},\
  \bibinfo {pages} {042114} (\bibinfo {year} {2020})}\BibitemShut {NoStop}%
\bibitem [{\citenamefont {Paga}\ \emph {et~al.}(2021)\citenamefont {Paga},
  \citenamefont {Zhai}, \citenamefont {Baity-Jesi}, \citenamefont {Calore},
  \citenamefont {Cruz}, \citenamefont {Fernandez}, \citenamefont {Gil-Narvion},
  \citenamefont {Gonzalez-Adalid~Pemartin}, \citenamefont {Gordillo-Guerrero},
  \citenamefont {I{\~{n}}iguez}, \citenamefont {Maiorano}, \citenamefont
  {Marinari}, \citenamefont {Martin-Mayor}, \citenamefont {Moreno-Gordo},
  \citenamefont {Mu{\~{n}}oz-Sudupe}, \citenamefont {Navarro}, \citenamefont
  {Orbach}, \citenamefont {Parisi}, \citenamefont {Perez-Gaviro}, \citenamefont
  {Ricci-Tersenghi}, \citenamefont {Ruiz-Lorenzo}, \citenamefont {Schifano},
  \citenamefont {Schlagel}, \citenamefont {Seoane}, \citenamefont {Tarancon},
  \citenamefont {Tripiccione},\ and\ \citenamefont {Yllanes}}]{paga21}%
  \BibitemOpen
  \bibfield  {author} {\bibinfo {author} {\bibfnamefont {I.}~\bibnamefont
  {Paga}}, \bibinfo {author} {\bibfnamefont {Q.}~\bibnamefont {Zhai}}, \bibinfo
  {author} {\bibfnamefont {M.}~\bibnamefont {Baity-Jesi}}, \bibinfo {author}
  {\bibfnamefont {E.}~\bibnamefont {Calore}}, \bibinfo {author} {\bibfnamefont
  {A.}~\bibnamefont {Cruz}}, \bibinfo {author} {\bibfnamefont {L.~A.}\
  \bibnamefont {Fernandez}}, \bibinfo {author} {\bibfnamefont {J.}~\bibnamefont
  {Gil-Narvion}}, \bibinfo {author} {\bibfnamefont {I.}~\bibnamefont
  {Gonzalez-Adalid~Pemartin}}, \bibinfo {author} {\bibfnamefont
  {A.}~\bibnamefont {Gordillo-Guerrero}}, \bibinfo {author} {\bibfnamefont
  {D.}~\bibnamefont {I{\~{n}}iguez}}, \bibinfo {author} {\bibfnamefont
  {A.}~\bibnamefont {Maiorano}}, \bibinfo {author} {\bibfnamefont
  {E.}~\bibnamefont {Marinari}}, \bibinfo {author} {\bibfnamefont
  {V.}~\bibnamefont {Martin-Mayor}}, \bibinfo {author} {\bibfnamefont
  {J.}~\bibnamefont {Moreno-Gordo}}, \bibinfo {author} {\bibfnamefont
  {A.}~\bibnamefont {Mu{\~{n}}oz-Sudupe}}, \bibinfo {author} {\bibfnamefont
  {D.}~\bibnamefont {Navarro}}, \bibinfo {author} {\bibfnamefont {R.~L.}\
  \bibnamefont {Orbach}}, \bibinfo {author} {\bibfnamefont {G.}~\bibnamefont
  {Parisi}}, \bibinfo {author} {\bibfnamefont {S.}~\bibnamefont
  {Perez-Gaviro}}, \bibinfo {author} {\bibfnamefont {F.}~\bibnamefont
  {Ricci-Tersenghi}}, \bibinfo {author} {\bibfnamefont {J.~J.}\ \bibnamefont
  {Ruiz-Lorenzo}}, \bibinfo {author} {\bibfnamefont {S.}~\bibnamefont
  {Schifano}}, \bibinfo {author} {\bibfnamefont {D.~L.}\ \bibnamefont
  {Schlagel}}, \bibinfo {author} {\bibfnamefont {B.}~\bibnamefont {Seoane}},
  \bibinfo {author} {\bibfnamefont {A.}~\bibnamefont {Tarancon}}, \bibinfo
  {author} {\bibfnamefont {R.}~\bibnamefont {Tripiccione}},\ and\ \bibinfo
  {author} {\bibfnamefont {D.}~\bibnamefont {Yllanes}},\ }\href {\doibase
  10.1088/1742-5468/abdfca} {\bibfield  {journal} {\bibinfo  {journal} {Journal
  of Statistical Mechanics: Theory and Experiment}\ }\textbf {\bibinfo {volume}
  {2021}},\ \bibinfo {pages} {033301} (\bibinfo {year} {2021})}\BibitemShut
  {NoStop}%
\bibitem [{\citenamefont {Suzuki}(1977)}]{suzuki77}%
  \BibitemOpen
  \bibfield  {author} {\bibinfo {author} {\bibfnamefont {M.}~\bibnamefont
  {Suzuki}},\ }\href@noop {} {\bibfield  {journal} {\bibinfo  {journal} {Prog.
  Theor. Phys.}\ }\textbf {\bibinfo {volume} {58}},\ \bibinfo {pages} {1151}
  (\bibinfo {year} {1977})}\BibitemShut {NoStop}%
\bibitem [{\citenamefont {Gunnarsson}\ \emph {et~al.}(1991)\citenamefont
  {Gunnarsson}, \citenamefont {Svedlindh}, \citenamefont {Nordblad},
  \citenamefont {Lundgren}, \citenamefont {Aruga},\ and\ \citenamefont
  {Ito}}]{gunnarsson91}%
  \BibitemOpen
  \bibfield  {author} {\bibinfo {author} {\bibfnamefont {K.}~\bibnamefont
  {Gunnarsson}}, \bibinfo {author} {\bibfnamefont {P.}~\bibnamefont
  {Svedlindh}}, \bibinfo {author} {\bibfnamefont {P.}~\bibnamefont {Nordblad}},
  \bibinfo {author} {\bibfnamefont {L.}~\bibnamefont {Lundgren}}, \bibinfo
  {author} {\bibfnamefont {H.}~\bibnamefont {Aruga}},\ and\ \bibinfo {author}
  {\bibfnamefont {A.}~\bibnamefont {Ito}},\ }\href {\doibase
  10.1103/PhysRevB.43.8199} {\bibfield  {journal} {\bibinfo  {journal} {Phys.
  Rev. B}\ }\textbf {\bibinfo {volume} {43}},\ \bibinfo {pages} {8199}
  (\bibinfo {year} {1991})}\BibitemShut {NoStop}%
\bibitem [{\citenamefont {Hasenbusch}\ \emph {et~al.}(2008)\citenamefont
  {Hasenbusch}, \citenamefont {Pelissetto},\ and\ \citenamefont
  {Vicari}}]{hasenbusch2008}%
  \BibitemOpen
  \bibfield  {author} {\bibinfo {author} {\bibfnamefont {M.}~\bibnamefont
  {Hasenbusch}}, \bibinfo {author} {\bibfnamefont {A.}~\bibnamefont
  {Pelissetto}},\ and\ \bibinfo {author} {\bibfnamefont {E.}~\bibnamefont
  {Vicari}},\ }\href {\doibase 10.1103/PhysRevB.78.214205} {\bibfield
  {journal} {\bibinfo  {journal} {Phys. Rev. B}\ }\textbf {\bibinfo {volume}
  {78}},\ \bibinfo {pages} {214205} (\bibinfo {year} {2008})}\BibitemShut
  {NoStop}%
\bibitem [{\citenamefont {Baity-Jesi}\ \emph {et~al.}(2013)\citenamefont
  {Baity-Jesi}, \citenamefont {Ba\~nos}, \citenamefont {Cruz}, \citenamefont
  {Fernandez}, \citenamefont {Gil-Narvion}, \citenamefont {Gordillo-Guerrero},
  \citenamefont {I\~niguez}, \citenamefont {Maiorano}, \citenamefont
  {Mantovani}, \citenamefont {Marinari}, \citenamefont {Martin-Mayor},
  \citenamefont {Monforte-Garcia}, \citenamefont {Sudupe}, \citenamefont
  {Navarro}, \citenamefont {Parisi}, \citenamefont {Perez-Gaviro},
  \citenamefont {Pivanti}, \citenamefont {Ricci-Tersenghi}, \citenamefont
  {Ruiz-Lorenzo}, \citenamefont {Schifano}, \citenamefont {Seoane},
  \citenamefont {Tarancon}, \citenamefont {Tripiccione},\ and\ \citenamefont
  {Yllanes}}]{baity-jesi13}%
  \BibitemOpen
  \bibfield  {author} {\bibinfo {author} {\bibfnamefont {M.}~\bibnamefont
  {Baity-Jesi}}, \bibinfo {author} {\bibfnamefont {R.~A.}\ \bibnamefont
  {Ba\~nos}}, \bibinfo {author} {\bibfnamefont {A.}~\bibnamefont {Cruz}},
  \bibinfo {author} {\bibfnamefont {L.~A.}\ \bibnamefont {Fernandez}}, \bibinfo
  {author} {\bibfnamefont {J.~M.}\ \bibnamefont {Gil-Narvion}}, \bibinfo
  {author} {\bibfnamefont {A.}~\bibnamefont {Gordillo-Guerrero}}, \bibinfo
  {author} {\bibfnamefont {D.}~\bibnamefont {I\~niguez}}, \bibinfo {author}
  {\bibfnamefont {A.}~\bibnamefont {Maiorano}}, \bibinfo {author}
  {\bibfnamefont {F.}~\bibnamefont {Mantovani}}, \bibinfo {author}
  {\bibfnamefont {E.}~\bibnamefont {Marinari}}, \bibinfo {author}
  {\bibfnamefont {V.}~\bibnamefont {Martin-Mayor}}, \bibinfo {author}
  {\bibfnamefont {J.}~\bibnamefont {Monforte-Garcia}}, \bibinfo {author}
  {\bibfnamefont {A.~M.~n.}\ \bibnamefont {Sudupe}}, \bibinfo {author}
  {\bibfnamefont {D.}~\bibnamefont {Navarro}}, \bibinfo {author} {\bibfnamefont
  {G.}~\bibnamefont {Parisi}}, \bibinfo {author} {\bibfnamefont
  {S.}~\bibnamefont {Perez-Gaviro}}, \bibinfo {author} {\bibfnamefont
  {M.}~\bibnamefont {Pivanti}}, \bibinfo {author} {\bibfnamefont
  {F.}~\bibnamefont {Ricci-Tersenghi}}, \bibinfo {author} {\bibfnamefont
  {J.~J.}\ \bibnamefont {Ruiz-Lorenzo}}, \bibinfo {author} {\bibfnamefont
  {S.~F.}\ \bibnamefont {Schifano}}, \bibinfo {author} {\bibfnamefont
  {B.}~\bibnamefont {Seoane}}, \bibinfo {author} {\bibfnamefont
  {A.}~\bibnamefont {Tarancon}}, \bibinfo {author} {\bibfnamefont
  {R.}~\bibnamefont {Tripiccione}},\ and\ \bibinfo {author} {\bibfnamefont
  {D.}~\bibnamefont {Yllanes}} (\bibinfo {collaboration} {Janus
  Collaboration}),\ }\href {\doibase 10.1103/PhysRevB.88.224416} {\bibfield
  {journal} {\bibinfo  {journal} {Phys. Rev. B}\ }\textbf {\bibinfo {volume}
  {88}},\ \bibinfo {pages} {224416} (\bibinfo {year} {2013})}\BibitemShut
  {NoStop}%
\bibitem [{\citenamefont {Cannella}\ and\ \citenamefont
  {Mydosh}(1972)}]{cannella72}%
  \BibitemOpen
  \bibfield  {author} {\bibinfo {author} {\bibfnamefont {V.}~\bibnamefont
  {Cannella}}\ and\ \bibinfo {author} {\bibfnamefont {J.~A.}\ \bibnamefont
  {Mydosh}},\ }\href {\doibase 10.1103/PhysRevB.6.4220} {\bibfield  {journal}
  {\bibinfo  {journal} {Phys. Rev. B}\ }\textbf {\bibinfo {volume} {6}},\
  \bibinfo {pages} {4220} (\bibinfo {year} {1972})}\BibitemShut {NoStop}%
\bibitem [{\citenamefont {Gingras}\ \emph {et~al.}(1997)\citenamefont
  {Gingras}, \citenamefont {Stager}, \citenamefont {Raju}, \citenamefont
  {Gaulin},\ and\ \citenamefont {Greedan}}]{gingras97}%
  \BibitemOpen
  \bibfield  {author} {\bibinfo {author} {\bibfnamefont {M.~J.~P.}\
  \bibnamefont {Gingras}}, \bibinfo {author} {\bibfnamefont {C.~V.}\
  \bibnamefont {Stager}}, \bibinfo {author} {\bibfnamefont {N.~P.}\
  \bibnamefont {Raju}}, \bibinfo {author} {\bibfnamefont {B.~D.}\ \bibnamefont
  {Gaulin}},\ and\ \bibinfo {author} {\bibfnamefont {J.~E.}\ \bibnamefont
  {Greedan}},\ }\href {\doibase 10.1103/PhysRevLett.78.947} {\bibfield
  {journal} {\bibinfo  {journal} {Phys. Rev. Lett.}\ }\textbf {\bibinfo
  {volume} {78}},\ \bibinfo {pages} {947} (\bibinfo {year} {1997})}\BibitemShut
  {NoStop}%
\bibitem [{\citenamefont {Kadowaki}\ and\ \citenamefont
  {Nishimori}(1998)}]{annealing}%
  \BibitemOpen
  \bibfield  {author} {\bibinfo {author} {\bibfnamefont {T.}~\bibnamefont
  {Kadowaki}}\ and\ \bibinfo {author} {\bibfnamefont {H.}~\bibnamefont
  {Nishimori}},\ }\href {\doibase 10.1103/PhysRevE.58.5355} {\bibfield
  {journal} {\bibinfo  {journal} {Phys. Rev. E}\ }\textbf {\bibinfo {volume}
  {58}},\ \bibinfo {pages} {5355} (\bibinfo {year} {1998})}\BibitemShut
  {NoStop}%
\bibitem [{\citenamefont {Isono}\ \emph {et~al.}(2013)\citenamefont {Isono},
  \citenamefont {Kamo}, \citenamefont {Ueda}, \citenamefont {Takahashi},
  \citenamefont {Nakao}, \citenamefont {Kumai}, \citenamefont {Nakao},
  \citenamefont {Kobayashi}, \citenamefont {Murakami},\ and\ \citenamefont
  {Mori}}]{hydrogen}%
  \BibitemOpen
  \bibfield  {author} {\bibinfo {author} {\bibfnamefont {T.}~\bibnamefont
  {Isono}}, \bibinfo {author} {\bibfnamefont {H.}~\bibnamefont {Kamo}},
  \bibinfo {author} {\bibfnamefont {A.}~\bibnamefont {Ueda}}, \bibinfo {author}
  {\bibfnamefont {K.}~\bibnamefont {Takahashi}}, \bibinfo {author}
  {\bibfnamefont {A.}~\bibnamefont {Nakao}}, \bibinfo {author} {\bibfnamefont
  {R.}~\bibnamefont {Kumai}}, \bibinfo {author} {\bibfnamefont
  {H.}~\bibnamefont {Nakao}}, \bibinfo {author} {\bibfnamefont
  {K.}~\bibnamefont {Kobayashi}}, \bibinfo {author} {\bibfnamefont
  {Y.}~\bibnamefont {Murakami}},\ and\ \bibinfo {author} {\bibfnamefont
  {H.}~\bibnamefont {Mori}},\ }\href@noop {} {\bibfield  {journal} {\bibinfo
  {journal} {Nature Communications}\ }\textbf {\bibinfo {volume} {4}},\
  \bibinfo {pages} {1344} (\bibinfo {year} {2013})}\BibitemShut {NoStop}%
\bibitem [{\citenamefont {Hotta}(2010)}]{ch}%
  \BibitemOpen
  \bibfield  {author} {\bibinfo {author} {\bibfnamefont {C.}~\bibnamefont
  {Hotta}},\ }\href {\doibase 10.1103/PhysRevB.82.241104} {\bibfield  {journal}
  {\bibinfo  {journal} {Phys. Rev. B}\ }\textbf {\bibinfo {volume} {82}},\
  \bibinfo {pages} {241104} (\bibinfo {year} {2010})}\BibitemShut {NoStop}%
\bibitem [{\citenamefont {Naka}\ and\ \citenamefont {Ishihara}(2010)}]{naka}%
  \BibitemOpen
  \bibfield  {author} {\bibinfo {author} {\bibfnamefont {M.}~\bibnamefont
  {Naka}}\ and\ \bibinfo {author} {\bibfnamefont {S.}~\bibnamefont
  {Ishihara}},\ }\href@noop {} {\bibfield  {journal} {\bibinfo  {journal} {J.
  Phys. Soc. Jpn.}\ }\textbf {\bibinfo {volume} {79}},\ \bibinfo {pages}
  {063707} (\bibinfo {year} {2010})}\BibitemShut {NoStop}%
\bibitem [{\citenamefont {Lunkenheimer}\ \emph {et~al.}(2012)\citenamefont
  {Lunkenheimer}, \citenamefont {M\"uller}, \citenamefont {Krohns},
  \citenamefont {Schrettle}, \citenamefont {Loidl}, \citenamefont {Hartmann},
  \citenamefont {Rommel}, \citenamefont {de~Souza}, \citenamefont {Hotta},
  \citenamefont {Schlueter},\ and\ \citenamefont {M.}}]{peter}%
  \BibitemOpen
  \bibfield  {author} {\bibinfo {author} {\bibfnamefont {P.}~\bibnamefont
  {Lunkenheimer}}, \bibinfo {author} {\bibfnamefont {J.~S.}\ \bibnamefont
  {M\"uller}}, \bibinfo {author} {\bibfnamefont {S.}~\bibnamefont {Krohns}},
  \bibinfo {author} {\bibfnamefont {F.}~\bibnamefont {Schrettle}}, \bibinfo
  {author} {\bibfnamefont {A.}~\bibnamefont {Loidl}}, \bibinfo {author}
  {\bibfnamefont {B.}~\bibnamefont {Hartmann}}, \bibinfo {author}
  {\bibfnamefont {R.}~\bibnamefont {Rommel}}, \bibinfo {author} {\bibfnamefont
  {M.}~\bibnamefont {de~Souza}}, \bibinfo {author} {\bibfnamefont
  {C.}~\bibnamefont {Hotta}}, \bibinfo {author} {\bibfnamefont
  {J.}~\bibnamefont {Schlueter}},\ and\ \bibinfo {author} {\bibfnamefont
  {L.}~\bibnamefont {M.}},\ }\href@noop {} {\bibfield  {journal} {\bibinfo
  {journal} {Nature Materials}\ }\textbf {\bibinfo {volume} {11}},\ \bibinfo
  {pages} {755} (\bibinfo {year} {2012})}\BibitemShut {NoStop}%
\bibitem [{\citenamefont {Tokura}(2006)}]{Tokura}%
  \BibitemOpen
  \bibfield  {author} {\bibinfo {author} {\bibfnamefont {Y.}~\bibnamefont
  {Tokura}},\ }\href@noop {} {\bibfield  {journal} {\bibinfo  {journal} {Rep.
  Prog. Phys.}\ }\textbf {\bibinfo {volume} {6}},\ \bibinfo {pages} {200}
  (\bibinfo {year} {2006})}\BibitemShut {NoStop}%
\bibitem [{\citenamefont {Harris}\ \emph {et~al.}(1997)\citenamefont {Harris},
  \citenamefont {Bramwell}, \citenamefont {McMorrow}, \citenamefont {Zeiske},\
  and\ \citenamefont {Godfrey}}]{spinice}%
  \BibitemOpen
  \bibfield  {author} {\bibinfo {author} {\bibfnamefont {M.~J.}\ \bibnamefont
  {Harris}}, \bibinfo {author} {\bibfnamefont {S.~T.}\ \bibnamefont
  {Bramwell}}, \bibinfo {author} {\bibfnamefont {D.~F.}\ \bibnamefont
  {McMorrow}}, \bibinfo {author} {\bibfnamefont {T.}~\bibnamefont {Zeiske}},\
  and\ \bibinfo {author} {\bibfnamefont {K.~W.}\ \bibnamefont {Godfrey}},\
  }\href {\doibase 10.1103/PhysRevLett.79.2554} {\bibfield  {journal} {\bibinfo
   {journal} {Phys. Rev. Lett.}\ }\textbf {\bibinfo {volume} {79}},\ \bibinfo
  {pages} {2554} (\bibinfo {year} {1997})}\BibitemShut {NoStop}%
\bibitem [{\citenamefont {Ramirez}\ \emph {et~al.}(1999)\citenamefont
  {Ramirez}, \citenamefont {Hayashi}, \citenamefont {Cava}, \citenamefont
  {Siddharthan},\ and\ \citenamefont {Shastry}}]{spinice2}%
  \BibitemOpen
  \bibfield  {author} {\bibinfo {author} {\bibfnamefont {A.~P.}\ \bibnamefont
  {Ramirez}}, \bibinfo {author} {\bibfnamefont {A.}~\bibnamefont {Hayashi}},
  \bibinfo {author} {\bibfnamefont {R.~J.}\ \bibnamefont {Cava}}, \bibinfo
  {author} {\bibfnamefont {R.}~\bibnamefont {Siddharthan}},\ and\ \bibinfo
  {author} {\bibfnamefont {B.~S.}\ \bibnamefont {Shastry}},\ }\href@noop {}
  {\bibfield  {journal} {\bibinfo  {journal} {Nature}\ }\textbf {\bibinfo
  {volume} {399}},\ \bibinfo {pages} {333} (\bibinfo {year}
  {1999})}\BibitemShut {NoStop}%
\bibitem [{\citenamefont {Balents}(2010)}]{sl}%
  \BibitemOpen
  \bibfield  {author} {\bibinfo {author} {\bibfnamefont {L.}~\bibnamefont
  {Balents}},\ }\href@noop {} {\bibfield  {journal} {\bibinfo  {journal}
  {Nature}\ }\textbf {\bibinfo {volume} {464}},\ \bibinfo {pages} {199}
  (\bibinfo {year} {2010})}\BibitemShut {NoStop}%
\bibitem [{\citenamefont {Imada}(1987{\natexlab{a}})}]{imada87}%
  \BibitemOpen
  \bibfield  {author} {\bibinfo {author} {\bibfnamefont {M.}~\bibnamefont
  {Imada}},\ }\href@noop {} {\bibfield  {journal} {\bibinfo  {journal} {J.
  Phys. Soc. Jpn.}\ }\textbf {\bibinfo {volume} {56}},\ \bibinfo {pages} {881}
  (\bibinfo {year} {1987}{\natexlab{a}})}\BibitemShut {NoStop}%
\bibitem [{\citenamefont {Vojta}(2010)}]{vojta10}%
  \BibitemOpen
  \bibfield  {author} {\bibinfo {author} {\bibfnamefont {T.}~\bibnamefont
  {Vojta}},\ }\href@noop {} {\bibfield  {journal} {\bibinfo  {journal} {J. Low
  Temp. Phys.}\ }\textbf {\bibinfo {volume} {161}},\ \bibinfo {pages} {299}
  (\bibinfo {year} {2010})}\BibitemShut {NoStop}%
\bibitem [{\citenamefont {Watanabe}\ \emph {et~al.}(2014)\citenamefont
  {Watanabe}, \citenamefont {Kawamura}, \citenamefont {Nakano},\ and\
  \citenamefont {Sakai}}]{kawamura14-1}%
  \BibitemOpen
  \bibfield  {author} {\bibinfo {author} {\bibfnamefont {K.}~\bibnamefont
  {Watanabe}}, \bibinfo {author} {\bibfnamefont {H.}~\bibnamefont {Kawamura}},
  \bibinfo {author} {\bibfnamefont {H.}~\bibnamefont {Nakano}},\ and\ \bibinfo
  {author} {\bibfnamefont {T.}~\bibnamefont {Sakai}},\ }\href@noop {}
  {\bibfield  {journal} {\bibinfo  {journal} {J. Phys. Soc. Jpn.}\ }\textbf
  {\bibinfo {volume} {83}},\ \bibinfo {pages} {034714} (\bibinfo {year}
  {2014})}\BibitemShut {NoStop}%
\bibitem [{\citenamefont {Shimokawa}\ \emph {et~al.}(2015)\citenamefont
  {Shimokawa}, \citenamefont {Watanabe},\ and\ \citenamefont
  {Kawamura}}]{kawamura15}%
  \BibitemOpen
  \bibfield  {author} {\bibinfo {author} {\bibfnamefont {T.}~\bibnamefont
  {Shimokawa}}, \bibinfo {author} {\bibfnamefont {K.}~\bibnamefont
  {Watanabe}},\ and\ \bibinfo {author} {\bibfnamefont {H.}~\bibnamefont
  {Kawamura}},\ }\href {\doibase 10.1103/PhysRevB.92.134407} {\bibfield
  {journal} {\bibinfo  {journal} {Phys. Rev. B}\ }\textbf {\bibinfo {volume}
  {92}},\ \bibinfo {pages} {134407} (\bibinfo {year} {2015})}\BibitemShut
  {NoStop}%
\bibitem [{\citenamefont {Kimchi}\ \emph {et~al.}(2018)\citenamefont {Kimchi},
  \citenamefont {Nahum},\ and\ \citenamefont {Senthil}}]{kimuchi18}%
  \BibitemOpen
  \bibfield  {author} {\bibinfo {author} {\bibfnamefont {I.}~\bibnamefont
  {Kimchi}}, \bibinfo {author} {\bibfnamefont {A.}~\bibnamefont {Nahum}},\ and\
  \bibinfo {author} {\bibfnamefont {T.}~\bibnamefont {Senthil}},\ }\href
  {\doibase 10.1103/PhysRevX.8.031028} {\bibfield  {journal} {\bibinfo
  {journal} {Phys. Rev. X}\ }\textbf {\bibinfo {volume} {8}},\ \bibinfo {pages}
  {031028} (\bibinfo {year} {2018})}\BibitemShut {NoStop}%
\bibitem [{\citenamefont {Liu}\ \emph {et~al.}(2018)\citenamefont {Liu},
  \citenamefont {Shao}, \citenamefont {Lin}, \citenamefont {Guo},\ and\
  \citenamefont {Sandvik}}]{shao18}%
  \BibitemOpen
  \bibfield  {author} {\bibinfo {author} {\bibfnamefont {L.}~\bibnamefont
  {Liu}}, \bibinfo {author} {\bibfnamefont {H.}~\bibnamefont {Shao}}, \bibinfo
  {author} {\bibfnamefont {Y.-C.}\ \bibnamefont {Lin}}, \bibinfo {author}
  {\bibfnamefont {W.}~\bibnamefont {Guo}},\ and\ \bibinfo {author}
  {\bibfnamefont {A.~W.}\ \bibnamefont {Sandvik}},\ }\href {\doibase
  10.1103/PhysRevX.8.041040} {\bibfield  {journal} {\bibinfo  {journal} {Phys.
  Rev. X}\ }\textbf {\bibinfo {volume} {8}},\ \bibinfo {pages} {041040}
  (\bibinfo {year} {2018})}\BibitemShut {NoStop}%
\bibitem [{\citenamefont {Wu}\ and\ \citenamefont {Gong}(2019)}]{sheng19}%
  \BibitemOpen
  \bibfield  {author} {\bibinfo {author} {\bibfnamefont {H.-Q.}\ \bibnamefont
  {Wu}}\ and\ \bibinfo {author} {\bibfnamefont {D.~N.}\ \bibnamefont {Gong},
  \bibfnamefont {S.-S.and~Sheng}},\ }\href {\doibase
  10.1103/PhysRevB.99.085141} {\bibfield  {journal} {\bibinfo  {journal} {Phys.
  Rev. B}\ }\textbf {\bibinfo {volume} {99}},\ \bibinfo {pages} {085141}
  (\bibinfo {year} {2019})}\BibitemShut {NoStop}%
\bibitem [{\citenamefont {Hasenbusch}\ \emph {et~al.}(2007)\citenamefont
  {Hasenbusch}, \citenamefont {Toldin}, \citenamefont {Pelissetto},\ and\
  \citenamefont {Vicari}}]{Hasenbusch2007}%
  \BibitemOpen
  \bibfield  {author} {\bibinfo {author} {\bibfnamefont {M.}~\bibnamefont
  {Hasenbusch}}, \bibinfo {author} {\bibfnamefont {F.~P.}\ \bibnamefont
  {Toldin}}, \bibinfo {author} {\bibfnamefont {A.}~\bibnamefont {Pelissetto}},\
  and\ \bibinfo {author} {\bibfnamefont {E.}~\bibnamefont {Vicari}},\ }\href
  {\doibase 10.1103/PhysRevB.76.094402} {\bibfield  {journal} {\bibinfo
  {journal} {Phys. Rev. B}\ }\textbf {\bibinfo {volume} {76}},\ \bibinfo
  {pages} {094402} (\bibinfo {year} {2007})}\BibitemShut {NoStop}%
\bibitem [{\citenamefont {Wannier}(1950)}]{wannier}%
  \BibitemOpen
  \bibfield  {author} {\bibinfo {author} {\bibfnamefont {G.~H.}\ \bibnamefont
  {Wannier}},\ }\href {\doibase 10.1103/PhysRev.79.357} {\bibfield  {journal}
  {\bibinfo  {journal} {Phys. Rev.}\ }\textbf {\bibinfo {volume} {79}},\
  \bibinfo {pages} {357} (\bibinfo {year} {1950})}\BibitemShut {NoStop}%
\bibitem [{\citenamefont {Villain}(1980)}]{villain80}%
  \BibitemOpen
  \bibfield  {author} {\bibinfo {author} {\bibfnamefont {J.}~\bibnamefont
  {Villain}},\ }\href@noop {} {\bibfield  {journal} {\bibinfo  {journal} {J.
  Phys. France}\ }\textbf {\bibinfo {volume} {41}},\ \bibinfo {pages} {1263}
  (\bibinfo {year} {1980})}\BibitemShut {NoStop}%
\bibitem [{\citenamefont {Gull}\ \emph {et~al.}(2011)\citenamefont {Gull},
  \citenamefont {Millis}, \citenamefont {Lichtenstein}, \citenamefont
  {Rubtsov}, \citenamefont {Troyer},\ and\ \citenamefont {Werner}}]{ct-qmc}%
  \BibitemOpen
  \bibfield  {author} {\bibinfo {author} {\bibfnamefont {E.}~\bibnamefont
  {Gull}}, \bibinfo {author} {\bibfnamefont {A.~J.}\ \bibnamefont {Millis}},
  \bibinfo {author} {\bibfnamefont {A.~I.}\ \bibnamefont {Lichtenstein}},
  \bibinfo {author} {\bibfnamefont {A.~N.}\ \bibnamefont {Rubtsov}}, \bibinfo
  {author} {\bibfnamefont {M.}~\bibnamefont {Troyer}},\ and\ \bibinfo {author}
  {\bibfnamefont {P.}~\bibnamefont {Werner}},\ }\href {\doibase
  10.1103/RevModPhys.83.349} {\bibfield  {journal} {\bibinfo  {journal} {Rev.
  Mod. Phys.}\ }\textbf {\bibinfo {volume} {83}},\ \bibinfo {pages} {349}
  (\bibinfo {year} {2011})}\BibitemShut {NoStop}%
\bibitem [{\citenamefont {Mitsumoto}\ and\ \citenamefont
  {Kawamura}(2021)}]{mitsumoto21}%
  \BibitemOpen
  \bibfield  {author} {\bibinfo {author} {\bibfnamefont {K.}~\bibnamefont
  {Mitsumoto}}\ and\ \bibinfo {author} {\bibfnamefont {H.}~\bibnamefont
  {Kawamura}},\ }\href {\doibase 10.1103/PhysRevB.104.184432} {\bibfield
  {journal} {\bibinfo  {journal} {Phys. Rev. B}\ }\textbf {\bibinfo {volume}
  {104}},\ \bibinfo {pages} {184432} (\bibinfo {year} {2021})}\BibitemShut
  {NoStop}%
\bibitem [{\citenamefont {Isakov}\ and\ \citenamefont
  {Moessner}(2003)}]{isakov03}%
  \BibitemOpen
  \bibfield  {author} {\bibinfo {author} {\bibfnamefont {S.~V.}\ \bibnamefont
  {Isakov}}\ and\ \bibinfo {author} {\bibfnamefont {R.}~\bibnamefont
  {Moessner}},\ }\href {\doibase 10.1103/PhysRevB.68.104409} {\bibfield
  {journal} {\bibinfo  {journal} {Phys. Rev. B}\ }\textbf {\bibinfo {volume}
  {68}},\ \bibinfo {pages} {104409} (\bibinfo {year} {2003})}\BibitemShut
  {NoStop}%
\bibitem [{\citenamefont {Imada}(1987{\natexlab{b}})}]{imada87S}%
  \BibitemOpen
  \bibfield  {author} {\bibinfo {author} {\bibfnamefont {M.}~\bibnamefont
  {Imada}},\ }\href@noop {} {\bibfield  {journal} {\bibinfo  {journal} {M.
  Suzuki (eds) Quantum Monte Carlo Methods in Equilibrium and Nonequilibrium
  Systems, Springer, Berlin, Heidelberg}\ ,\ \bibinfo {pages} {125}} (\bibinfo
  {year} {1987}{\natexlab{b}})}\BibitemShut {NoStop}%
\bibitem [{\citenamefont {Moessner}\ \emph {et~al.}(2000)\citenamefont
  {Moessner}, \citenamefont {Sondhi},\ and\ \citenamefont
  {Chandra}}]{moessner00}%
  \BibitemOpen
  \bibfield  {author} {\bibinfo {author} {\bibfnamefont {R.}~\bibnamefont
  {Moessner}}, \bibinfo {author} {\bibfnamefont {S.~L.}\ \bibnamefont
  {Sondhi}},\ and\ \bibinfo {author} {\bibfnamefont {P.}~\bibnamefont
  {Chandra}},\ }\href {\doibase 10.1103/PhysRevLett.84.4457} {\bibfield
  {journal} {\bibinfo  {journal} {Phys. Rev. Lett.}\ }\textbf {\bibinfo
  {volume} {84}},\ \bibinfo {pages} {4457} (\bibinfo {year}
  {2000})}\BibitemShut {NoStop}%
\bibitem [{\citenamefont {Blankschtein}\ \emph {et~al.}(1984)\citenamefont
  {Blankschtein}, \citenamefont {Ma}, \citenamefont {Berker}, \citenamefont
  {Grest},\ and\ \citenamefont {Soukoulis}}]{blankschtein84}%
  \BibitemOpen
  \bibfield  {author} {\bibinfo {author} {\bibfnamefont {D.}~\bibnamefont
  {Blankschtein}}, \bibinfo {author} {\bibfnamefont {M.}~\bibnamefont {Ma}},
  \bibinfo {author} {\bibfnamefont {A.~N.}\ \bibnamefont {Berker}}, \bibinfo
  {author} {\bibfnamefont {G.~S.}\ \bibnamefont {Grest}},\ and\ \bibinfo
  {author} {\bibfnamefont {C.~M.}\ \bibnamefont {Soukoulis}},\ }\href {\doibase
  10.1103/PhysRevB.29.5250} {\bibfield  {journal} {\bibinfo  {journal} {Phys.
  Rev. B}\ }\textbf {\bibinfo {volume} {29}},\ \bibinfo {pages} {5250}
  (\bibinfo {year} {1984})}\BibitemShut {NoStop}%
\bibitem [{Note1()}]{Note1}%
  \BibitemOpen
  \bibinfo {note} {Since for $R=0$ we have translational symmetry, we
  identified A/B/C sublattices for each replica to take the proper overlap
  between the three sublattices.}\BibitemShut {Stop}%
\bibitem [{\citenamefont {Marinari}\ \emph {et~al.}(1998)\citenamefont
  {Marinari}, \citenamefont {Parisi},\ and\ \citenamefont
  {Ruiz-Lorenzo}}]{marinari98}%
  \BibitemOpen
  \bibfield  {author} {\bibinfo {author} {\bibfnamefont {E.}~\bibnamefont
  {Marinari}}, \bibinfo {author} {\bibfnamefont {G.}~\bibnamefont {Parisi}},\
  and\ \bibinfo {author} {\bibfnamefont {J.~J.}\ \bibnamefont {Ruiz-Lorenzo}},\
  }\href {\doibase 10.1103/PhysRevB.58.14852} {\bibfield  {journal} {\bibinfo
  {journal} {Phys. Rev. B}\ }\textbf {\bibinfo {volume} {58}},\ \bibinfo
  {pages} {14852} (\bibinfo {year} {1998})}\BibitemShut {NoStop}%
\bibitem [{\citenamefont {Katzgraber}\ \emph {et~al.}(2001)\citenamefont
  {Katzgraber}, \citenamefont {Palassini},\ and\ \citenamefont
  {Young}}]{katzgraber01}%
  \BibitemOpen
  \bibfield  {author} {\bibinfo {author} {\bibfnamefont {H.~G.}\ \bibnamefont
  {Katzgraber}}, \bibinfo {author} {\bibfnamefont {M.}~\bibnamefont
  {Palassini}},\ and\ \bibinfo {author} {\bibfnamefont {A.~P.}\ \bibnamefont
  {Young}},\ }\href {\doibase 10.1103/PhysRevB.63.184422} {\bibfield  {journal}
  {\bibinfo  {journal} {Phys. Rev. B}\ }\textbf {\bibinfo {volume} {63}},\
  \bibinfo {pages} {184422} (\bibinfo {year} {2001})}\BibitemShut {NoStop}%
\bibitem [{\citenamefont {Castellani}\ and\ \citenamefont
  {Cavagna}(2005)}]{castellani2005}%
  \BibitemOpen
  \bibfield  {author} {\bibinfo {author} {\bibfnamefont {T.}~\bibnamefont
  {Castellani}}\ and\ \bibinfo {author} {\bibfnamefont {A.}~\bibnamefont
  {Cavagna}},\ }\href {\doibase 10.1088/1742-5468/2005/05/p05012} {\bibfield
  {journal} {\bibinfo  {journal} {Journal of Statistical Mechanics: Theory and
  Experiment}\ }\textbf {\bibinfo {volume} {2005}},\ \bibinfo {pages} {P05012}
  (\bibinfo {year} {2005})}\BibitemShut {NoStop}%
\bibitem [{\citenamefont {Kaul}(2015)}]{kaul15}%
  \BibitemOpen
  \bibfield  {author} {\bibinfo {author} {\bibfnamefont {R.~K.}\ \bibnamefont
  {Kaul}},\ }\href {\doibase 10.1103/PhysRevLett.115.157202} {\bibfield
  {journal} {\bibinfo  {journal} {Phys. Rev. Lett.}\ }\textbf {\bibinfo
  {volume} {115}},\ \bibinfo {pages} {157202} (\bibinfo {year}
  {2015})}\BibitemShut {NoStop}%
\bibitem [{\citenamefont {Shirakura}\ \emph {et~al.}(2014)\citenamefont
  {Shirakura}, \citenamefont {Matsubara},\ and\ \citenamefont
  {Suzuki}}]{matsubara14}%
  \BibitemOpen
  \bibfield  {author} {\bibinfo {author} {\bibfnamefont {T.}~\bibnamefont
  {Shirakura}}, \bibinfo {author} {\bibfnamefont {F.}~\bibnamefont
  {Matsubara}},\ and\ \bibinfo {author} {\bibfnamefont {N.}~\bibnamefont
  {Suzuki}},\ }\href {\doibase 10.1103/PhysRevB.90.144410} {\bibfield
  {journal} {\bibinfo  {journal} {Phys. Rev. B}\ }\textbf {\bibinfo {volume}
  {90}},\ \bibinfo {pages} {144410} (\bibinfo {year} {2014})}\BibitemShut
  {NoStop}%
\bibitem [{\citenamefont {Guo}\ \emph {et~al.}(1994)\citenamefont {Guo},
  \citenamefont {Bhatt},\ and\ \citenamefont {Huse}}]{guo94}%
  \BibitemOpen
  \bibfield  {author} {\bibinfo {author} {\bibfnamefont {M.}~\bibnamefont
  {Guo}}, \bibinfo {author} {\bibfnamefont {R.~N.}\ \bibnamefont {Bhatt}},\
  and\ \bibinfo {author} {\bibfnamefont {D.~A.}\ \bibnamefont {Huse}},\ }\href
  {\doibase 10.1103/PhysRevLett.72.4137} {\bibfield  {journal} {\bibinfo
  {journal} {Phys. Rev. Lett.}\ }\textbf {\bibinfo {volume} {72}},\ \bibinfo
  {pages} {4137} (\bibinfo {year} {1994})}\BibitemShut {NoStop}%
\bibitem [{\citenamefont {Cugliandolo}\ \emph {et~al.}(1997)\citenamefont
  {Cugliandolo}, \citenamefont {Kurchan},\ and\ \citenamefont
  {Peliti}}]{cugliandolo1997}%
  \BibitemOpen
  \bibfield  {author} {\bibinfo {author} {\bibfnamefont {L.~F.}\ \bibnamefont
  {Cugliandolo}}, \bibinfo {author} {\bibfnamefont {J.}~\bibnamefont
  {Kurchan}},\ and\ \bibinfo {author} {\bibfnamefont {L.}~\bibnamefont
  {Peliti}},\ }\href {\doibase 10.1103/PhysRevE.55.3898} {\bibfield  {journal}
  {\bibinfo  {journal} {Phys. Rev. E}\ }\textbf {\bibinfo {volume} {55}},\
  \bibinfo {pages} {3898} (\bibinfo {year} {1997})}\BibitemShut {NoStop}%
\bibitem [{\citenamefont {Fisher}\ and\ \citenamefont {Huse}(1988)}]{fisher88}%
  \BibitemOpen
  \bibfield  {author} {\bibinfo {author} {\bibfnamefont {D.~S.}\ \bibnamefont
  {Fisher}}\ and\ \bibinfo {author} {\bibfnamefont {D.~A.}\ \bibnamefont
  {Huse}},\ }\href {\doibase 10.1103/PhysRevB.38.386} {\bibfield  {journal}
  {\bibinfo  {journal} {Phys. Rev. B}\ }\textbf {\bibinfo {volume} {38}},\
  \bibinfo {pages} {386} (\bibinfo {year} {1988})}\BibitemShut {NoStop}%
\bibitem [{\citenamefont {Hartmann}\ and\ \citenamefont
  {Young}(2002)}]{hartmann2002}%
  \BibitemOpen
  \bibfield  {author} {\bibinfo {author} {\bibfnamefont {A.~K.}\ \bibnamefont
  {Hartmann}}\ and\ \bibinfo {author} {\bibfnamefont {A.~P.}\ \bibnamefont
  {Young}},\ }\href {\doibase 10.1103/PhysRevB.66.094419} {\bibfield  {journal}
  {\bibinfo  {journal} {Phys. Rev. B}\ }\textbf {\bibinfo {volume} {66}},\
  \bibinfo {pages} {094419} (\bibinfo {year} {2002})}\BibitemShut {NoStop}%
\bibitem [{\citenamefont {Imry}\ and\ \citenamefont {Ma}(1975)}]{imry-ma}%
  \BibitemOpen
  \bibfield  {author} {\bibinfo {author} {\bibfnamefont {Y.}~\bibnamefont
  {Imry}}\ and\ \bibinfo {author} {\bibfnamefont {S.-K.}\ \bibnamefont {Ma}},\
  }\href {\doibase 10.1103/PhysRevLett.35.1399} {\bibfield  {journal} {\bibinfo
   {journal} {Phys. Rev. Lett.}\ }\textbf {\bibinfo {volume} {35}},\ \bibinfo
  {pages} {1399} (\bibinfo {year} {1975})}\BibitemShut {NoStop}%
\bibitem [{\citenamefont {Aizenman}\ and\ \citenamefont
  {Wehr}(1989)}]{aizenman89}%
  \BibitemOpen
  \bibfield  {author} {\bibinfo {author} {\bibfnamefont {M.}~\bibnamefont
  {Aizenman}}\ and\ \bibinfo {author} {\bibfnamefont {J.}~\bibnamefont
  {Wehr}},\ }\href {\doibase 10.1103/PhysRevLett.62.2503} {\bibfield  {journal}
  {\bibinfo  {journal} {Phys. Rev. Lett.}\ }\textbf {\bibinfo {volume} {62}},\
  \bibinfo {pages} {2503} (\bibinfo {year} {1989})}\BibitemShut {NoStop}%
\bibitem [{\citenamefont {Bray}\ and\ \citenamefont
  {Moore}(1985{\natexlab{b}})}]{bray-moore}%
  \BibitemOpen
  \bibfield  {author} {\bibinfo {author} {\bibfnamefont {A.~J.}\ \bibnamefont
  {Bray}}\ and\ \bibinfo {author} {\bibfnamefont {M.~A.}\ \bibnamefont
  {Moore}},\ }\href@noop {} {\bibfield  {journal} {\bibinfo  {journal} {J.
  Phys. C: Solid State Phys.}\ }\textbf {\bibinfo {volume} {18}},\ \bibinfo
  {pages} {L927} (\bibinfo {year} {1985}{\natexlab{b}})}\BibitemShut {NoStop}%
\bibitem [{\citenamefont {T.}\ and\ \citenamefont
  {Villain}(1988)}]{nattermann}%
  \BibitemOpen
  \bibfield  {author} {\bibinfo {author} {\bibfnamefont {N.}~\bibnamefont
  {T.}}\ and\ \bibinfo {author} {\bibfnamefont {J.}~\bibnamefont {Villain}},\
  }\href@noop {} {\bibfield  {journal} {\bibinfo  {journal} {Phase
  Transitions}\ }\textbf {\bibinfo {volume} {11}},\ \bibinfo {pages} {5}
  (\bibinfo {year} {1988})}\BibitemShut {NoStop}%
\bibitem [{\citenamefont {Bray}\ and\ \citenamefont
  {Moore}(1984)}]{bray-moore84}%
  \BibitemOpen
  \bibfield  {author} {\bibinfo {author} {\bibfnamefont {A.~J.}\ \bibnamefont
  {Bray}}\ and\ \bibinfo {author} {\bibfnamefont {M.~A.}\ \bibnamefont
  {Moore}},\ }\href {\doibase 10.1088/0022-3719/17/18/004} {\bibfield
  {journal} {\bibinfo  {journal} {Journal of Physics C: Solid State Physics}\
  }\textbf {\bibinfo {volume} {17}},\ \bibinfo {pages} {L463} (\bibinfo {year}
  {1984})}\BibitemShut {NoStop}%
\bibitem [{\citenamefont {Huse}\ and\ \citenamefont
  {Morgenstern}(1985)}]{huse85}%
  \BibitemOpen
  \bibfield  {author} {\bibinfo {author} {\bibfnamefont {D.~A.}\ \bibnamefont
  {Huse}}\ and\ \bibinfo {author} {\bibfnamefont {I.}~\bibnamefont
  {Morgenstern}},\ }\href {\doibase 10.1103/PhysRevB.32.3032} {\bibfield
  {journal} {\bibinfo  {journal} {Phys. Rev. B}\ }\textbf {\bibinfo {volume}
  {32}},\ \bibinfo {pages} {3032} (\bibinfo {year} {1985})}\BibitemShut
  {NoStop}%
\bibitem [{\citenamefont {McMillan}(1984)}]{mcmillan84}%
  \BibitemOpen
  \bibfield  {author} {\bibinfo {author} {\bibfnamefont {W.~L.}\ \bibnamefont
  {McMillan}},\ }\href {\doibase 10.1103/PhysRevB.30.476} {\bibfield  {journal}
  {\bibinfo  {journal} {Phys. Rev. B}\ }\textbf {\bibinfo {volume} {30}},\
  \bibinfo {pages} {476} (\bibinfo {year} {1984})}\BibitemShut {NoStop}%
\bibitem [{\citenamefont {Parisi}(1979)}]{parisi79}%
  \BibitemOpen
  \bibfield  {author} {\bibinfo {author} {\bibfnamefont {G.}~\bibnamefont
  {Parisi}},\ }\href {\doibase 10.1103/PhysRevLett.43.1754} {\bibfield
  {journal} {\bibinfo  {journal} {Phys. Rev. Lett.}\ }\textbf {\bibinfo
  {volume} {43}},\ \bibinfo {pages} {1754} (\bibinfo {year}
  {1979})}\BibitemShut {NoStop}%
\bibitem [{\citenamefont {Parisi}(1983)}]{parisi83}%
  \BibitemOpen
  \bibfield  {author} {\bibinfo {author} {\bibfnamefont {G.}~\bibnamefont
  {Parisi}},\ }\href {\doibase 10.1103/PhysRevLett.50.1946} {\bibfield
  {journal} {\bibinfo  {journal} {Phys. Rev. Lett.}\ }\textbf {\bibinfo
  {volume} {50}},\ \bibinfo {pages} {1946} (\bibinfo {year}
  {1983})}\BibitemShut {NoStop}%
\bibitem [{\citenamefont {Parisi}(1980{\natexlab{a}})}]{parisi80}%
  \BibitemOpen
  \bibfield  {author} {\bibinfo {author} {\bibfnamefont {G.}~\bibnamefont
  {Parisi}},\ }\href {\doibase 10.1088/0305-4470/13/3/042} {\bibfield
  {journal} {\bibinfo  {journal} {Journal of Physics A: Mathematical and
  General}\ }\textbf {\bibinfo {volume} {13}},\ \bibinfo {pages} {1101}
  (\bibinfo {year} {1980}{\natexlab{a}})}\BibitemShut {NoStop}%
\bibitem [{\citenamefont {Parisi}(1980{\natexlab{b}})}]{parisi80-2}%
  \BibitemOpen
  \bibfield  {author} {\bibinfo {author} {\bibfnamefont {G.}~\bibnamefont
  {Parisi}},\ }\href {\doibase 10.1088/0305-4470/13/5/047} {\bibfield
  {journal} {\bibinfo  {journal} {Journal of Physics A: Mathematical and
  General}\ }\textbf {\bibinfo {volume} {13}},\ \bibinfo {pages} {1887}
  (\bibinfo {year} {1980}{\natexlab{b}})}\BibitemShut {NoStop}%
\bibitem [{\citenamefont {Parisi}(1980{\natexlab{c}})}]{parisi80-3}%
  \BibitemOpen
  \bibfield  {author} {\bibinfo {author} {\bibfnamefont {G.}~\bibnamefont
  {Parisi}},\ }\href {\doibase 10.1088/0305-4470/13/4/009} {\bibfield
  {journal} {\bibinfo  {journal} {Journal of Physics A: Mathematical and
  General}\ }\textbf {\bibinfo {volume} {13}},\ \bibinfo {pages} {L115}
  (\bibinfo {year} {1980}{\natexlab{c}})}\BibitemShut {NoStop}%
\bibitem [{\citenamefont {Kirkpatrick}\ and\ \citenamefont
  {Sherrington}(1978)}]{SK1978}%
  \BibitemOpen
  \bibfield  {author} {\bibinfo {author} {\bibfnamefont {S.}~\bibnamefont
  {Kirkpatrick}}\ and\ \bibinfo {author} {\bibfnamefont {D.}~\bibnamefont
  {Sherrington}},\ }\href {\doibase 10.1103/PhysRevB.17.4384} {\bibfield
  {journal} {\bibinfo  {journal} {Phys. Rev. B}\ }\textbf {\bibinfo {volume}
  {17}},\ \bibinfo {pages} {4384} (\bibinfo {year} {1978})}\BibitemShut
  {NoStop}%
\bibitem [{\citenamefont {Sherrington}\ and\ \citenamefont
  {Kirkpatrick}(1975)}]{SK1975}%
  \BibitemOpen
  \bibfield  {author} {\bibinfo {author} {\bibfnamefont {D.}~\bibnamefont
  {Sherrington}}\ and\ \bibinfo {author} {\bibfnamefont {S.}~\bibnamefont
  {Kirkpatrick}},\ }\href {\doibase 10.1103/PhysRevLett.35.1792} {\bibfield
  {journal} {\bibinfo  {journal} {Phys. Rev. Lett.}\ }\textbf {\bibinfo
  {volume} {35}},\ \bibinfo {pages} {1792} (\bibinfo {year}
  {1975})}\BibitemShut {NoStop}%
\bibitem [{\citenamefont {Alvarez~Ba{\~{n}}os}\ \emph
  {et~al.}(2010)\citenamefont {Alvarez~Ba{\~{n}}os}, \citenamefont {Cruz},
  \citenamefont {Fernandez}, \citenamefont {Gil-Narvion}, \citenamefont
  {Gordillo-Guerrero}, \citenamefont {Guidetti}, \citenamefont {Maiorano},
  \citenamefont {Mantovani}, \citenamefont {Marinari}, \citenamefont
  {Martin-Mayor}, \citenamefont {Monforte-Garcia}, \citenamefont
  {Mu{\~{n}}oz~Sudupe}, \citenamefont {Navarro}, \citenamefont {Parisi},
  \citenamefont {Perez-Gaviro}, \citenamefont {Ruiz-Lorenzo}, \citenamefont
  {Schifano}, \citenamefont {Seoane}, \citenamefont {Tarancon}, \citenamefont
  {Tripiccione},\ and\ \citenamefont {Yllanes}}]{alvarez_ba_os2010}%
  \BibitemOpen
  \bibfield  {author} {\bibinfo {author} {\bibfnamefont {R.}~\bibnamefont
  {Alvarez~Ba{\~{n}}os}}, \bibinfo {author} {\bibfnamefont {A.}~\bibnamefont
  {Cruz}}, \bibinfo {author} {\bibfnamefont {L.~A.}\ \bibnamefont {Fernandez}},
  \bibinfo {author} {\bibfnamefont {J.~M.}\ \bibnamefont {Gil-Narvion}},
  \bibinfo {author} {\bibfnamefont {A.}~\bibnamefont {Gordillo-Guerrero}},
  \bibinfo {author} {\bibfnamefont {M.}~\bibnamefont {Guidetti}}, \bibinfo
  {author} {\bibfnamefont {A.}~\bibnamefont {Maiorano}}, \bibinfo {author}
  {\bibfnamefont {F.}~\bibnamefont {Mantovani}}, \bibinfo {author}
  {\bibfnamefont {E.}~\bibnamefont {Marinari}}, \bibinfo {author}
  {\bibfnamefont {V.}~\bibnamefont {Martin-Mayor}}, \bibinfo {author}
  {\bibfnamefont {J.}~\bibnamefont {Monforte-Garcia}}, \bibinfo {author}
  {\bibfnamefont {A.}~\bibnamefont {Mu{\~{n}}oz~Sudupe}}, \bibinfo {author}
  {\bibfnamefont {D.}~\bibnamefont {Navarro}}, \bibinfo {author} {\bibfnamefont
  {G.}~\bibnamefont {Parisi}}, \bibinfo {author} {\bibfnamefont
  {S.}~\bibnamefont {Perez-Gaviro}}, \bibinfo {author} {\bibfnamefont {J.~J.}\
  \bibnamefont {Ruiz-Lorenzo}}, \bibinfo {author} {\bibfnamefont {S.~F.}\
  \bibnamefont {Schifano}}, \bibinfo {author} {\bibfnamefont {B.}~\bibnamefont
  {Seoane}}, \bibinfo {author} {\bibfnamefont {A.}~\bibnamefont {Tarancon}},
  \bibinfo {author} {\bibfnamefont {R.}~\bibnamefont {Tripiccione}},\ and\
  \bibinfo {author} {\bibfnamefont {D.}~\bibnamefont {Yllanes}},\ }\href
  {\doibase 10.1088/1742-5468/2010/06/p06026} {\bibfield  {journal} {\bibinfo
  {journal} {Journal of Statistical Mechanics: Theory and Experiment}\ }\textbf
  {\bibinfo {volume} {2010}},\ \bibinfo {pages} {P06026} (\bibinfo {year}
  {2010})}\BibitemShut {NoStop}%
\bibitem [{\citenamefont {Dutta}\ \emph {et~al.}(2015)\citenamefont {Dutta},
  \citenamefont {Aeppli}, \citenamefont {Chakrabarti},\ and\ \citenamefont
  {Divakaran}}]{dutta}%
  \BibitemOpen
  \bibfield  {author} {\bibinfo {author} {\bibfnamefont {A.}~\bibnamefont
  {Dutta}}, \bibinfo {author} {\bibfnamefont {G.}~\bibnamefont {Aeppli}},
  \bibinfo {author} {\bibfnamefont {B.~K.}\ \bibnamefont {Chakrabarti}},\ and\
  \bibinfo {author} {\bibfnamefont {U.}~\bibnamefont {Divakaran}},\ }\href@noop
  {} {\  (\bibinfo {year} {2015})}\BibitemShut {NoStop}%
\bibitem [{\citenamefont {Kitaev}(2006)}]{kitaev2006}%
  \BibitemOpen
  \bibfield  {author} {\bibinfo {author} {\bibfnamefont {A.}~\bibnamefont
  {Kitaev}},\ }\href {\doibase 10.1016/j.aop.2005.10.005} {\bibfield  {journal}
  {\bibinfo  {journal} {Ann. Phys.}\ }\textbf {\bibinfo {volume} {321}},\
  \bibinfo {pages} {2} (\bibinfo {year} {2006})}\BibitemShut {NoStop}%
\bibitem [{\citenamefont {Liao}\ \emph {et~al.}(2017)\citenamefont {Liao},
  \citenamefont {Xie}, \citenamefont {Chen}, \citenamefont {Liu}, \citenamefont
  {Xie}, \citenamefont {Huang}, \citenamefont {Normand},\ and\ \citenamefont
  {Xiang}}]{liao2017}%
  \BibitemOpen
  \bibfield  {author} {\bibinfo {author} {\bibfnamefont {H.~J.}\ \bibnamefont
  {Liao}}, \bibinfo {author} {\bibfnamefont {Z.~Y.}\ \bibnamefont {Xie}},
  \bibinfo {author} {\bibfnamefont {J.}~\bibnamefont {Chen}}, \bibinfo {author}
  {\bibfnamefont {Z.~Y.}\ \bibnamefont {Liu}}, \bibinfo {author} {\bibfnamefont
  {H.~D.}\ \bibnamefont {Xie}}, \bibinfo {author} {\bibfnamefont {R.~Z.}\
  \bibnamefont {Huang}}, \bibinfo {author} {\bibfnamefont {B.}~\bibnamefont
  {Normand}},\ and\ \bibinfo {author} {\bibfnamefont {T.}~\bibnamefont
  {Xiang}},\ }\href {\doibase 10.1103/PhysRevLett.118.137202} {\bibfield
  {journal} {\bibinfo  {journal} {Phys. Rev. Lett.}\ }\textbf {\bibinfo
  {volume} {118}},\ \bibinfo {pages} {137202} (\bibinfo {year}
  {2017})}\BibitemShut {NoStop}%
\bibitem [{\citenamefont {Yan}\ \emph {et~al.}(2011)\citenamefont {Yan},
  \citenamefont {Huse},\ and\ \citenamefont {White}}]{yan2011}%
  \BibitemOpen
  \bibfield  {author} {\bibinfo {author} {\bibfnamefont {S.}~\bibnamefont
  {Yan}}, \bibinfo {author} {\bibfnamefont {D.~A.}\ \bibnamefont {Huse}},\ and\
  \bibinfo {author} {\bibfnamefont {S.~R.}\ \bibnamefont {White}},\ }\href
  {\doibase 10.1126/science.1201080} {\bibfield  {journal} {\bibinfo  {journal}
  {Science}\ }\textbf {\bibinfo {volume} {332}},\ \bibinfo {pages} {1173}
  (\bibinfo {year} {2011})}\BibitemShut {NoStop}%
\bibitem [{\citenamefont {Depenbrock}\ \emph {et~al.}(2012)\citenamefont
  {Depenbrock}, \citenamefont {McCulloch},\ and\ \citenamefont
  {Schollw\"ock}}]{depenbrock2012}%
  \BibitemOpen
  \bibfield  {author} {\bibinfo {author} {\bibfnamefont {S.}~\bibnamefont
  {Depenbrock}}, \bibinfo {author} {\bibfnamefont {I.~P.}\ \bibnamefont
  {McCulloch}},\ and\ \bibinfo {author} {\bibfnamefont {U.}~\bibnamefont
  {Schollw\"ock}},\ }\href {\doibase 10.1103/PhysRevLett.109.067201} {\bibfield
   {journal} {\bibinfo  {journal} {Phys. Rev. Lett.}\ }\textbf {\bibinfo
  {volume} {109}},\ \bibinfo {pages} {067201} (\bibinfo {year}
  {2012})}\BibitemShut {NoStop}%
\bibitem [{\citenamefont {Nishimoto}\ \emph {et~al.}(2013)\citenamefont
  {Nishimoto}, \citenamefont {Shibata},\ and\ \citenamefont
  {Hotta}}]{nishimoto2013}%
  \BibitemOpen
  \bibfield  {author} {\bibinfo {author} {\bibfnamefont {S.}~\bibnamefont
  {Nishimoto}}, \bibinfo {author} {\bibfnamefont {N.}~\bibnamefont {Shibata}},\
  and\ \bibinfo {author} {\bibfnamefont {C.}~\bibnamefont {Hotta}},\ }\href
  {\doibase 10.1038/ncomms3287 (2013).} {\bibfield  {journal} {\bibinfo
  {journal} {Nature Communications}\ }\textbf {\bibinfo {volume} {4}},\
  \bibinfo {pages} {2287} (\bibinfo {year} {2013})}\BibitemShut {NoStop}%
\bibitem [{\citenamefont {Kaneko}\ \emph {et~al.}(2014)\citenamefont {Kaneko},
  \citenamefont {Morita},\ and\ \citenamefont {Imada}}]{kaneko2014}%
  \BibitemOpen
  \bibfield  {author} {\bibinfo {author} {\bibfnamefont {R.}~\bibnamefont
  {Kaneko}}, \bibinfo {author} {\bibfnamefont {S.}~\bibnamefont {Morita}},\
  and\ \bibinfo {author} {\bibfnamefont {M.}~\bibnamefont {Imada}},\ }\href
  {\doibase 10.7566/JPSJ.83.093707} {\bibfield  {journal} {\bibinfo  {journal}
  {J. Phys. Soc. Jpn.}\ }\textbf {\bibinfo {volume} {83}},\ \bibinfo {pages}
  {093707} (\bibinfo {year} {2014})}\BibitemShut {NoStop}%
\bibitem [{\citenamefont {Iqbal}\ \emph {et~al.}(2016)\citenamefont {Iqbal},
  \citenamefont {Hu}, \citenamefont {Thomale}, \citenamefont {Poilblanc},\ and\
  \citenamefont {Becca}}]{yasir2016}%
  \BibitemOpen
  \bibfield  {author} {\bibinfo {author} {\bibfnamefont {Y.}~\bibnamefont
  {Iqbal}}, \bibinfo {author} {\bibfnamefont {W.-J.}\ \bibnamefont {Hu}},
  \bibinfo {author} {\bibfnamefont {R.}~\bibnamefont {Thomale}}, \bibinfo
  {author} {\bibfnamefont {D.}~\bibnamefont {Poilblanc}},\ and\ \bibinfo
  {author} {\bibfnamefont {F.}~\bibnamefont {Becca}},\ }\href {\doibase
  10.1103/PhysRevB.93.144411} {\bibfield  {journal} {\bibinfo  {journal} {Phys.
  Rev. B}\ }\textbf {\bibinfo {volume} {93}},\ \bibinfo {pages} {144411}
  (\bibinfo {year} {2016})}\BibitemShut {NoStop}%
\bibitem [{\citenamefont {Hu}\ \emph {et~al.}(2019)\citenamefont {Hu},
  \citenamefont {Zhu}, \citenamefont {Eggert},\ and\ \citenamefont
  {He}}]{shijie2019}%
  \BibitemOpen
  \bibfield  {author} {\bibinfo {author} {\bibfnamefont {S.}~\bibnamefont
  {Hu}}, \bibinfo {author} {\bibfnamefont {W.}~\bibnamefont {Zhu}}, \bibinfo
  {author} {\bibfnamefont {S.}~\bibnamefont {Eggert}},\ and\ \bibinfo {author}
  {\bibfnamefont {Y.-C.}\ \bibnamefont {He}},\ }\href {\doibase
  10.1103/PhysRevLett.123.207203} {\bibfield  {journal} {\bibinfo  {journal}
  {Phys. Rev. Lett.}\ }\textbf {\bibinfo {volume} {123}},\ \bibinfo {pages}
  {207203} (\bibinfo {year} {2019})}\BibitemShut {NoStop}%
\bibitem [{\citenamefont {Hu}\ \emph {et~al.}(2013)\citenamefont {Hu},
  \citenamefont {Becca}, \citenamefont {Parola},\ and\ \citenamefont
  {Sorella}}]{becca2013}%
  \BibitemOpen
  \bibfield  {author} {\bibinfo {author} {\bibfnamefont {W.-J.}\ \bibnamefont
  {Hu}}, \bibinfo {author} {\bibfnamefont {F.}~\bibnamefont {Becca}}, \bibinfo
  {author} {\bibfnamefont {A.}~\bibnamefont {Parola}},\ and\ \bibinfo {author}
  {\bibfnamefont {S.}~\bibnamefont {Sorella}},\ }\href {\doibase
  10.1103/PhysRevB.88.060402} {\bibfield  {journal} {\bibinfo  {journal} {Phys.
  Rev. B}\ }\textbf {\bibinfo {volume} {88}},\ \bibinfo {pages} {060402}
  (\bibinfo {year} {2013})}\BibitemShut {NoStop}%
\bibitem [{\citenamefont {Nomura}\ and\ \citenamefont
  {Imada}(2020)}]{nomura2020}%
  \BibitemOpen
  \bibfield  {author} {\bibinfo {author} {\bibfnamefont {Y.}~\bibnamefont
  {Nomura}}\ and\ \bibinfo {author} {\bibfnamefont {M.}~\bibnamefont {Imada}},\
  }\href@noop {} {\bibfield  {journal} {\bibinfo  {journal} {arXiv:2005.14142,
  Phys. Rev. X, in press}\ } (\bibinfo {year} {2020})}\BibitemShut {NoStop}%
\bibitem [{\citenamefont {Martin-Mayor}\ and\ \citenamefont
  {Hen}(2015)}]{martin-mayor15}%
  \BibitemOpen
  \bibfield  {author} {\bibinfo {author} {\bibfnamefont {V.}~\bibnamefont
  {Martin-Mayor}}\ and\ \bibinfo {author} {\bibfnamefont {I.}~\bibnamefont
  {Hen}},\ }\href {\doibase 10.1038/srep15324} {\bibfield  {journal} {\bibinfo
  {journal} {Scientific Reports}\ }\textbf {\bibinfo {volume} {5}},\ \bibinfo
  {pages} {15324} (\bibinfo {year} {2015})}\BibitemShut {NoStop}%
\bibitem [{\citenamefont {Katzgraber}\ \emph {et~al.}(2014)\citenamefont
  {Katzgraber}, \citenamefont {Hamze},\ and\ \citenamefont
  {Andrist}}]{katzgraber14}%
  \BibitemOpen
  \bibfield  {author} {\bibinfo {author} {\bibfnamefont {H.~G.}\ \bibnamefont
  {Katzgraber}}, \bibinfo {author} {\bibfnamefont {F.}~\bibnamefont {Hamze}},\
  and\ \bibinfo {author} {\bibfnamefont {R.~S.}\ \bibnamefont {Andrist}},\
  }\href {\doibase 10.1103/PhysRevX.4.021008} {\bibfield  {journal} {\bibinfo
  {journal} {Phys. Rev. X}\ }\textbf {\bibinfo {volume} {4}},\ \bibinfo {pages}
  {021008} (\bibinfo {year} {2014})}\BibitemShut {NoStop}%
\bibitem [{\citenamefont {Parisi}\ and\ \citenamefont
  {Virasoro}(1989)}]{parisi89}%
  \BibitemOpen
  \bibfield  {author} {\bibinfo {author} {\bibfnamefont {G.}~\bibnamefont
  {Parisi}}\ and\ \bibinfo {author} {\bibfnamefont {M.~A.}\ \bibnamefont
  {Virasoro}},\ }\href {\doibase 10.1051/jphys:0198900500220331700} {\bibfield
  {journal} {\bibinfo  {journal} {J. Phys. France}\ }\textbf {\bibinfo {volume}
  {50}},\ \bibinfo {pages} {3317} (\bibinfo {year} {1989})}\BibitemShut
  {NoStop}%
\end{thebibliography}%

\end{document}